\documentclass[draft]{agujournal2019}
\usepackage{url}
\usepackage{lineno}
\usepackage[inline]{trackchanges} 
\usepackage{soul}
\usepackage{amsmath}
\usepackage{xcolor}
\usepackage{subcaption}
\usepackage{mathtools}
\usepackage{natbib}

\journalname{Astrobiology}

\begin{document} 

   \title{Land Fraction Diversity on Earth-like Planets and Implications for their Habitability}

   \authors{Dennis H\"oning\affil{1,*}, Tilman Spohn\affil{2}}
   \affiliation{1}{Potsdam-Institute for Climate Impact Research, Potsdam, Germany}
   \affiliation{2}{International Space Science Institute, Bern, Switzerland}
   \affiliation{*}{Corresponding Author (dennis.hoening@pik-potsdam.de)}


 
\justifying
 
\begin{abstract}

A balanced ratio of ocean to land is believed to be essential for an Earth-like biosphere and one may conjecture that plate-tectonics planets should be similar in geological properties. After all, the volume of continental crust evolves towards an equilibrium between production and erosion. If the interior thermal states of Earth-sized exoplanets are similar to the Earth’s -- a straightforward assumption due to the temperature-dependence of mantle viscosity -- one might expect a similar equilibrium between continental production and erosion to establish and, hence, a similar land fraction. We will show that this conjecture is not likely to be true. Positive feedback associated with the coupled mantle water -- continental crust cycle may rather lead to a manifold of three possible planets, depending on their early history: a land planet, an ocean planet and a balanced Earth-like planet. In addition, thermal blanketing of the interior by the continents enhances the sensitivity of continental growth to its history and, eventually, to initial conditions. Much of the blanketing effect is however compensated by mantle depletion in radioactive elements. A model of the long-term carbonate-silicate cycle shows the land and the ocean planet to differ by about 5 K in average surface temperature. A larger continental surface fraction results both in higher weathering rates and enhanced outgassing, partly compensating each other. Still, the land planet is expected to have a substantially dryer, colder and harsher climate possibly with extended cold deserts in comparison with the ocean planet and with the present-day Earth. Using a model of balancing water availability and nutrients from continental crust weathering, we find  the bioproductivity and the biomass of both the land and ocean planet to be reduced by a third to half of Earth's. The biosphere on these planets might not be substantial enough to produce a supply of free oxygen.

\end{abstract}


\section{Introduction}
\label{sec1}
In the absence of observable unique biomarkers, the search for life beyond the solar system focuses on the search for habitable planets \citep[e.g.,][]{Segura:2010}. The concept of the circumstellar habitable zone (CHZ) around a host star has long been established \citep{hart:1979,Huggett:1995} and developed further (see e.g., \citealt{cockell2016} and  \citealt{Ramirez:2018} for reviews). The CHZ is defined as the zone of orbits in which a planet would have surface temperatures that allow the presence of liquid water, thus recognizing the importance of water for life as we know it \citep[e.g.,][]{Brack:2010}. The boundaries of the habitable zone depend on the luminosity of the star but also on the atmosphere pressure at the surface of the planet and on atmospheric chemistry, in particular, on the concentration of the atmosphere greenhouse gases carbon-dioxide and methane \citep[e.g.,][]{lammer2009}. More recent discussions of planetary habitability go beyond stellar luminosity and atmosphere chemistry to include other planetary properties such as the ratio between land and water covered surface areas, diversity of geologic provinces, magnetic fields, and processes such as plate tectonics that drive the planetary engine and its evolution \citep[e.g.,][]{Heller&Armstrong:2014, Schulze-Makuch:2020, Glaser:2020, Lingam:2018, lingam2019, lingam:2021}. 

Having both emerged continents and oceans on Earth is an outcome of plate tectonics. About 40\% of its surface is covered with continental crust of which 87\% is emerged; 13\% of the area is continental shelf area and covered by water. Emerged land provides easy access to solar energy and weathering of continental crust rock provides essential nutrients for the  biosphere. As a result, most of the biomass is found on continents and their shelves while the deep ocean biomass is much lower \citep[e.g.,][]{Behrenfeld:1997, Gray:1997, Kallmeyer:2012}, even though the net primary production NPP from oceans is similar to that from land \citep{Field:1998}.

The importance of a suitable balance between land and ocean covered surface areas for life as-we-know-it has recently been emphasized by \cite{lingam2019, lingam:2021} and \cite{Glaser:2020}. The required balance is a consequence of the need for water on the one hand and nutrients such as phosphorous and nitrogen on the other. On a land dominated planet with minimal water, the biological productivity should be low due to water limitations. Ocean-dominated planets could be stymied by the lack of nutrients. If phosphorous is the productivity limiting nutrient \citep[e.g.,][]{Kamerlin:2013} then the increased weatherability of subaerial continental rock by slightly acidic rain-water gives land surfaces an edge over the ocean floors. The build up of free oxygen in the atmosphere for more advanced life is likewise favoured by continental crust weathering that produces the sediments in which organic carbon is buried and removed from oxidization. \cite{lingam2019, lingam:2021} discuss a simple parameterization of the  NPP and the biomass as functions of the land-ocean surface ratio. They find both functions to be maximized at the present-day Earth ratio and to decrease by an order of magnitude as one goes to land dominated (1-2$\sigma$) or ocean dominated worlds. We note that the Earth-like land-ocean ratio would tend to maximise the extent of continental shelf areas where life will profit from solar insolation and the riverine entry of dissolved nutrients. \cite{lingam2019, lingam:2021} also find that oxygen can build up in worlds that have a water coverage between roughly 25 and 85\%. Thus, planets with a balance of landmasses and oceans might be potentially conducive to hosting rich biospheres. In addition to the land-ocean distribution, the surface temperature will matter for the bioproductivity as has been emphasised recently  by \cite{Heller&Armstrong:2014}, \cite{Lingam:2018} and \cite{Schulze-Makuch:2020}.

It is well established that the concentration of carbon-dioxide in the Earth's atmosphere is buffered by the long-term carbonate-silicate cycle \citep[e.g.,][]{Kasting:1993:1}, thereby stabilising the atmosphere temperature and clement conditions. Negative feedback in the cycle causes the atmospheric CO$_2$  to attain an equilibrium between carbon outgassing and recycling \cite[e.g.,][]{Lehmer:2020,Isson:2020}. Plate tectonics is a central element of the modern cycle, exchanging CO$_2$ between the atmosphere and the mantle reservoirs through volcanism and subduction and the rate of outgassing is a function of the interior properties of the planet. The surface temperature is buffered as the rate of removal of CO$_2$ from the atmosphere through rain and weathering of continental crust -- the latter a product of plate tectonics -- increases  with increasing temperature \citep[e.g.,][]{Walker:1981, Kasting:1993:1,Foley:2015}. Weathering of continental crust results in carbon fixation and its rate depends on the continental surface area and the topography. Water is the main agent for transporting CO$_2$ from the atmosphere to the surface and for transporting weathering products including nutrients from the continents to the oceans and eventually to the subduction zones, where CO$_2$ is then returned to the mantle. Like CO$_2$, water is cycled by plate tectonics through the mantle and mantle outgassing may have been a major source of the present ocean mass \citep[e.g.,][]{ElkinsTanton:2011}. The oceans provide the single most important thermal reservoir for buffering the surface temperature \citep[e.g.,][]{Bindoff:2007, Cheng:2019, vonSchuckmann:2021} as well as they provide a sink for atmospheric CO$_2$ on short timescales \citep[e.g.,][]{pierrehumbert2010}.

The operation of plate tectonics, like on present-day Earth, is vital to the long-term carbon cycle outlined above. Venus may have had surface recycling \citep[e.g.,][]{Crameri:2017} but its hypsometric curve is unimodal \citep[e.g.,][]{Rosenblatt:1994}. While there are units resembling terrestrial rifts and ridges such as Beta Regio, there is no clear indication of subduction zones and continental crust, although Ishtar Terra as a high elevation physiographic province  resembles  a terrestrial continent \citep{Ivanov:2011}. Rather, it seems more likely that Venus has a continuous lithosphere. While Venus may be debated \citep{davaille2017}, the smaller planets Mars and Mercury are widely agreed to have continuous outer shells termed stagnant-lid lithospheres. Planetary mantle convection modelling finds stagnant lids to form naturally as a consequence of a temperature dependent rheology of mantle rock \citep[e.g.,][]{Stein:2008} while subduction and plate-like behaviour require additional strain related mechanisms for which water may be important \citep[e.g.,][]{Bercovici:2003}.

Uncertainties in estimates of the interior properties of exoplanets have significant effects on assessments of their habitability \cite[e.g.,][]{Seales:2020}. For example, uncertainties in the concentrations of heat producing elements will naturally affect estimates of the outgassing rates and the evolution of the surface temperature \cite[e.g.,][]{ONeill:2020,oosterloo2021,Unterborn:2022}. It has even been suggested that there may be feedback between the surface temperature and  the tectonic regime (plate-tectonics versus stagnant lid) that could result in irreversible bifurcation between the two modes \citep{Weller:2015,Weller:2018}. Another key uncertainty in exoplanet evolution is the initial mantle temperature. Theoretical considerations and simple thermal history models suggest that differences in the initial mantle temperature should become insignificant after $\sim$1 Gyr of evolution due to the temperature-dependence of mantle viscosity \citep[e.g.,][]{Tozer:1967}, at least for planets less massive or as massive as  Earth \citep{kruijver:2021}. Therefore, one might conclude that the climate at 4.5 Gyr should be largely independent of the initial mantle temperature. This conjecture, however, neglects feedback associated with the growth of continental crust, which we will explore in this paper.

\begin{figure}
    \centering
    \includegraphics[width=\textwidth]{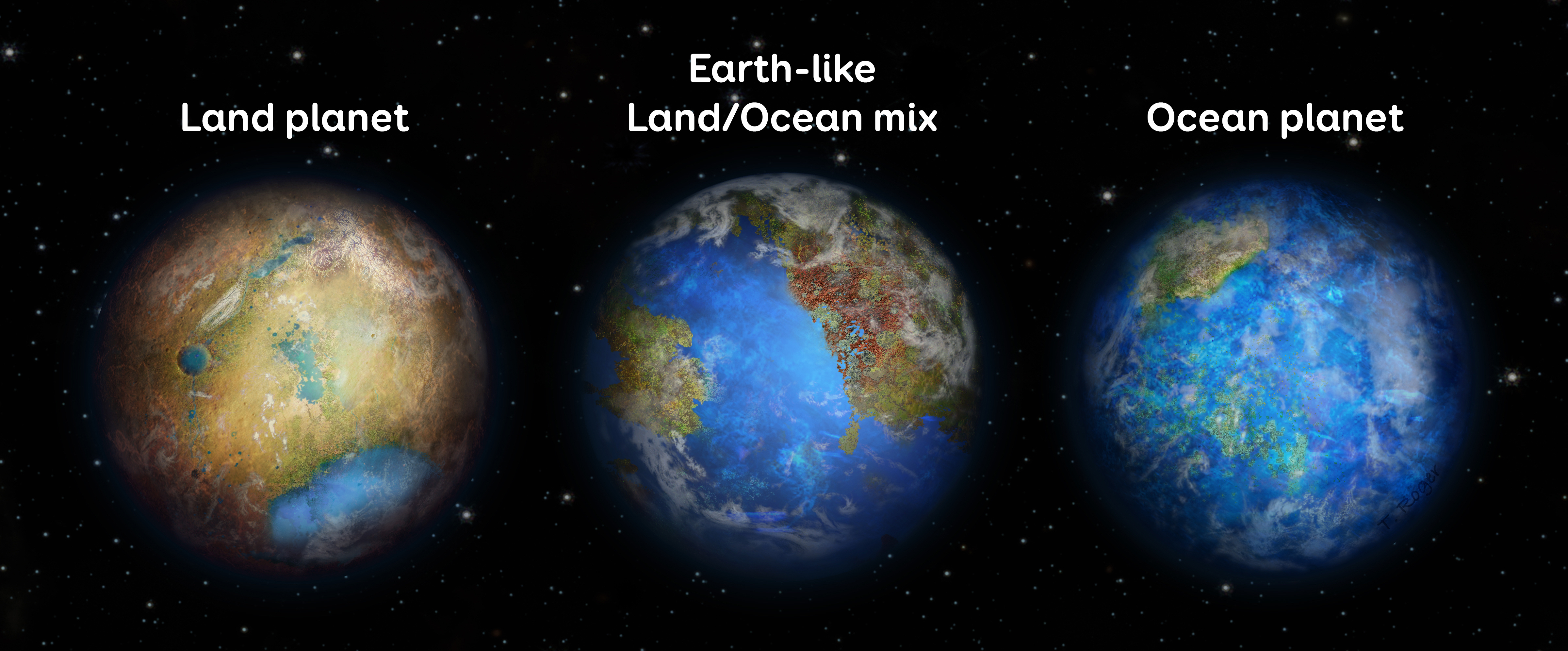}
    \caption{Graphical illustration of the \textit{Land Planet}, Earth-like land/continent mix, and \textit{Ocean Planet}. Credit: Europlanet 2024 RI/T. Roger}
    \label{fig:illustration}
\end{figure}

Even for Earth-sized planets with plate tectonics, it is not a given that the surface would have a balanced ocean to land coverage similar to Earth. \cite{Honing:2016} and \cite{Honing:2019a} have argued that -- depending on initial conditions -- the evolution of a plate tectonics Earth-like planet would rather result in a planet mostly covered by continents or mostly covered by oceans (Fig. \ref{fig:illustration}). In their analysis, the continental crust and the water cycles are feedback cycles with either one or three fixed points depending on whether negative feedback or positive feedback dominates, respectively. Of the three fixed points, two would be stable (pertaining to the continent and the ocean planet) and one conditionally stable (the Earth). These differences in the outcome of planetary evolution are likely to be mirrored in the climate and possibly the biomass. The present climate of such a planet would retain a memory of the planet's very early evolution, since the equilibrium between CO$_2$ outgassing and silicate weathering would depend on the fraction of emerged continents and on the thermal state of the planet's interior. Their rates of change in turn would depend on the properties of the coupled continent-water system.

The dependence of the strengths of positive and negative feedback on details of continental growth mechanisms complicates an extensive analysis of evolution pathways of (exo)planets \citep{Honing:2016}. Therefore, in the present paper we will parameterise feedback using four dimensionless quantities. We will show evolution pathways that depend on the assumed values of these quantities and then comprehensively discuss reasonable parameter choices and evolution scenarios. We extend our earlier models by analysing the effects of insulating continents delaying mantle cooling and the opposing effect of a depletion of mantle radioactive isotopes with continental growth. Furthermore, we calculate average surface temperatures for the model planets using a model of the carbonate-silicate cycle and estimate the  bioproductivity and -mass following \cite{lingam:2021}. We will present our models in a 3D phase space spanned by continental coverage, mantle water concentration and mantle temperature as variables. 

In the following section, we begin by discussing plate tectonics from the point of view of complex non-linear system theory. We will then discuss fixed points of the system and their stability depending on dimensionless parameters that describe the strengths of positive and negative feedback. Coupling the model to a thermal evolution model of the mantle, we will show how fixed points evolve with time and steer the evolution of the planet and its habitability (Section \ref{sec:results}). This section is followed by a discussion and conclusions.

\section[Mantle evolution and continental growth]{Mantle thermo-chemical evolution and continental growth as a nonlinear system}
\label{sec2}
Plate tectonics and its role in geology has been described in a large number of publications \citep[e.g.,][]{Sleep:2015}. In a nutshell, the Earth's surface is covered by seven major plates that move horizontally across the surface driven by solid-state convection in the mantle. The plates consist of oceanic crust and lithosphere and continental crust and lithosphere. Oceanic crust is produced by pressure release partial melting of mantle rock underneath mid-oceanic ridges and is returned to the mantle in subduction zones. The subduction zones are the places where modern plate tectonics produces continental crust by partially melting oceanic crust, subducted sediments, and mantle rock in the presence of subducted water reducing their solidus temperature \citep[e.g.,][]{campbell:1983}.

Continents, despite their longevity in comparison with oceanic crust,  are continuously weathered and eroded and the resulting sediments are transported to subduction zones by surface water flow. In addition, continental crust is eroded through friction between the continental keel and the subducting slab. The rates of subduction and crust production  are closely linked to the rate of mantle flow, which is governed by the rheology of mantle rock. On the scale of mantle flow, the rheology can be parameterised by an effective viscosity. The latter depends non-linearly on mantle temperature and on the concentration of water and -- to a lesser extent -- of carbon-dioxide in mantle rock. In addition, the mantle viscosity is a function of pressure, which should be considered for planets more massive than 2-3 Earth masses \citep[e.g.,][]{kruijver:2021}. Since we restrict our analysis to Earth-mass planets, we neglect this dependence here.

Sediments play important roles in subduction zone processes and in this paper, they play a decisive role: The erosion rate of the continents is proportional to their area and thus provides negative feedback limiting  their growth. As carriers of water, however, sediments contribute to water transport (Hacker 2008; Deschamps et al. 2013) in subduction zones and provide positive feedback increasing the continental growth rate. Moreover, \cite{sobolev:2019} argue that sediments are essential for keeping plate tectonics running by lubricating the subducting plate.

While mid-oceanic ridges are the most prominent source regions of mantle water entering the oceans and the atmosphere, subduction zones are the major sink regions for surface water. Today, about $1.8\cdot10^{12}$ kg of water enter subduction zones each year mainly in hydrous minerals in sediments and hydrated basaltic crust \citep{jarrard:2003}. Significant, but difficult to constrain \citep[e.g.,][]{Honing:2014} quantities of the subducted water are expelled at shallow depth, the remaining water -- partly stored in serpentinites and other hydrous minerals -- reaches the source region of continental crust rock where hydrous minerals break down and release water that is consumed upon melting. The remainder of the water is subducted deeper into the mantle where it is eventually incorporated into nominally anhydrous minerals \citep{bell:1992}. As a consequence, mantle viscosity is reduced \citep{hirth:2003} and the rate of mantle convection flow increased. The water is transported in the convection flow towards mid-ocean ridges where part of it is outgassed into the atmosphere and oceans.

Feedback associated with the growth of continental crust can also originate from its thermal properties. The continental heat flow per unit area of about 70 mW/m$^2$ is known to be significantly smaller than the oceanic heat flow of about 105 mW/m$^2$ \citep[e.g.,][]{Jaupart:2015}. Moreover, the basal heat flow per unit area from the mantle into stable continental crust is only about 15 mW/m$^2$. These observations suggest that the continents act as thermal blankets of the mantle reducing the heat flow and the mantle cooling rate and diverting heat flow from sub-continental to sub-oceanic areas. The growth of continents diverting heat flow to oceanic areas can cause immediate positive feedback \citep{Honing:2019a}. But even a reduction of the overall cooling rate introduces a sensitivity of the continental growth rate to its past history.

An opposing effect to thermal blanketing of continental crust is the redistribution of heat sources during crust growth, which may provide negative feedback. Radiogenic elements, because of their large ionic radii and their valence states, are enriched in melts producing continental crust. The concentration of heat sources in present-day continental crust is debated, however. Estimates are based on measured heat flow values and concentrations in mantle and crustal rock \citep[e.g.,][]{jaupart:2016}. Uncertainties are due to uncertainties in the ratio between surface heat flow and mantle radiogenic heat production (Urey ratio) and the density of heat sources in the lower mantle. But it is estimated that the modern continental crust  contains between 25 and 40\% of the total planetary inventory of heat sources \citep{dye:2012,Jaupart:2015,jaupart:2016}. The transfer of heat sources from the mantle to the growing continental crust effectively cools the mantle \citep[e.g.,][]{Spohn:1991} and provides negative feedback by slowing down mantle convection and thus the production rate of new continental crust.

\begin{figure}
    \centering
    \includegraphics[width=\textwidth]{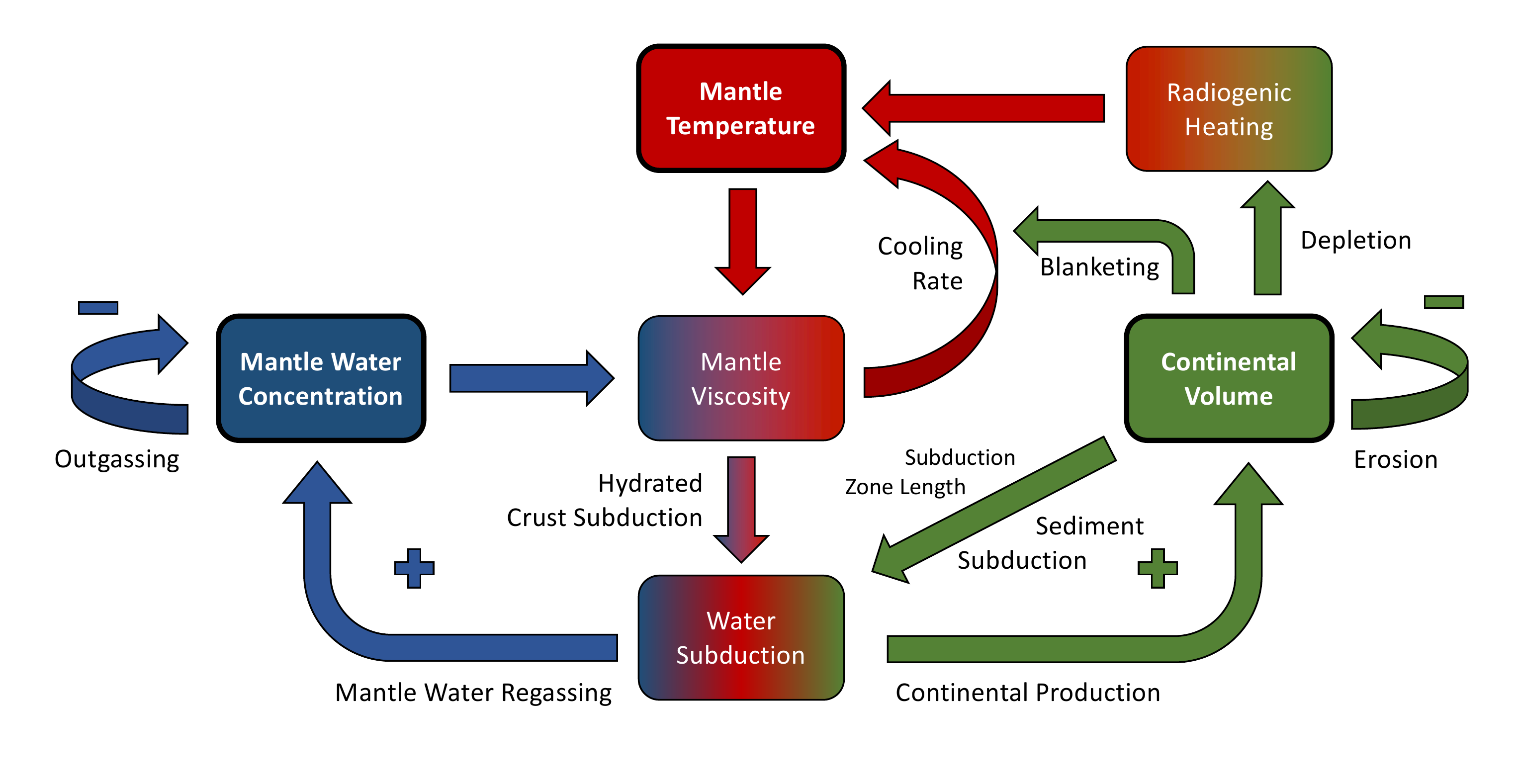}
    \caption{Graphical illustration of the feedback cycles linking continental coverage (green, compare Eq. \ref{eq:continent}), mantle water concentration (blue, compare Eq. \ref{eq:water}), and mantle temperature (red, compare Eq. \ref{eq:temperature_new}). The evolution of the former two is coupled through the subduction of water. Positive feedback -- indicated by '+' signs -- results from the subduction of water reducing the mantle viscosity and increasing the plate speed and the crust production rate. Increasing the continental coverage results in an increased production rate of sediments feeding additional water and rock into the melt zone from which continental crust originates. Water outgassing, surface erosion and cooling dampen the positive feedback and are indicated by '-' signs. Continental crust growth feeds back further on the mantle temperature through thermal blanketing and transfer of radiogenic elements from the mantle to the crust. For further explanation see text.}
    \label{fig:flowchart}
\end{figure}

The above description of plate tectonics suggests that the continental crust volume, the mantle water concentration and mantle temperature are linked through non-linear feedback cycles as illustrated in Figure \ref{fig:flowchart}.  An increase in mantle water concentration (blue) leads to a reduction of the mantle viscosity and thereby to an increase in the convection speed. The increased subduction rate will increase the mantle water concentration establishing a positive feedback. The latter may be balanced by negative feedback associated with mantle water outgassing. The water cycle is linked to the continental crust cycle (green) through the water subduction rate. An increase in continental coverage increases the weathering and erosion rates and -- assuming that the increased sedimentary volume will be subducted and contribute to water subduction -- to increased rates of mantle water regassing and continental crust production. This feedback may be balanced by continental loss through erosion. The thermal evolution of the mantle (red) is linked to the mantle water cycle via the mantle viscosity and to the continental cycle via thermal blanketing and mantle depletion of radiogenic elements. The role of sediments in water subduction and the insulating effect of continents determine the strength of positive feedback in the coupled system.

The mantle temperature is subject to the negative feedback provided by the temperature dependence of the mantle viscosity and known as the Tozer principle \citep{Tozer:1967}. Accordingly, an increase in mantle temperature will decrease the viscosity which will result in an increased convection and mantle cooling rate -- all else being equal. There is no positive feedback mechanism that could lead to runaway heating as long as the mantle heat source density decreases with time but continental crust growth and the mantle water budget can feedback onto the evolution of the mantle temperature. An increased degassing of the mantle will increase the mantle viscosity and result in reduced mantle cooling -- with effects on crust growth and the water budget. Likewise, an increase in continental surface area can reduce the surface heat flow and the cooling rate -- with possible readjustments of the convection system that may lead to adjustments of the plate speed. But only in an extreme case of thermal blanketing can continental growth result in temporary mantle heating, as we will show further below.

Mathematically, the non-linear feedback system can be described by the following set of differential equations. In writing Eqs. (\ref{eq:continent}) and (\ref{eq:water}) we follow \cite{Honing:2016} but we simplify their equations while keeping the most important feedback terms. We complement these equations by an energy balance equation that includes the effects of continents (Eq.\,\ref{eq:temperature_new}).

\begin{linenomath}
\begin{equation}
   \frac{\dot{A}}{\dot{A}_{l,E}} =\underbrace{\left[\xi_1 A^*+(1-\xi_1)\, v_p^* \right]\, L_{oo,oc}^* f_r}_\text{continental production} - \underbrace{\left[\xi_2 A^*+(1-\xi_2)\,L_{oc}^*\right]\,}_\text{erosion},
   \label{eq:continent}
\end{equation}
\end{linenomath}
\begin{linenomath}
\begin{equation}
   \frac{\dot{w}}{\dot{w}_{l,E}} =\underbrace{\left[\xi_1 A^*+(1-\xi_1)v_p^*\right]L_{oo,oc}^* f_r}_\text{regassing} - \underbrace{w^* v_p^*}_\text{outgassing},
   \label{eq:water}
\end{equation}
\end{linenomath}
\begin{linenomath}
\begin{equation}
\label{eq:temperature_new}
\begin{split}
\dot{T} \, \rho_m c_p V_m &=-q_{m,E} \underbrace{\Delta T_u^{*(1+\beta)} \sqrt{v_p^*} \left[(1-A)^*\xi_3+A^*(1-\xi_3)\right]}_\text{rel. surface heat flow}\\&+Q_{A_E}\underbrace{[1+ \xi_4 (1-A^*)]}_\text{rel. mantle heat prod.} + q_{core}.
\end{split}
\end{equation}
\end{linenomath}

$A$, $w$ and $T$ are the state variables of the system. $A$ is the continental surface fraction (combined surface area of the continents divided by the surface area of the planet), $w$ is the mantle water concentration, and $T$ the mantle temperature; 
$\dot{A}$, $\dot{w}$, and $\dot{T}$ are their rates of change with time, respectively. $\dot{A}_{l,E}$ and $\dot{w}_{l,E}$ are the present-day rates of continental loss by erosion and mantle water loss by outgassing (in mass per time), respectively, and $f_r$ is the present-day ratio between loss and gain rates ($f_r=1$, for equilibrium between continental erosion and production and between mantle water gain and loss). We generally use the asterisk to denote variables scaled to present-day Earth values and the index $E$ to denote the latter. $v_p(T,w)$ is the plate speed and a function of $T$ and $w$. It is taken to be proportional to the mantle convection flow speed. $L_{oo,oc} (A)$ is the combined length of ocean-ocean subduction zones $L_{oo}$ (such as the Marianas) and ocean-continent subduction zones $L_{oc}$ (such as the subduction zone at the Peru-Chile trench) and is a function of $A$. The length of subduction zones is calculated from the the total length of continental margins as a function of continental coverage (see \ref{subduction_zones} for details). $\Delta T_u$ is the temperature difference across the upper thermal boundary layer of mantle convection, $q_{m,E}$ is the present-day heat flow from the mantle, $Q_{A_E}$ is the radiogenic heat production rate as a function of time of the present-day depleted mantle (compare Eq. \ref{eq:heatprod} in Section \ref{sec:interior}). The depletion of mantle heat sources due to continental crust growth is modelled by the term  $1+\xi_4 (1-A^*)$, where $\xi_4$ is the present-day share of the Earth's inventory of radiogenic elements that is in the continental crust. $q_{core}$ is the heat flow from the core. $\rho_m$, $c_p$, and $V_m$ are the mantle density, specific heat capacity, and volume, respectively. $\xi_1$ to $\xi_4$ are parameters describing the relative weight of terms in the balance equations (compare Tab. \ref{tab:xi}). As we will discuss in more detail below, $\xi_1$ and $\xi_2$ weigh the strengths of positive and negative feedback in continental growth, and $f_r$ shifts the fixed points in the phase plane.

Eq. \ref{eq:continent} calculates the growth rate of continental crust area from a balance of continental crust production and loss through erosion. One part of the growth is directly proportional to the continental surface area. This part is a consequence of the effect of sediments on continental crust production and provides positive feedback as illustrated in Figure \ref{fig:flowchart}. Its relative weight is given by the dimensionless factor $\xi_1$, which may take values between 0 and 1. A second contribution to the growth rate is proportional to the plate speed, which depends on the water concentration in the mantle and the mantle temperature, thereby coupling Eq. \ref{eq:continent} to Eqs. \ref{eq:water} and \ref{eq:temperature_new}. Its relative contribution is given by $(1-\xi_1)$. Loss by erosion is composed of a surface erosion part depending on the continental surface area and provides negative feedback to the system. The remaining part of continental erosion is related to the friction in subduction zones (subduction erosion) and depends on the total length of ocean-continent subduction zones. It should be noted that the latter part is a function of the continental surface area. 

The term describing mantle water regassing (first term on the RHS of  Eq. \ref{eq:water}) mirrors the production term in Eq. \ref{eq:continent}, as both are related to the subduction of water.
The regassing term altogether provides non-linear positive feedback to the rate of change of mantle water concentration through the dependence of the plate speed on the latter (compare Fig. \ref{fig:flowchart}). Mantle water outgassing described by the second term on the RHS of Eq. \ref{eq:water} is directly proportional to $w^*$, which contributes negative feedback to the change of the mantle water concentration. A step-by-step derivation of Eqs. \ref{eq:continent} and \ref{eq:water} is given in \ref{continents_water}.

Eqs. \ref{eq:continent} and \ref{eq:water} are coupled to the energy equation (Eq. \ref{eq:temperature_new}) via the plate speed. The latter equation balances the surface heat flow with the heat production rate in the mantle by radiogenic decay and the heat flow from the core to calculate the rate of change of mantle internal energy. The surface heat flow is coupled to the water balance through the non-linear dependence of the mantle convection speed on the water concentration (\ref{sec:interior}) and is calculated in terms of the present-day mantle heat flow $-q_{m,E}$. Furthermore, it is coupled to the surface area of insulating continental crust through the term $(1-A)^*\xi_3+A^*(1-\xi_3)$. The term is written such that it will equal unity for $A$ equal 0.4, the value for the present-day Earth, independent of the choice of $\xi_3$.

$\xi_3$ is defined as the present-day ratio between the oceanic heat flow and the mantle heat flow. It is a measure of the efficiency of the thermal blanketing of the mantle by continental crust. We discuss a case of perfectly insulating continents, for which $\xi_3=1$, as well as a case for which $\xi_3=0.85$ and for which the percentage of mantle heat flow passing through continental crust is about 15\% \citep{jaupart:2016}. Neglecting continental insulation altogether requires setting $\xi_3=0.6$, which equals the share of the surface area of oceanic crust. A detailed derivation is given in Section \ref{sec:blanketing}. Other parameter values pertaining to the present-day Earth are collected in the tables in \ref{sec:tables}.

\begin{table}
\caption[]{Parameters $\xi_i$ that control the feedback in the coupled system. These parameters are defined for the present-day Earth and kept constant throughout the evolution. We explore wide ranges of values of $\xi_i$ since these carry large uncertainties.}

\label{tab:xi}
\centering
\begin{tabular}{l p{9cm}l l}
\hline
Symbol & 
Definition & Explored Values \\
\hline
$\xi_1$ & Present-day fraction of the continental production rate that is directly proportional to the area of the continents & 1/3, 2/3\\
$\xi_2$ & Present-day fraction of the continental erosion rate that is  directly proportional to the area of the continents & 1/3, 2/3\\
$\xi_3$ & Present-day share of heat flow through oceanic crust & 0.6, 0.85, 1\\
$\xi_4$ & Present-day share of heat sources in continental crust & 0, 1/4\\
$f_r$ & Present-day ratio between the rate of continental production and that of continental erosion & 1, 0.775\\
\hline
\end{tabular}
\end{table}

\section{Results}
\label{sec:results}

The state variables $A$, $w$ and $T$ span a phase space -- a space in which all possible states of the system are represented by unique combinations of the state variables -- in which solutions can be discussed. Equilibrium conditions within the phase space are described by $\dot{A}=0$ and $\dot{w}=0$. These conditions define two surfaces in phase space that intersect in lines towards which the system will tend to evolve.

We begin by introducing the continent-water system in a phase plane evaluated at a representative present-day mantle temperature (Section \ref{sec3}). Following this, we discuss how the phase planes and the state variables evolve with temperature in 3D phase space in Section \ref{subsec:noblanket}. In addition, we show time histories of selected quantities. Implications for the long-term carbon cycle are presented in Section \ref{sec4} and the effect of thermal blanketing on the system is shown in Section \ref{sec5}.

\subsection[Phase planes, fixed points, and stability]{Phase planes at constant present-day mantle temperature, fixed points, and stability
}
\label{sec3}

\begin{figure}
   \centering
   \includegraphics[width=\textwidth]{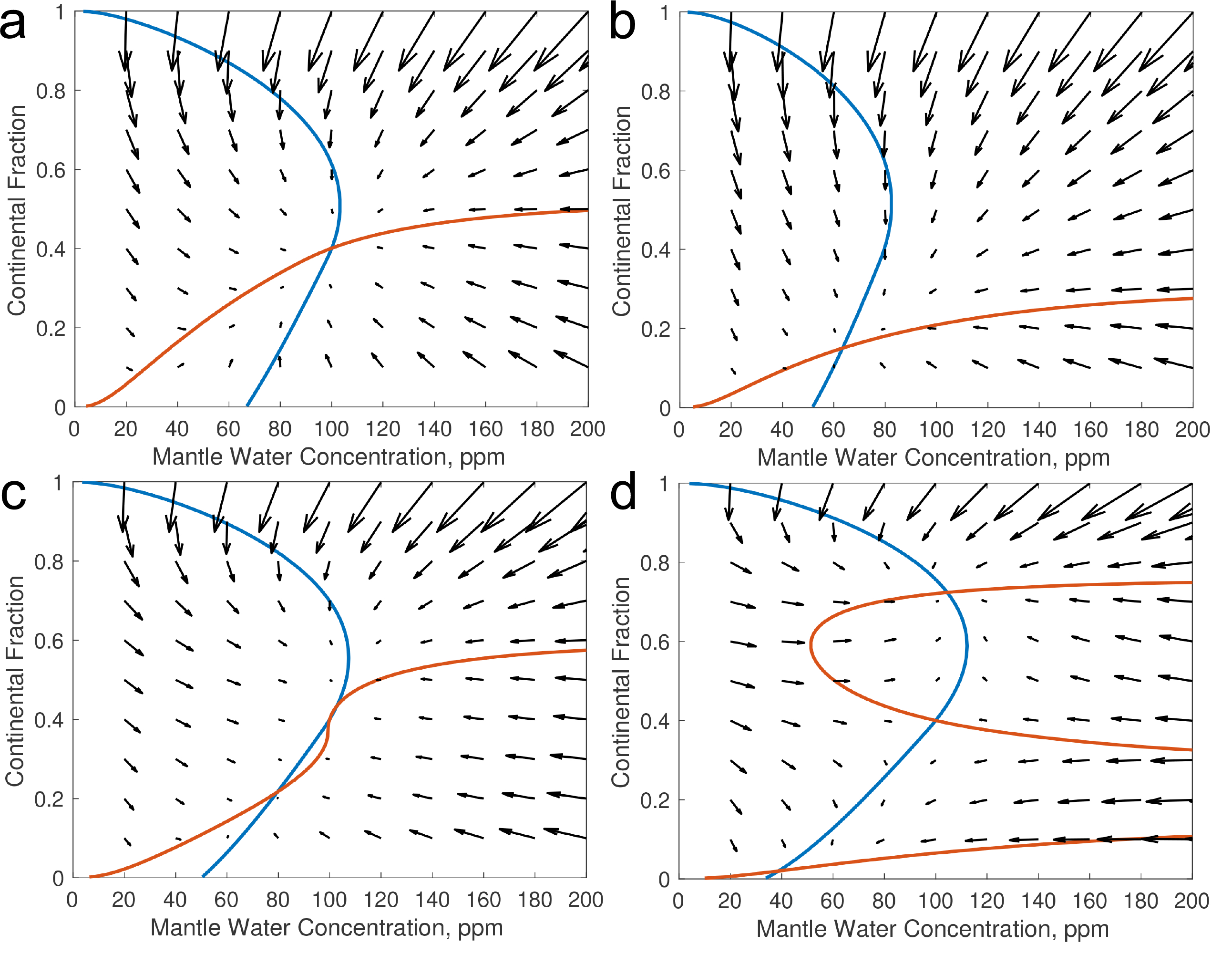}
   \caption{Phase planes defined by the mantle water concentration and continental coverage 
   at the present-day mantle temperature. Shown are local arrows pointing into the directions $dA/dw$ of evolution of $A$ and $w$. In addition, graphs of the integrals of $\dot{A} = 0$ (red) and $\dot{w} = 0$ (blue) are shown. Parameters values are: (a) $\xi_1$ = 1/3, $\xi_2$ = 2/3, $f_r$ = 1; (b) $\xi_1$ = 1/3, $\xi_2$ = 2/3, $f_r$ = 0.775; (c) $\xi_1 = \xi_2 = 0.5$, $f_r$ = 1; (d) $\xi_1$ = 2/3, $\xi_2$ = 1/3, $f_r$ = 1. For further discussion see the text.}
\label{fig:quiver}
\end{figure}

Figure \ref{fig:quiver} presents phase planes spanned by $A$ and $w$ at a constant representative present-day mantle temperature of 1750 K for various combinations of values of $\xi_i$ and $f_r$. The exact choice of the value of the present-day temperature is not critical and does not affect our results in any significant way because the mantle viscosity is scaled accordingly to reach this temperature at 4.5 Gyr. While values of $\xi_i$ parameterize the strength of the feedback, $f_r$ describes the ratio between the present-day Earth gain rates (continental production and mantle water regassing) and loss rates (continental erosion and mantle water outgassing) and therefore shifts the fixed points in the phase plane.

Fig. \ref{fig:quiver} displays local arrows pointing into the directions $dA/dw$ of evolution of $A$ and $w$ and graphs of the integrals of $\dot{A} = 0$ (red) and $\dot{w} = 0$ (blue). In panels a and b, with $\xi_1 < \xi_2$ these intersect in a single fixed point  marking a single stable solution to the system of Eqs. \ref{eq:continent}--\ref{eq:temperature_new} at constant mantle temperature $T$. The integral of $\dot{A}=0$ is a single valued function $A(w)$ under these conditions. Increasing the water content in the mantle will increase the growth rate through its dependence on $v_p(T,w)$. The increase is  balanced by the erosion rate increasing with the continental surface area. The integral $w(A)$ of $\dot{w}=0$ is a single valued function for all parameter combinations considered with a maximum value around 50\% continental coverage. The rate of water release first increases with mantle water concentration and is balanced by increasing water consumption in subduction zones. At around 50\% continental surface area, the curve bends over as continents begin to dominate on the surface. For geometrical reasons, ocean-continent subduction zones and ocean-ocean subduction zones will shrink in length while the length of continent-continent collision zones without subduction will increase. This will reduce the potential to subduct water. The mantle water consumption rate will then decrease with increasing $A$ and will need to be balanced, for $\dot{w}=0$, by a decreasing outgassing rate requiring decreasing values of $w$. The maximum of $w(A)$ mainly results from the variation of the total length of ocean-continent type subduction zones $L_{oc}$ with $A$, which has a maximum at $A \approx$ 0.37 (see \ref{subduction_zones}). The maximum of $w(A)$ occurs at somewhat larger value of $A$ because the effect of $L_{oc}$ decreasing with $A$ for $A >$ 0.37 is partly balanced by the sediment subduction flux increasing with $A$ and the effect of the latter on mantle water regassing.

At $\xi_1 \geq \xi_2$, the integral $A(w)$ of $\dot{A}=0$ becomes a multivalued function of the mantle water concentration (compare Fig. \ref{fig:quiver}, panels c and d). At small values of $A$, $v_p(T,w)$ increasing with $w$ dominates crustal growth as before and $A(w)$ has a positive slope on this first branch of the S-shaped curve. At a specific value of $A$, decreasing with increasing $\xi_1$, the graph of $A(w)$ bends over and its slope becomes negative. This occurs because the growth rate increasing with $\xi_1 A$ can only be balanced by erosion if $v_p(T,w)$ and hence $w$ decrease with increasing $A$. The increase of $A(w)$ with negative slope continues until at some value of $A$, which is $\sim0.6$ in Fig. \ref{fig:quiver} panel d, the decreasing length of ocean-continent subduction zones must be compensated by increasing values of $w$ to keep $\dot{A} = 0$. The critical value of $A$ at which the graph of the integral of $\dot{A}=0$ resumes a positive slope increases with increasing values of $\xi_1$.

The graphs of the integrals of $\dot{A} = 0$, $A(w)$ and $\dot{w} = 0$, $w(A)$, intersect at three points (panel d) for $\xi_1 > \xi_2$. Of these, two represent stable equilibria at small and large continental coverage, respectively. $A(w)$ has positive slopes at these points. The intersection point between the two -- at $A = 0.4$ and $w = 100$ppm here -- is a saddle point, stable with respect to perturbations in $w$ but unstable with respect to perturbations in $A$. $A(w)$ has a negative slope at this point. In panel c, we plot a transitory case for which $\xi_1 = \xi_2=0.5$, $f_r$=1. Here, the two curves just touch at $A = 0.4$ and $w = 100$ppm with $dA/dw = dw/dA$ and there is a stable equilibrium point at $A = 0.25$ and $w = 80$ppm. It should be noted that $\xi_1=\xi_2=0.5$ marks a bifurcation for the solutions of Eqs. \ref{eq:continent} -- \ref{eq:water} \citep{Honing:2019a}.

\begin{figure}
   \centering
  \includegraphics[clip,width=\columnwidth]{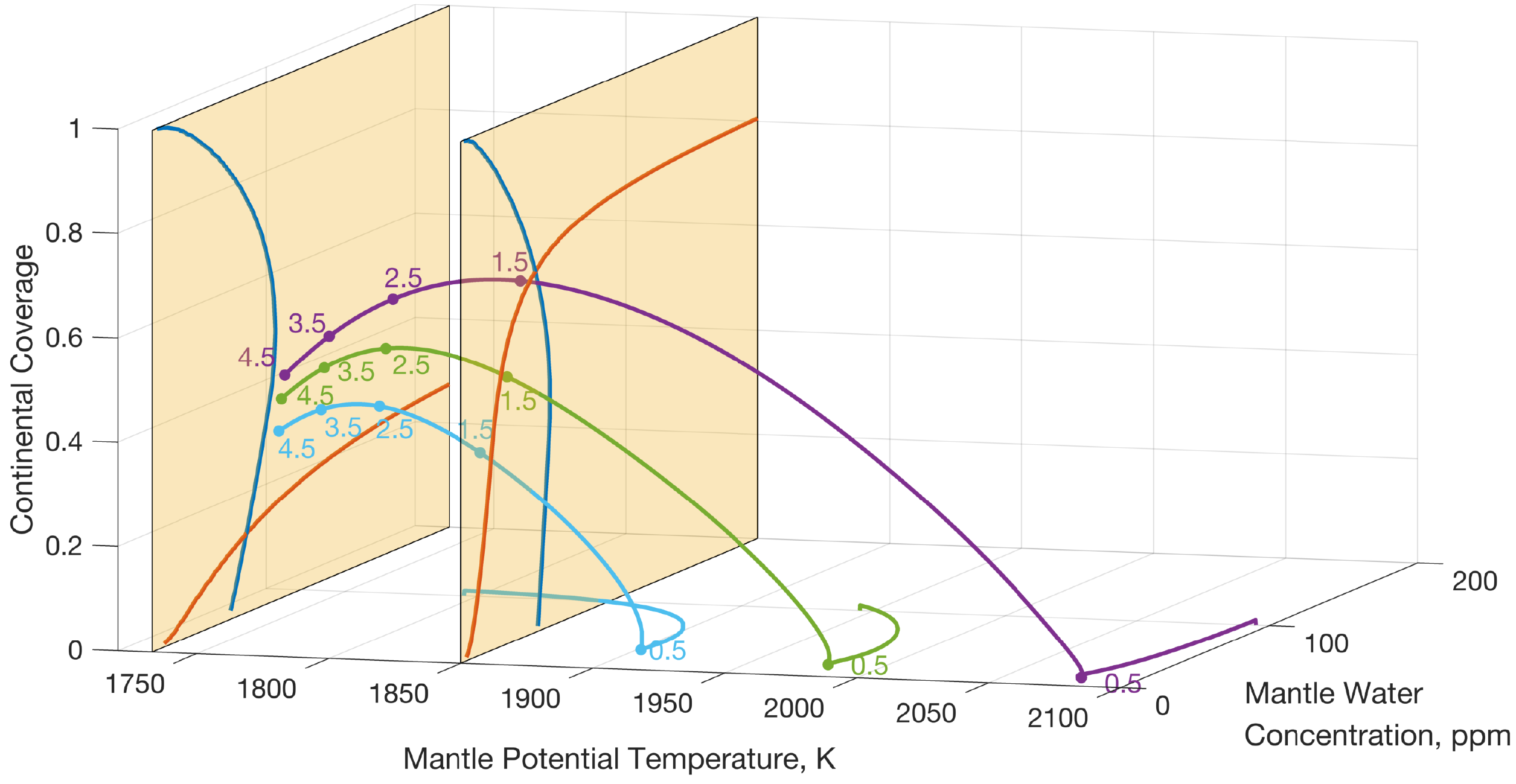}%
  
  \includegraphics[clip,width=\columnwidth]{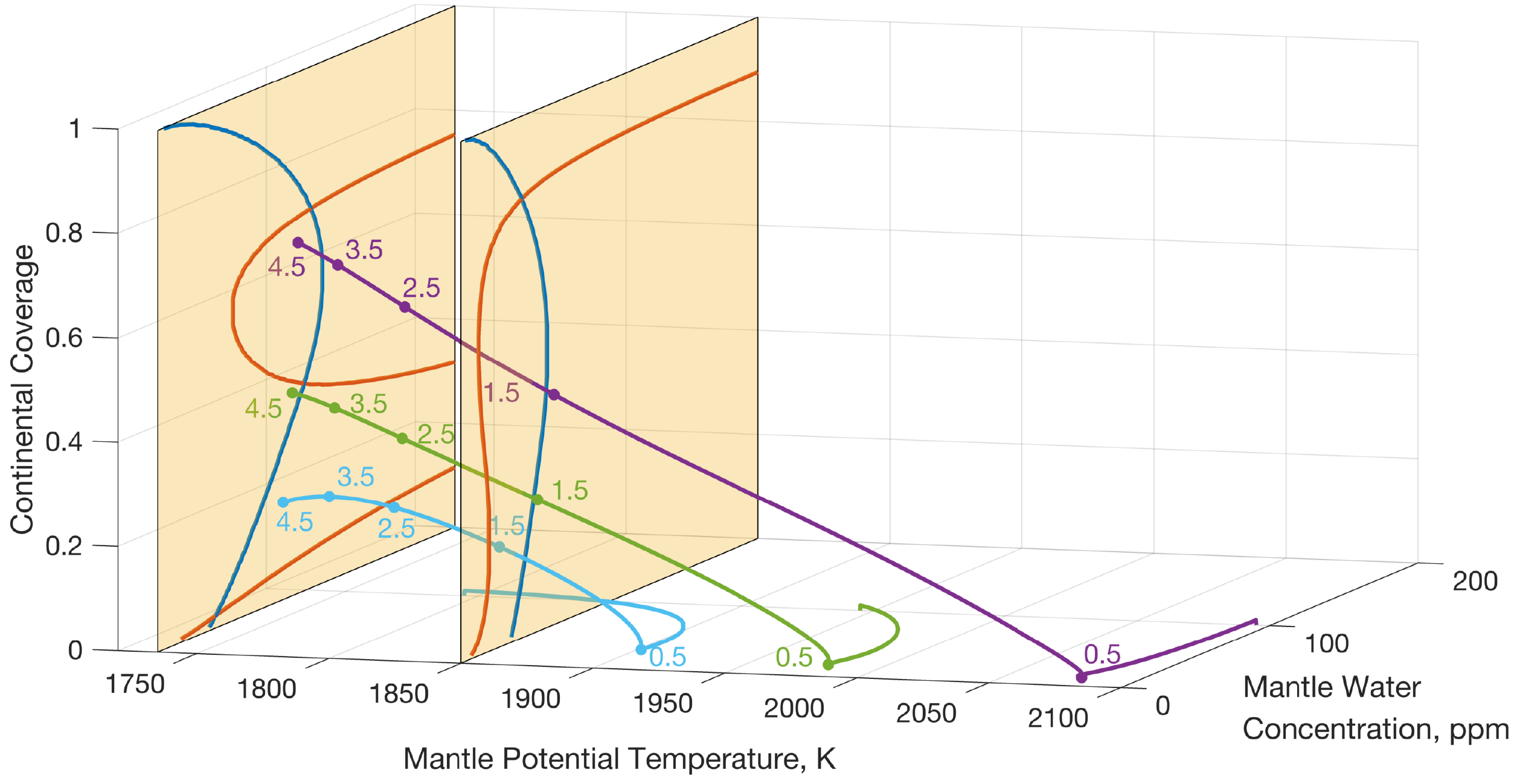}%
   \caption{Evolution models without thermal blanketing and continental crust enrichment in radiogenic elements. Shown are trajectories in the $A,w,T$ phase space along with $A,w$ phase planes at constant mantle temperatures of 1850 K and 1733 K. The latter temperature is the present-day mantle temperature in all models shown here. Time in units of Gyr is marked by dots and labelled. The model shown in the top panel has the same values of $\xi_1=1/3$, $\xi_2=2/3$, and $f_r=0.775$ as the model in panel b of Fig. \ref{fig:quiver}. In the bottom panel $\xi_1=2/3$, $\xi_2=1/3$, and $f_r=1$ are chosen as in panel d of Fig.  \ref{fig:quiver}. In both panels $\xi_3=0.6$, $\xi_4=0$ are chosen such that the terms related to $A$ in Eq. \ref{eq:temperature_new} cancel. The initial mantle temperature $T_0$ for the green curve has been chosen such that the a present-day continental coverage of 0.4 is obtained; the purple curve starts at a 150 K higher mantle temperature and the cyan curve at a 150 K lower temperature. The onset time $t_p$ of continental growth is 0.5 Gyr.}
\label{fig:3dplot}
\end{figure}
\subsection{Mantle cooling without thermal blanketing}
\label{subsec:noblanket}

\begin{figure*}
    \centering
    \begin{subfigure}[b]{0.45\textwidth}
        \centering
        \includegraphics[width=\textwidth]{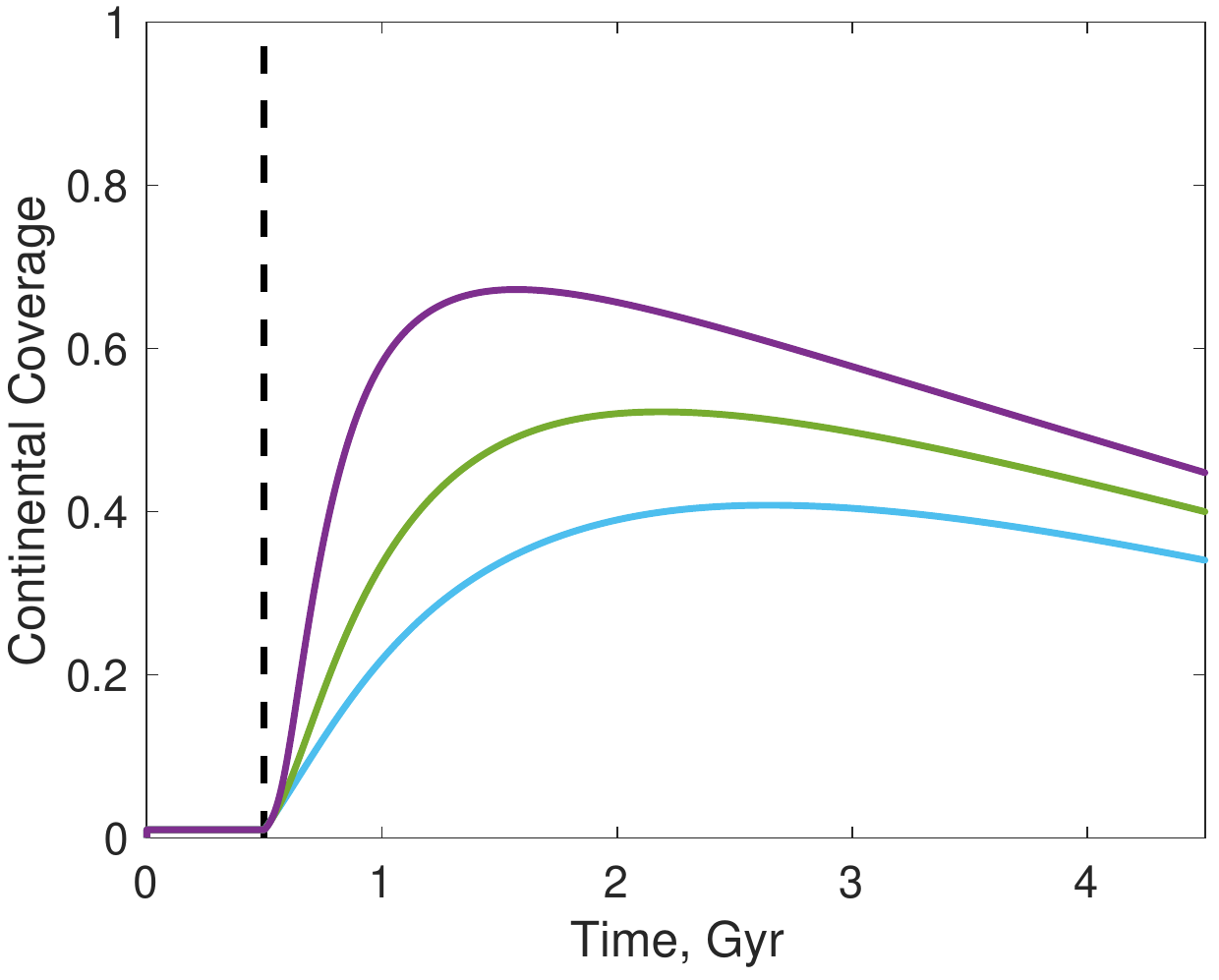}
    \end{subfigure}
    \hfill
    \begin{subfigure}[b]{0.45\textwidth}  
        \centering 
        \includegraphics[width=\textwidth]{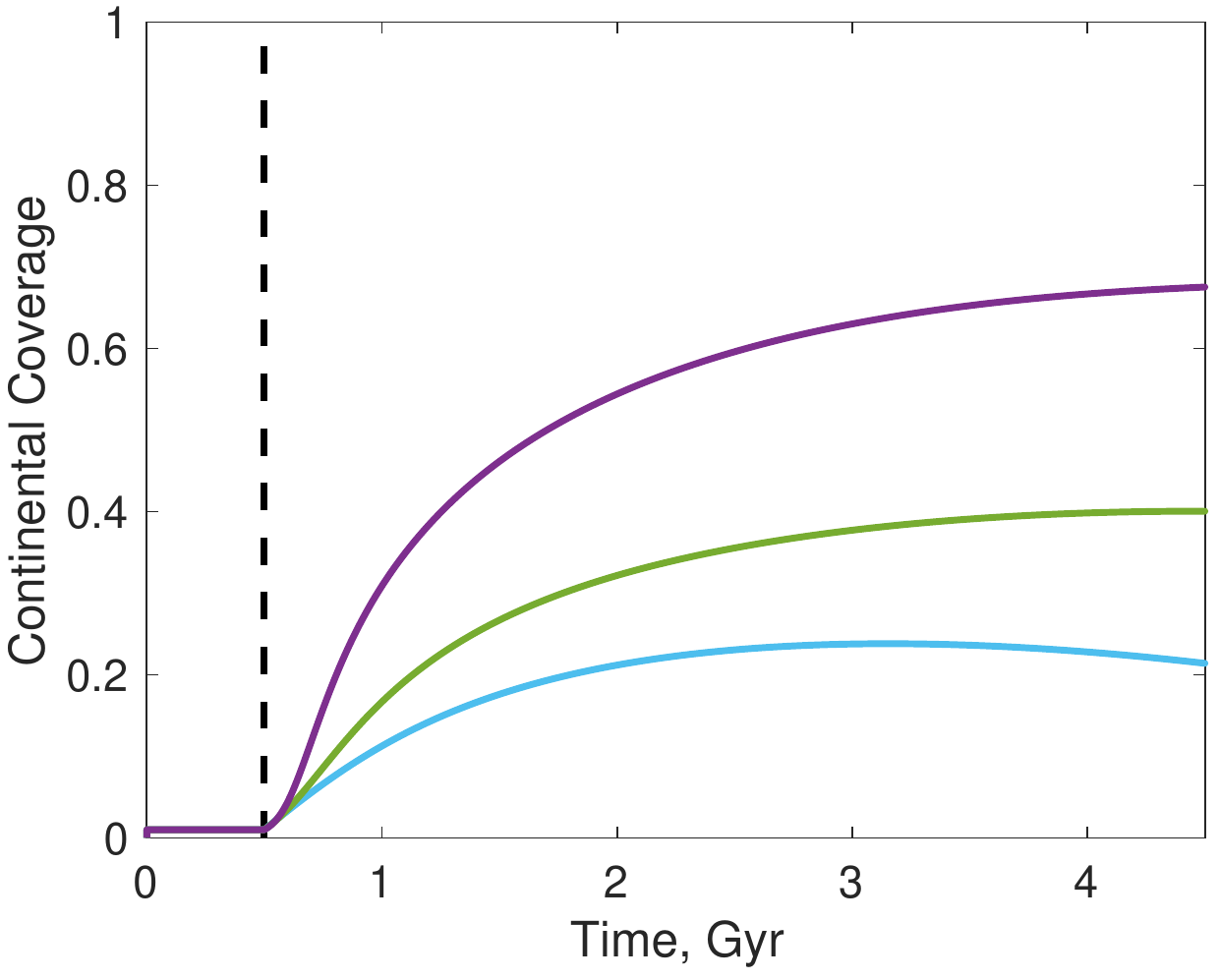}
    \end{subfigure}
    \begin{subfigure}[b]{0.45\textwidth}   
        \centering 
        \includegraphics[width=\textwidth]{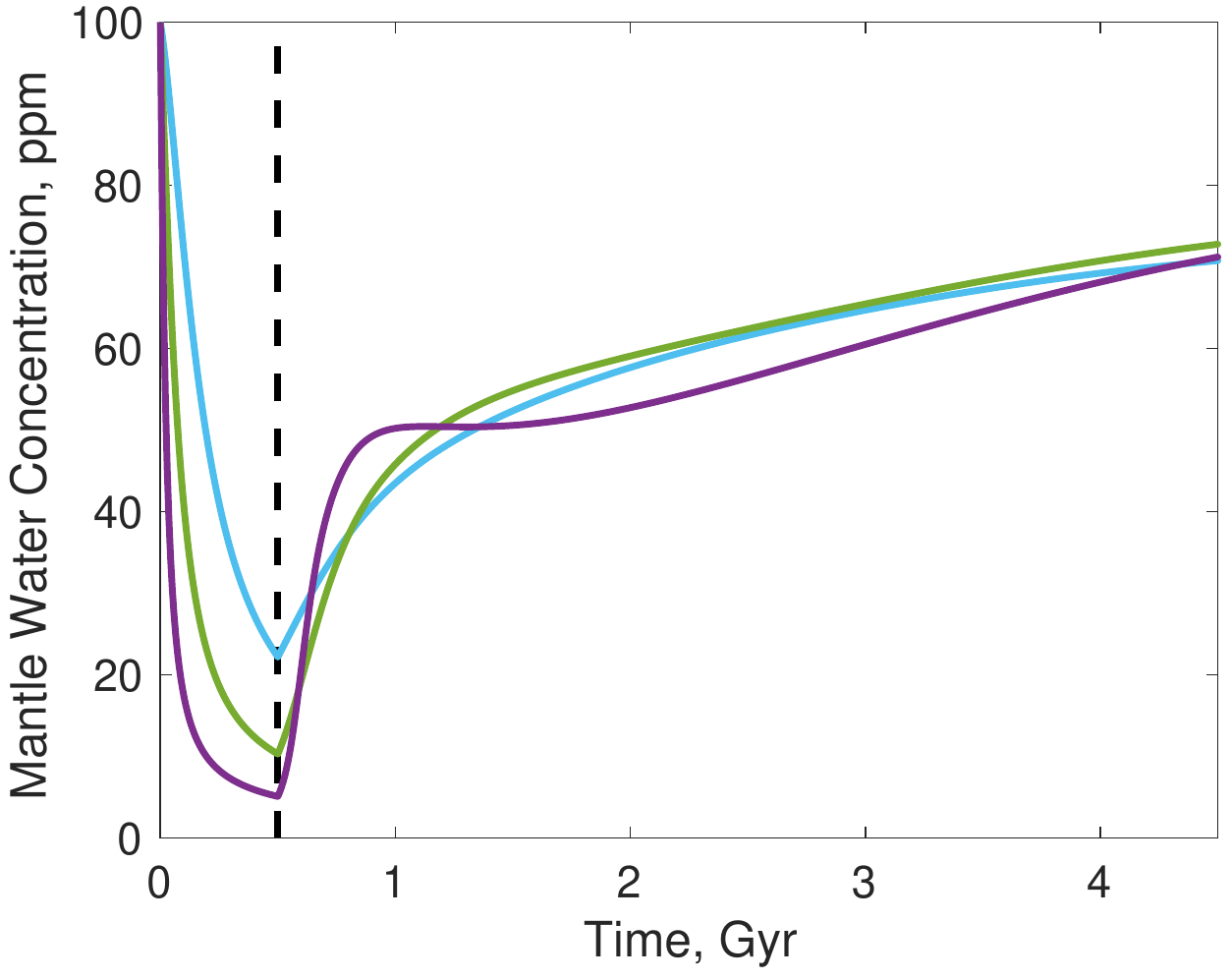}
    \end{subfigure}
    \hfill
    \begin{subfigure}[b]{0.45\textwidth}   
        \centering 
        \includegraphics[width=\textwidth]{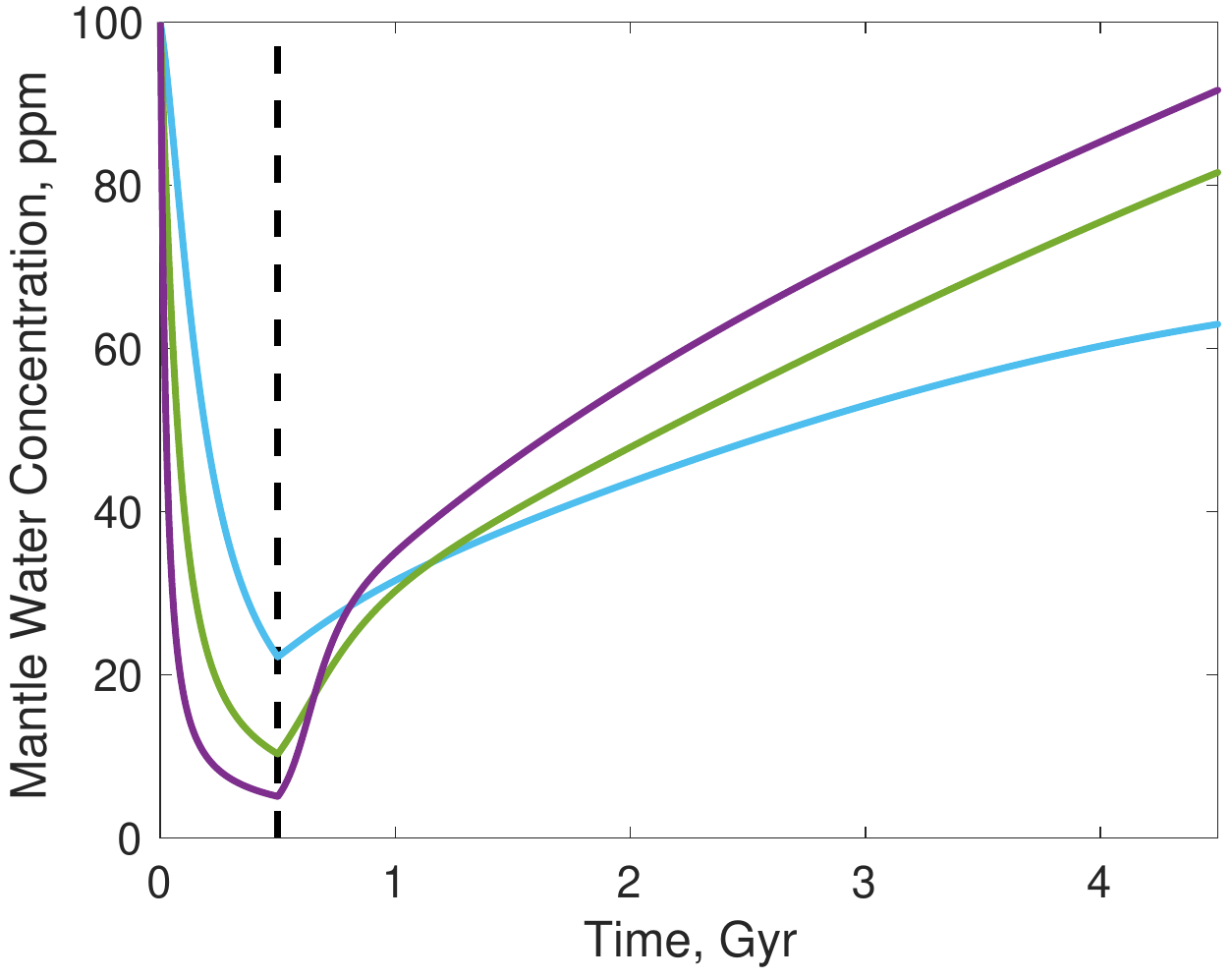}
    \end{subfigure}
    \begin{subfigure}[b]{0.45\textwidth}   
        \centering 
        \includegraphics[width=\textwidth]{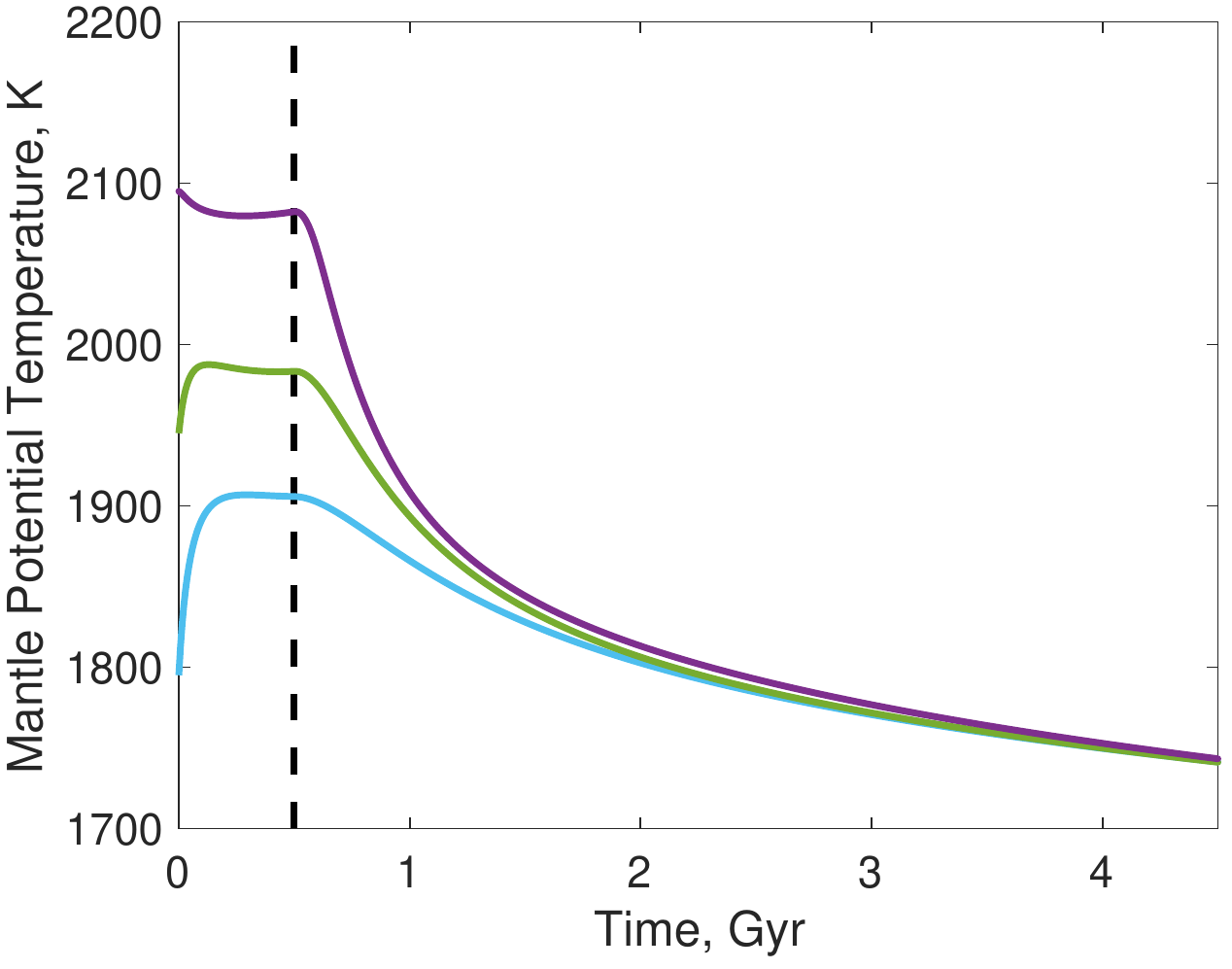}
    \end{subfigure}
    \hfill
    \begin{subfigure}[b]{0.45\textwidth}   
        \centering 
        \includegraphics[width=\textwidth]{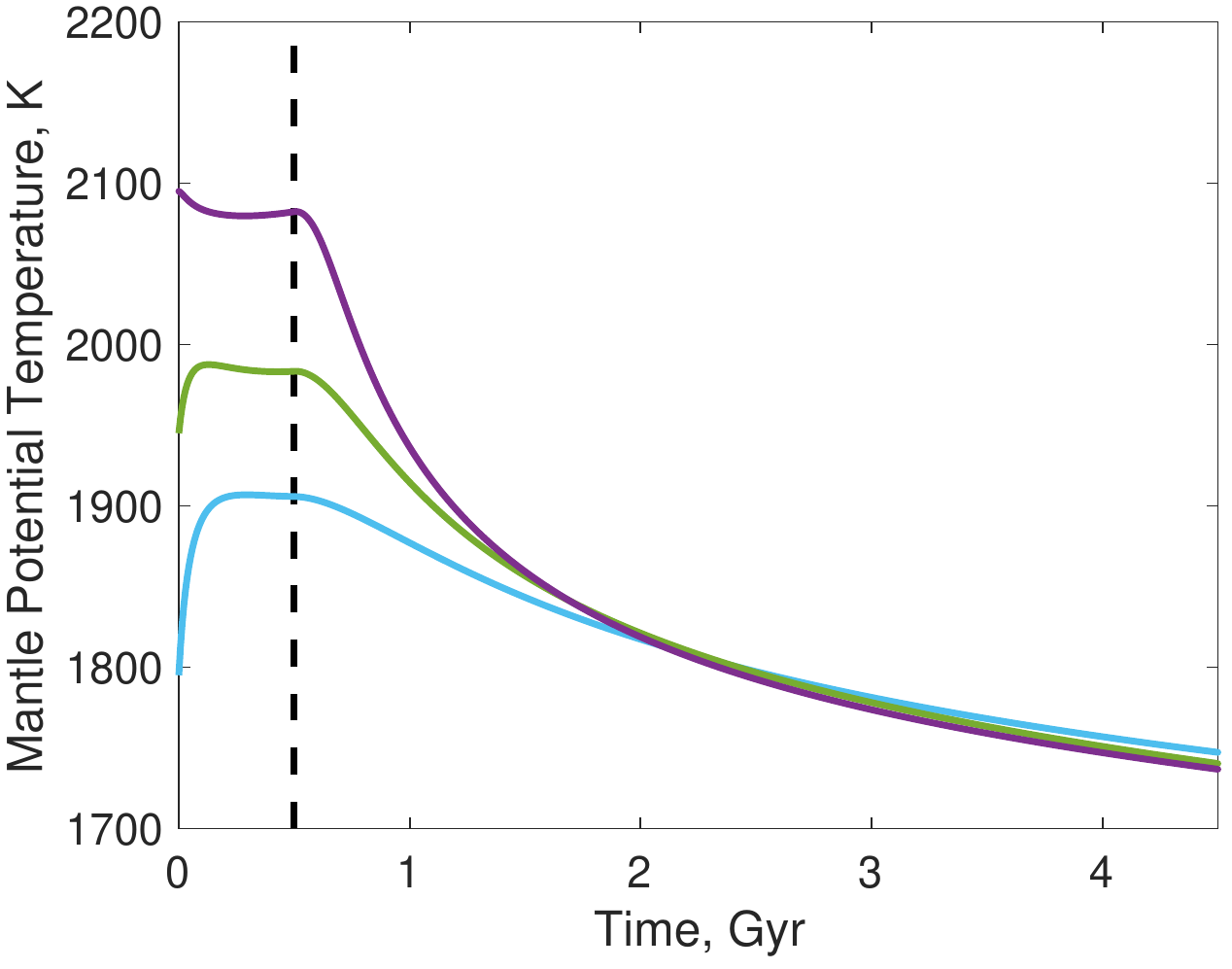}
    \end{subfigure}
    \begin{subfigure}[b]{0.45\textwidth}   
        \centering 
        \includegraphics[width=\textwidth]{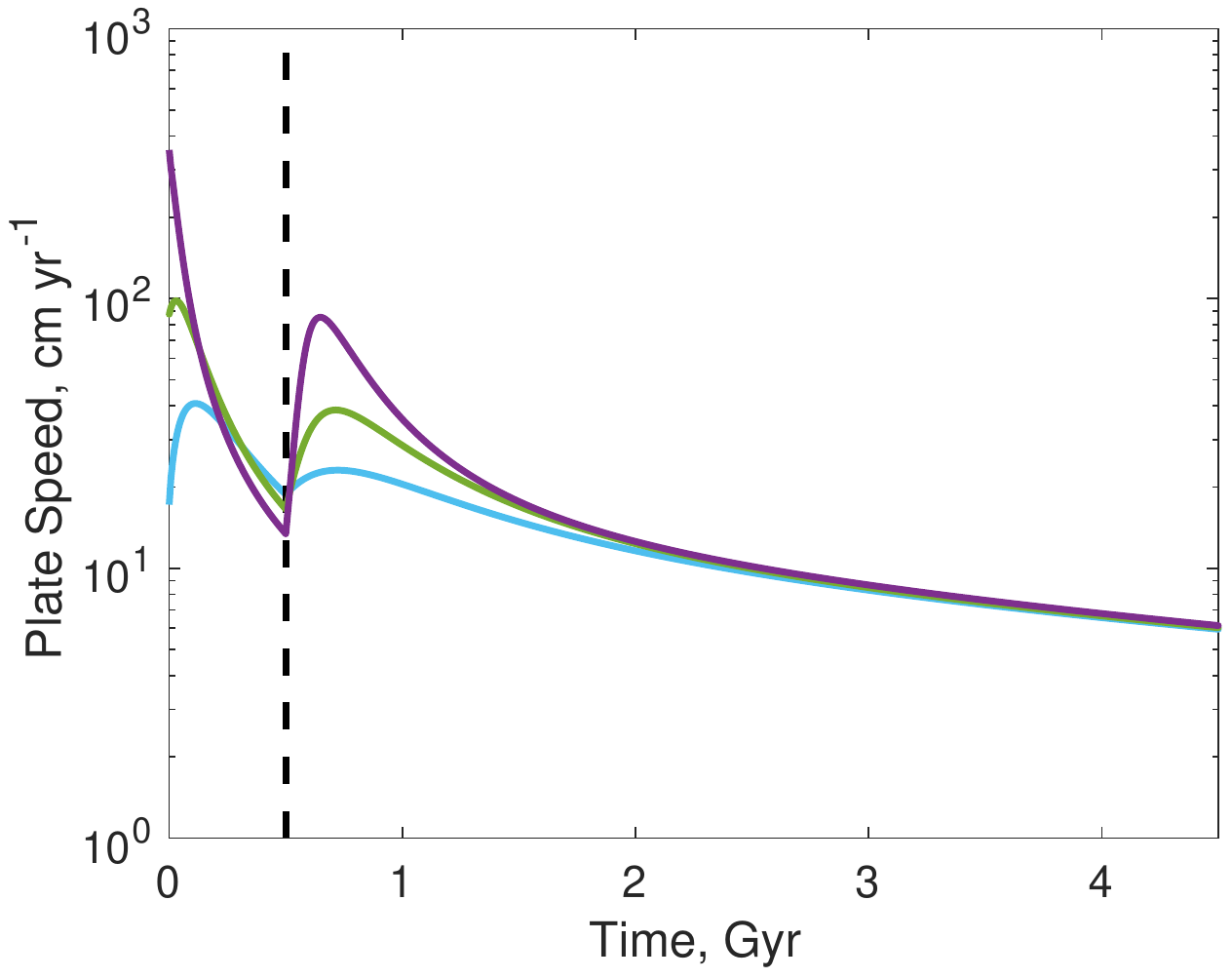}
    \end{subfigure}
    \hfill
    \begin{subfigure}[b]{0.45\textwidth}   
    \centering 
        \includegraphics[width=\textwidth]{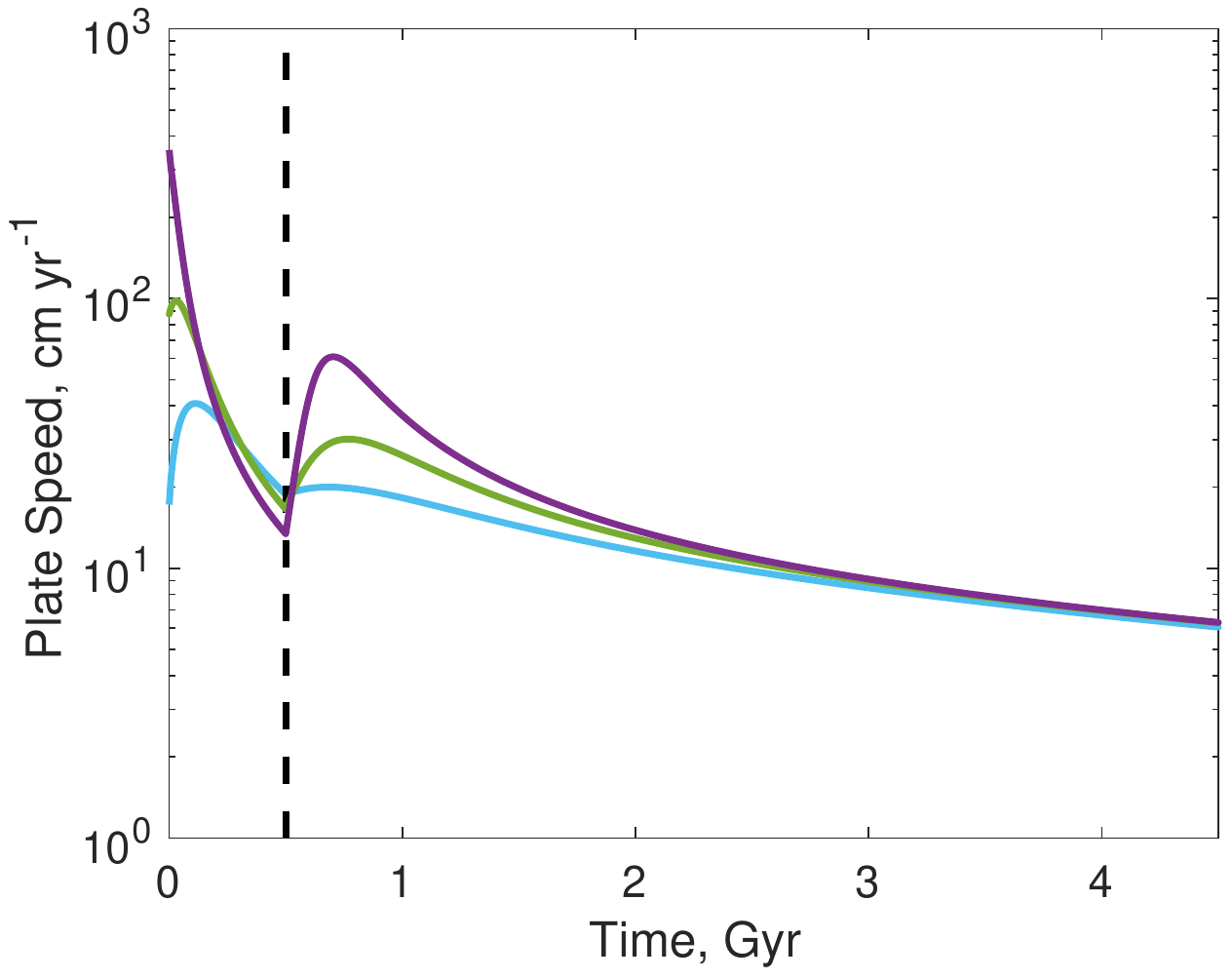}
    \end{subfigure}
    \caption
    {Continental coverage, mantle water concentration and temperature and plate speed over time (top to bottom) for the models of Fig. \ref{fig:3dplot}. Left: Negative feedback dominating, $\xi_1 < \xi_2$. Right: Positive feedback dominating, $\xi_1 > \xi_2$.}
    \label{fig:interior}
\end{figure*}

In the following, we will present numerical solutions to Eqs. \ref{eq:continent} -- \ref{eq:temperature_new} in the $A, w, T$-phase space (Fig. \ref{fig:3dplot}). We also show phase planes at the present day mantle temperature and at 1850 K, respectively, with graphs of the integrals of $\dot{A}=0$ and $\dot{w}=0$. The latter phase planes can be compared with those in Fig. \ref{fig:quiver} (in particular panels b and d). The parameters $\xi_3 = 0.6$ and $\xi_4 =0$ have been chosen such that the respective terms cancel in Eq. \ref{eq:temperature_new} as thermal blanketing and radiogenic depletion are neglected. In Fig. \ref{fig:interior} we show the state variables for the models as functions of time along with the plate speed $v_p$.

Model calculations are started at $t = 0$  by integrating Eqs. \ref{eq:water} and \ref{eq:temperature_new} with a given initial temperature $T_0$ and water content $w_0 = 100$ ppm. At a later time $t = t_p$ according to the assumed onset of plate tectonics -- or more specifically, the onset of continental growth by subduction zone melting --  we include Eq. \ref{eq:continent} in the integration with $A_0=0$ as starting value. The onset time $t_p$ is 0.5 Gyr for the models in Figs. \ref{fig:3dplot} and \ref{fig:interior}. Note that the onset of continental growth by subduction zone melting needs to be started that early as we grow the entire continental surface area by that mechanism. The starting values could be easily adjusted if plate tectonics is taken to start with the Proterozoic at $t_p = 2.5$ Gyr and the Archean continents had grown by a different mechanism to some area $A_0 > 0$ at the time.

We present three models in each panel of Fig. \ref{fig:3dplot} that differ by the chosen initial mantle temperature, colour coded cyan, green and purple. The initial mantle temperature for the models belonging to the green curves (1945 K) in both panels was chosen such that a continental coverage of 40\% as on present-day Earth results after 4.5 Gyr. For the models belonging to the purple curves, a 150 K higher initial temperature was chosen; for the models belonging to the cyan curves the initial temperature  was chosen 150 K lower. Otherwise the models have the same parameter values. We note that at given values of $\xi_1$, $\xi_2$, and $f_r$, the values of $T_0$ and $t_p$ can be traded against each other to arrive at the Earth's present-day continental coverage. For instance, we could have used a lower initial mantle temperature in combination with a later onset time $t_p$. We could also have set the value of $f_r>1$, i.e. forced the present-day Earth continental growth rate to exceed the erosion rate, which would, for example, have allowed a smaller value of $T_0$.

Fig. \ref{fig:3dplot} shows how the number of fixed points and their locations in the phase planes change with temperature. The phase planes at 1850 K ($t \approx 1.5$ Ga) in the top and bottom panels resemble each other by both showing a single fixed point. This can be understood as resulting from the strong temperature dependence of the viscosity and hence the flow speed $v_p$ which causes the term proportional to $v_p$ to dominate the production rate in Eq. \ref{eq:continent}, almost regardless of the chosen value of $\xi_1$. It is interesting to note that the graph of the integral of $\dot{A} = 0$ in the bottom panel already begins to attain an S-shape which it does not in the top panel. However, the single fixed point in the top panel with $\xi_1 < \xi_2$ at 1850 K is at a notably smaller continental coverage, thereby reflecting the stronger dependence of the erosion rate on $A$ under these conditions. At $A<0.6$, the two integral curves $\dot{A}=0$ (red) and  $\dot{w} = 0$ (blue) are almost independent of $w$ again reflecting the strong dependence of continental growth and mantle water hydration on temperature through $v_p(T,w)$. The importance  of the mantle water concentration increases with decreasing temperature, in particular at values of $A>0.6$ for which the length of ocean-continent subduction zones substantially decreases with increasing $A$.

All models in Figs. \ref{fig:3dplot} and \ref{fig:interior} begin with rapid degassing of water accompanied by an adjustment of the mantle temperature depending on the chosen value of $T_0$ even before continents start to grow. This adjustment to initial conditions is well known from thermal evolution models \citep[e.g.,][] {Breuer:2015}. The mantle then cools and temperatures converge until differences become insignificant after about 2 Gyr of evolution, although the models with $\xi_1>\xi_2$ cool a bit slower reaching 1880 K about 200 Myr later. The plate speed shown in the bottom panels of Fig. \ref{fig:interior} evolves similarly; after adjustment to initial conditions differences become small after about 2 Gyr of evolution. Again, the models with $\xi_1>\xi_2$ evolve less rapidly with a similar delay as for the mantle temperature. 

Differences between models with dominating positive versus negative feedback, respectively, are more significant for the mantle water concentration and, even more so, for the surface coverage by continental crust. With negative feedback dominating, $\xi_1 < \xi_2$, the mantle water concentration adjusts rapidly to initial conditions and increases steadily after a little more than 1 Gyr almost independent of the choice of initial mantle temperature to reach a present day value of about 70 ppm, the equilibrium value being about 65 ppm. For positive feedback dominating, $\xi_1 > \xi_2$, the initial adjustment of the variables to their initial conditions at times smaller than 0.5 Gyr corresponds to the one in the model with dominating negative feedback. But beyond 0.5 Gyr, the evolution paths diverge with greater and widening differences in mantle water concentrations and continental coverage. This is caused by the emergence of the three fixed points and the evolution paths tending towards them. 

Figs. \ref{fig:3dplot} and \ref{fig:interior} illustrate how positive feedback causes the present-day value of continental coverage to depend on its early evolution. Although mantle temperature values converge after a few billion years of evolution and differences in initial conditions disappear as predicted by the Tozer principle \citep{Tozer:1967}, differences in the early growth history of continental crust get emphasised if the feedback is positive as well as differences in the net mantle regassing rate persist. If negative feedback dominates, however, both continental coverage and mantle water concentration tend to converge towards their single equilibrium fixed point values. It should be noted that these equilibrium values evolve on their own with time or temperature while the state variables follow suit, albeit with some delay in time. It should further be noted that the continental surface area is predicted to shrink if negative feedback dominates while it is predicted to grow slightly or keep its level for the past 2 Gyr if positive feedback dominates. Moreover, we find a large surface coverage by continents to imply a wet mantle while a small coverage is related to a comparatively dry mantle. This is due to the fact that both gain rates -- continental production and mantle water regassing -- depend on the same physical parameters that control the transport of water in subduction zones. Because we neglect the loss of water due to atmospheric escape, the complement of the mantle water is to be found at the surface as ocean water.

\subsection{Implications for the long-term carbon cycle}
\label{sec4}

\begin{figure*}
    \centering
    \begin{subfigure}[b]{0.475\textwidth}
        \centering
        \includegraphics[width=\textwidth]{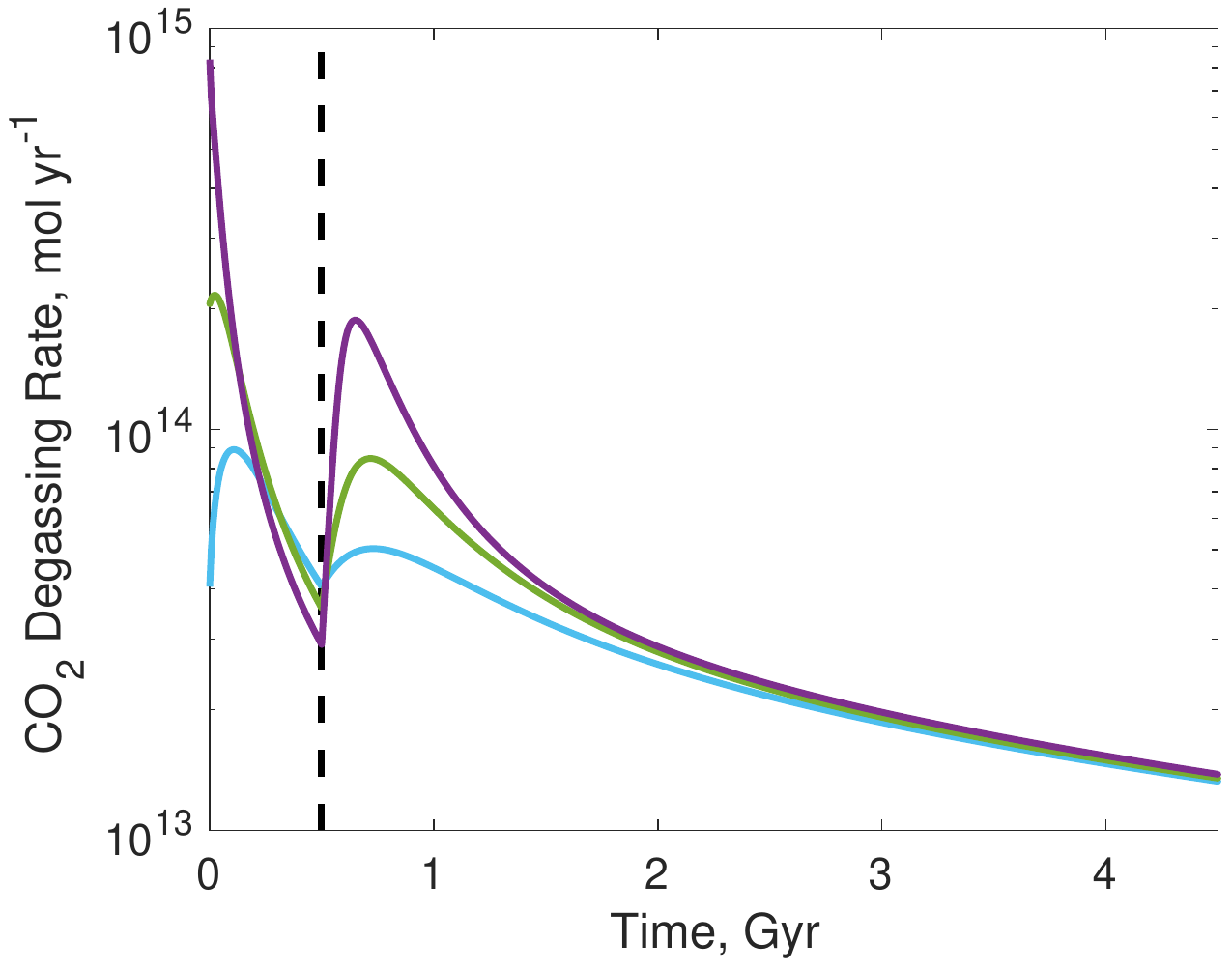}
    \end{subfigure}
    \hfill
    \begin{subfigure}[b]{0.475\textwidth}  
        \centering 
        \includegraphics[width=\textwidth]{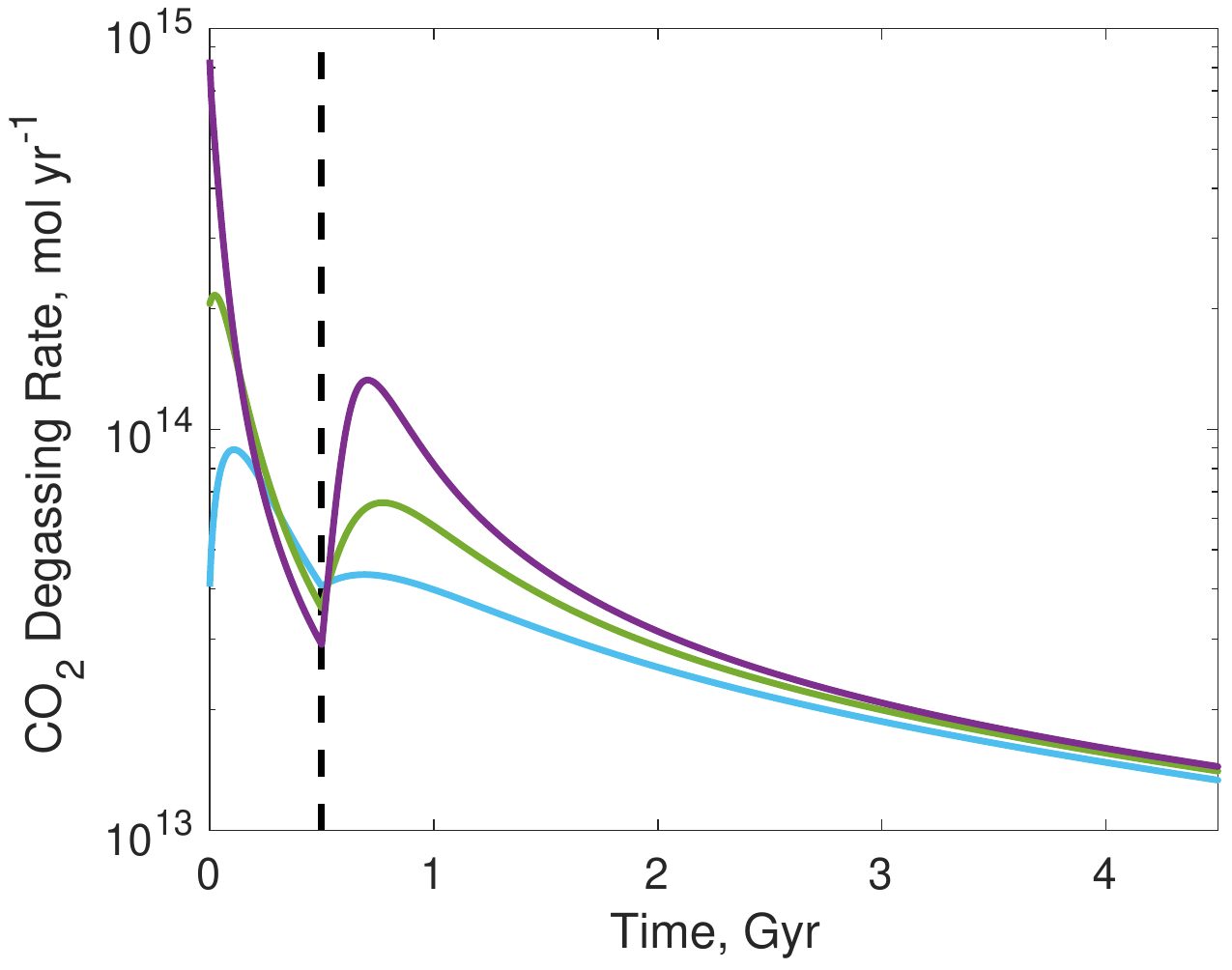}
    \end{subfigure}
    \begin{subfigure}[b]{0.475\textwidth}   
        \centering 
        \includegraphics[width=\textwidth]{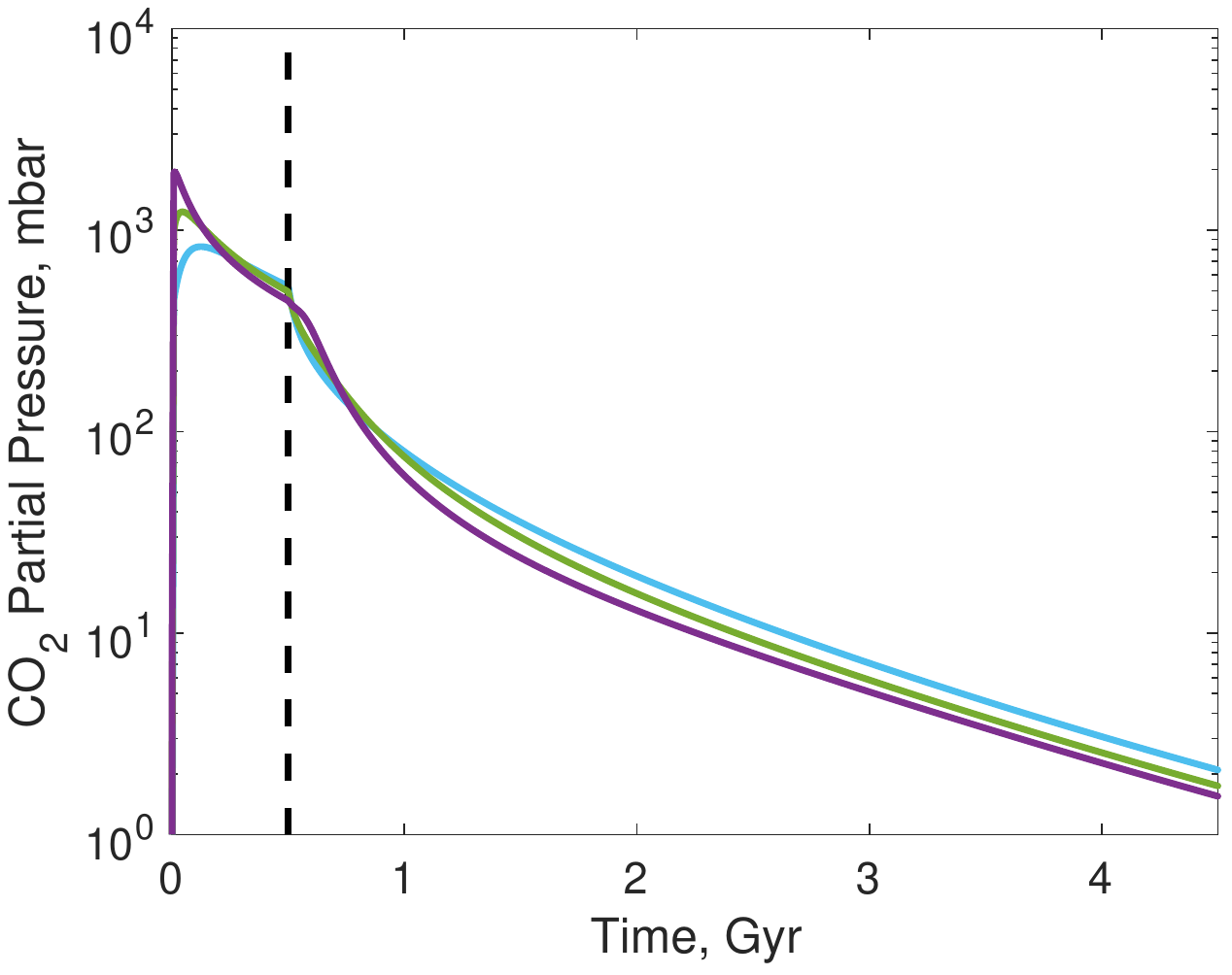}
    \end{subfigure}
    \hfill
    \begin{subfigure}[b]{0.475\textwidth}   
        \centering 
        \includegraphics[width=\textwidth]{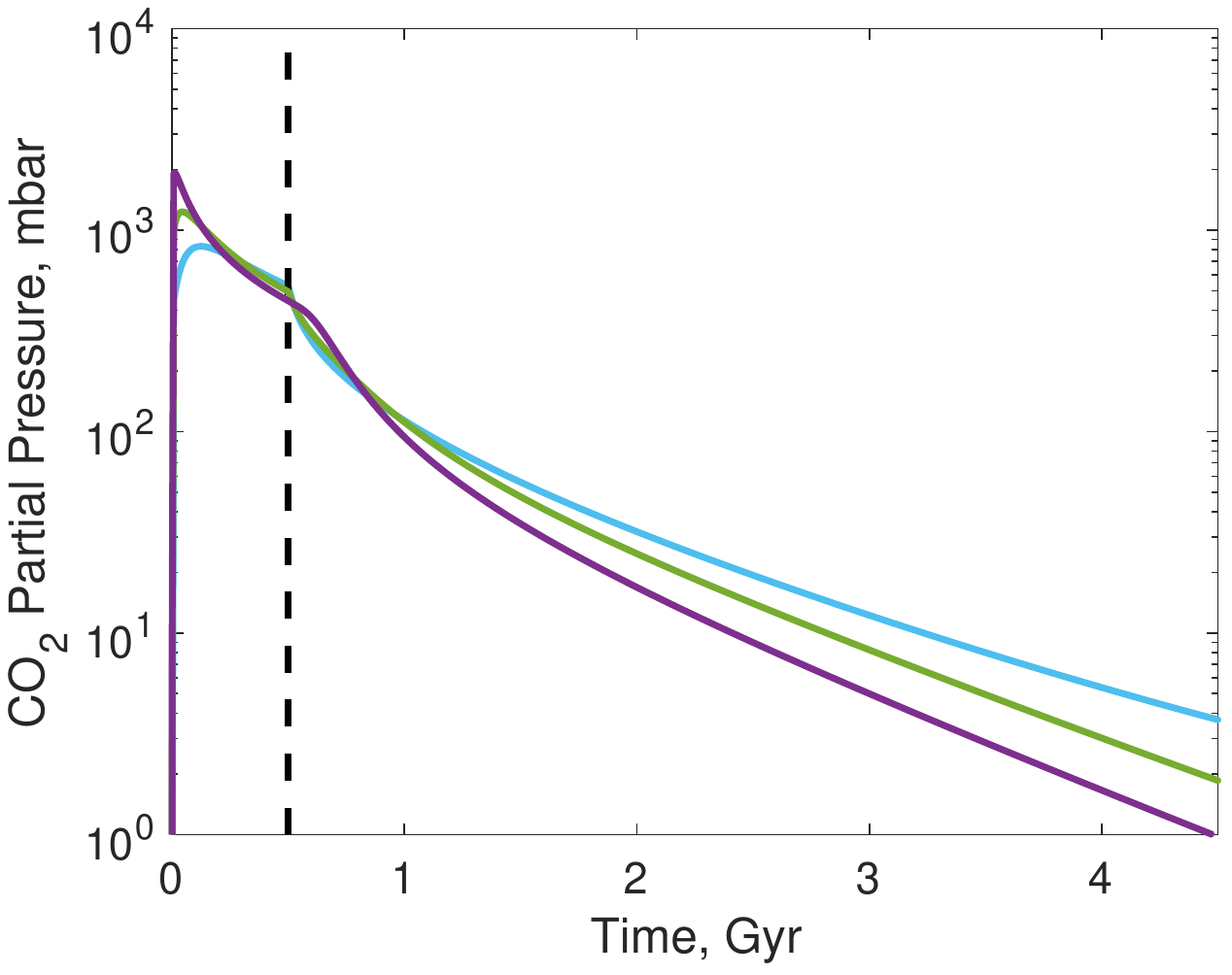}
    \end{subfigure}
    \begin{subfigure}[b]{0.475\textwidth}   
        \centering 
        \includegraphics[width=\textwidth]{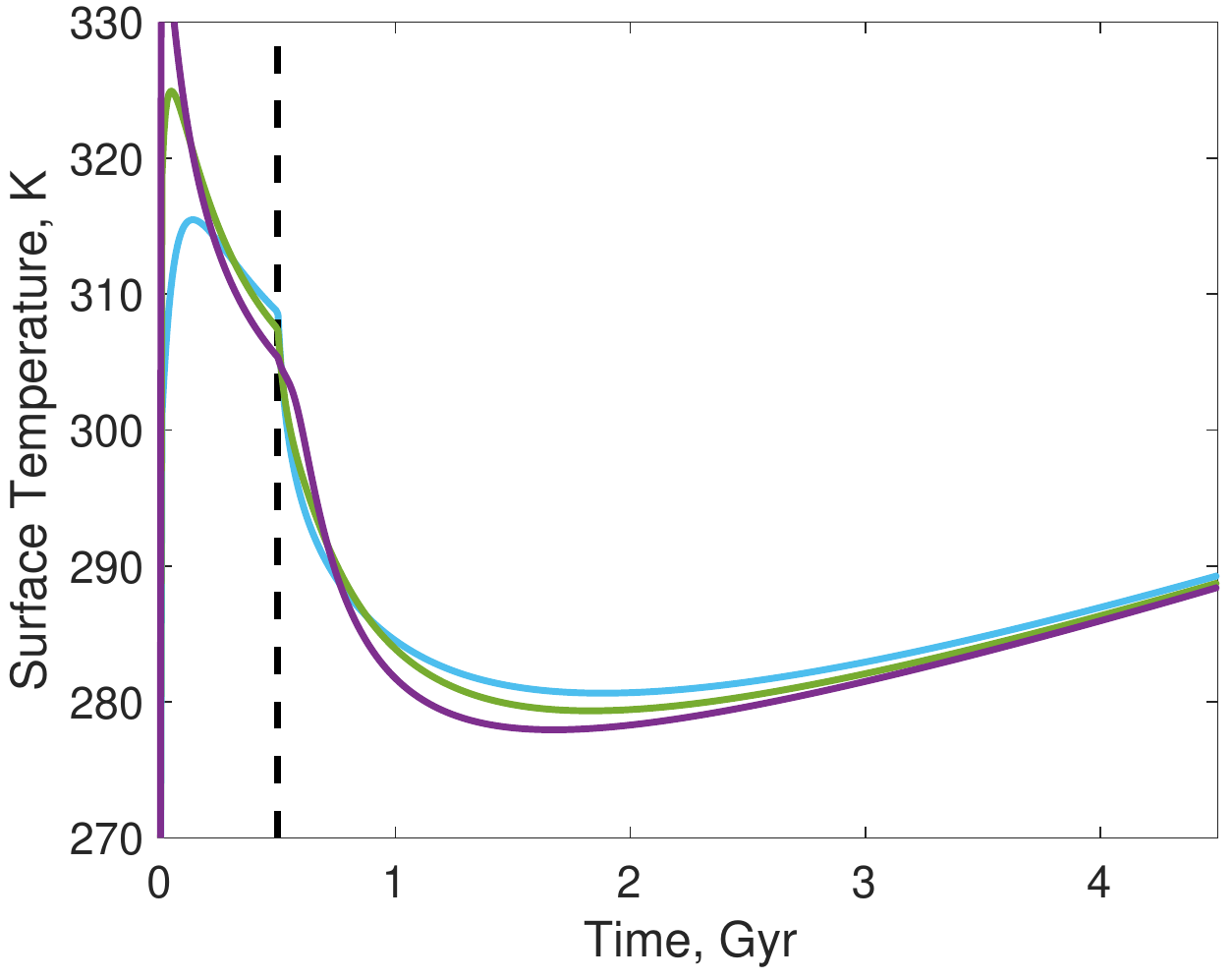}
    \end{subfigure}
    \hfill
    \begin{subfigure}[b]{0.475\textwidth}   
        \centering 
        \includegraphics[width=\textwidth]{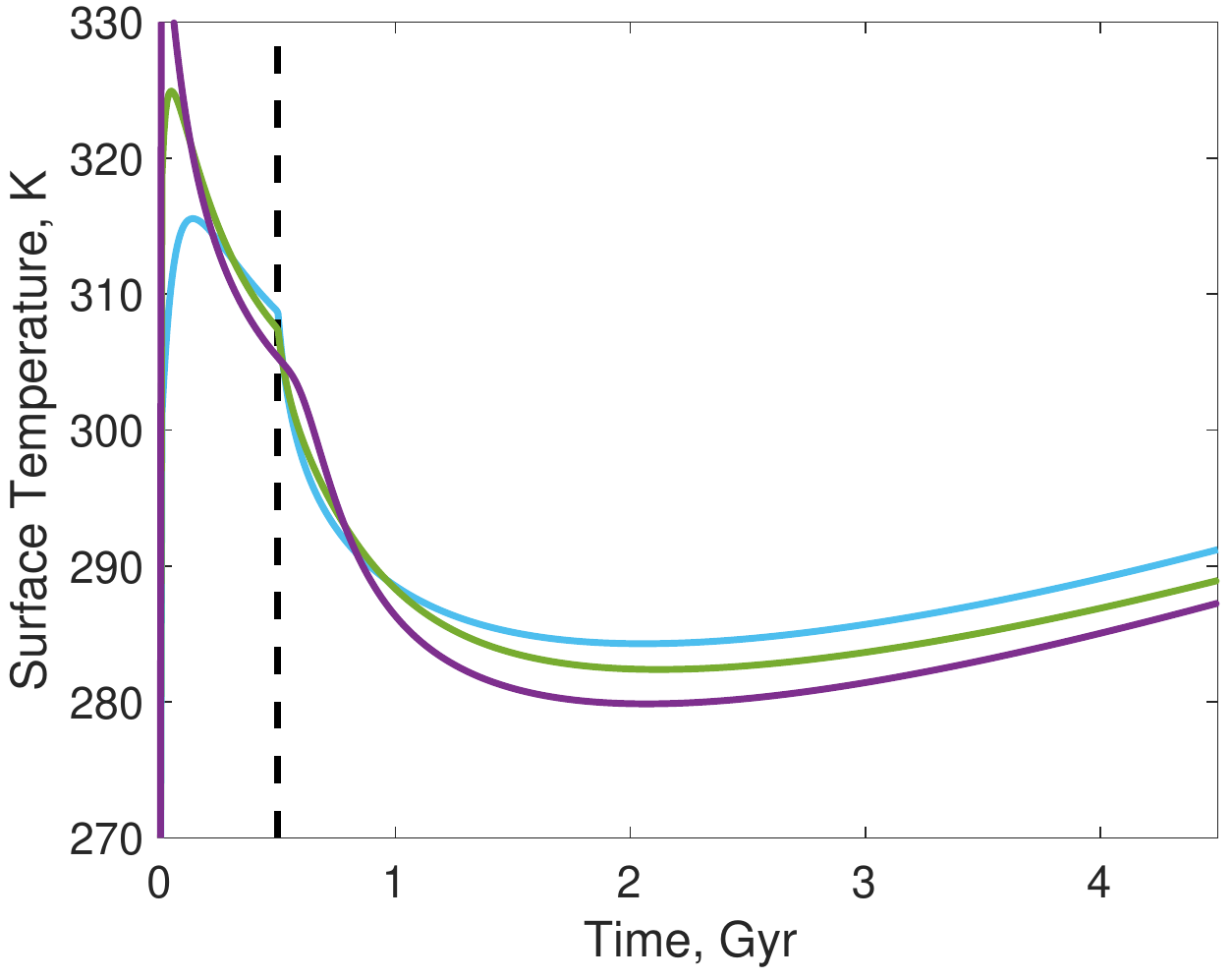}
    \end{subfigure}
    \caption{CO$_2$ degassing rate, CO$_2$ partial pressure and surface temperature as functions of time. The color code is the same as in Figures \ref{fig:3dplot} and \ref{fig:interior}. Left: Negative feedback dominating. Right: Positive feedback dominating.
    }
    \label{fig:climate}
\end{figure*}

The surface area of continental crust emerged above sea level has a direct impact on the climate because the rates of erosion and silicate weathering increase with increasing land area \citep[e.g.,][]{Foley:2015,graham2020,levenson2021}. A planet with a larger area of emerged continents will thus have a lower concentration of CO$_2$ in its atmosphere (all other factors being the same) and a lower surface temperature as compared with a planet with a smaller land area. For details of the calculation of the CO$_2$ degassing rate and partial pressure and the surface temperature see \ref{sec:carboncycle}. In Fig. \ref{fig:climate}, we plot the CO$_2$ degassing rate, partial pressure, and surface temperature as functions of time for the models with dominating negative (left panels) and positive (right panels) feedback shown in Figs. \ref{fig:3dplot} and \ref{fig:interior}. While differences in the degassing rate decrease with time and are negligible after 4.5 Gyr, the partial pressure of CO$_2$ is lower for the model starting with 150 K higher initial mantle temperature, in particular, if positive feedback dominates. And vice versa for the model starting with the lower initial temperature. This observation is consistent with the differences in continental surface area. All in all, with dominating positive feedback, differences in the surface temperature by about 5 K result after 4.5 Gyr even though all other parameter values are the same.

\subsection[Cooling with thermal blanketing]{Cooling with thermal blanketing and mantle depletion in radiogenic elements}
\label{sec5}

\begin{figure}
   \centering
  \includegraphics[clip,width=\columnwidth]{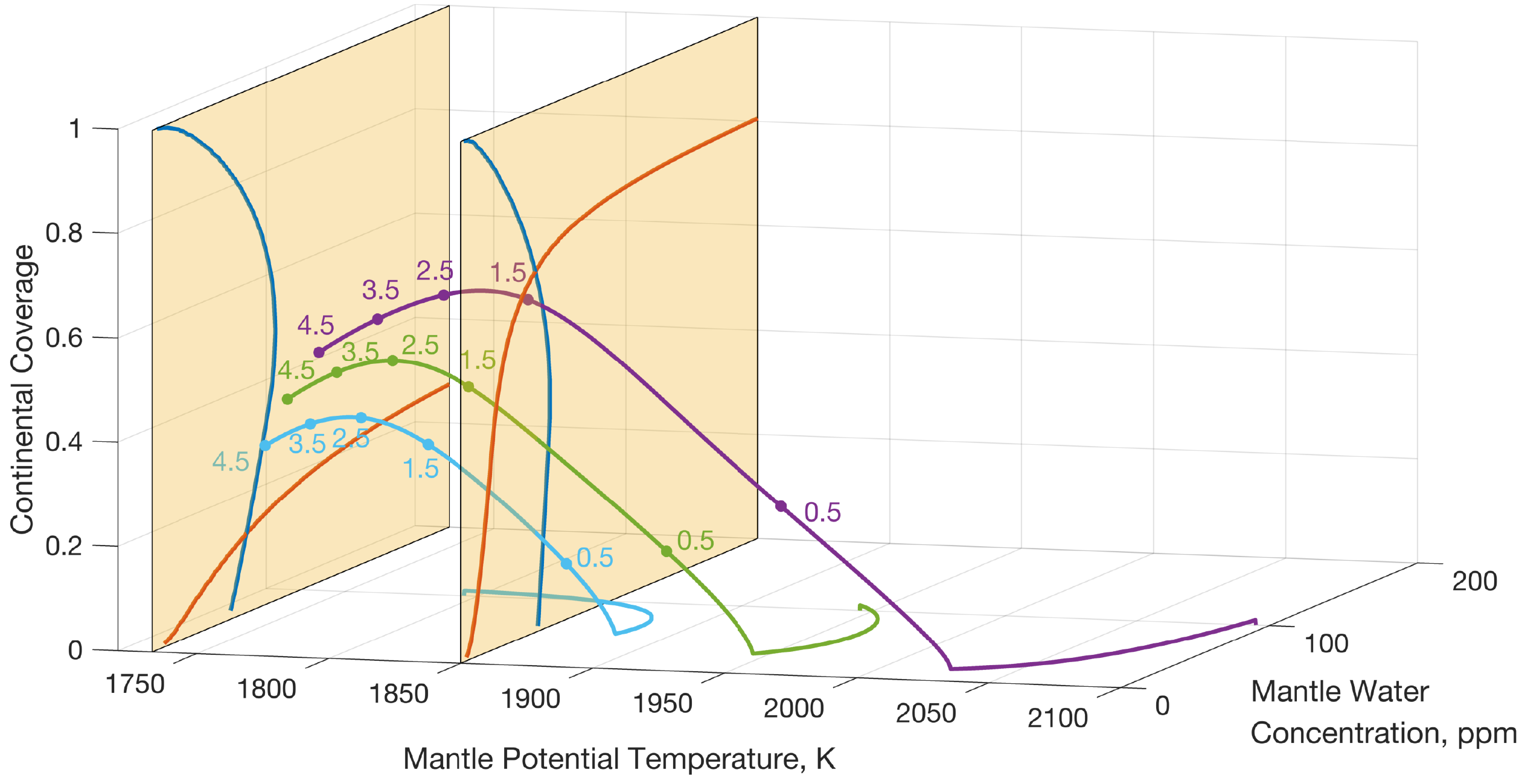}%
  
  \includegraphics[clip,width=\columnwidth]{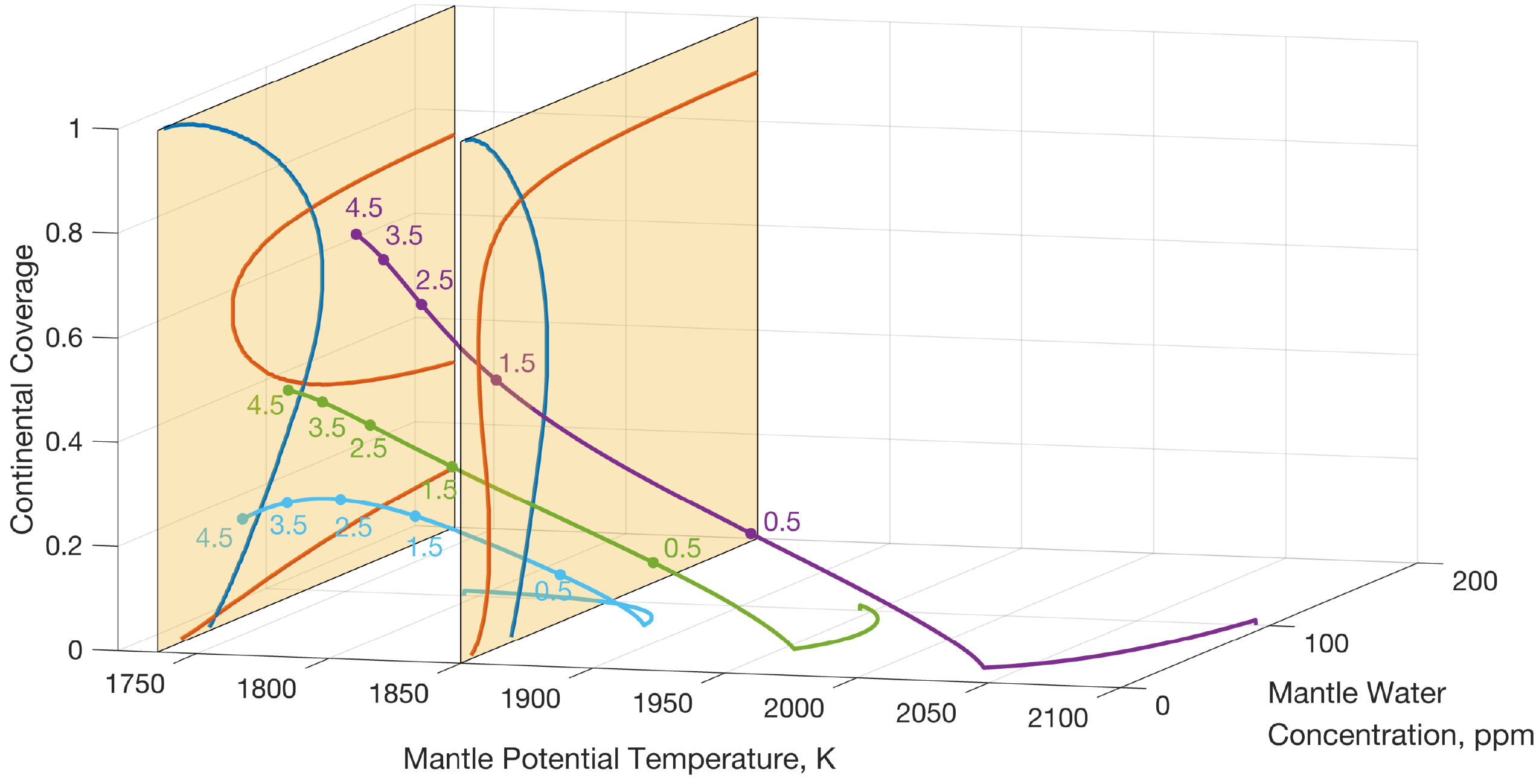}%
   \caption{Same as Fig. \ref{fig:3dplot} but with continents that are enriched in radiogenic elements and that thermally blanket the mantle. Top: $\xi_1 =1/3$, $\xi_2=2/3$, $\xi_3=0.85$ and $\xi_4=1/4$, $f_r=0.775$, onset time $t_p = 0.253$ Gyr; bottom: $\xi_1=2/3$, $\xi_2=1/3$, $\xi_3=0.85$ and $\xi_4=1/4$, $f_r=1$, onset time $t_p =  0.161$ Gyr.
   }
\label{fig:3dplot_insul}
\end{figure}

Thermal blanketing of the mantle by the continental crust and the depletion of the mantle in radiogenic heat sources transferred to the crust can provide additional feedback between the crust growth rate and its surface area. In Fig. \ref{fig:3dplot_insul}, we plot results for models with the same parameter values as in Fig. \ref{fig:3dplot} but account for thermal blanketing by setting $\xi_3 = 0.85$ and $\xi_4=1/4$ in Eq. \ref{eq:temperature_new}. The onset time $t_p$ of continental growth by subduction zone melting for the green model was adjusted such that the continental surface fraction grows to 0.4 in 4.5 Gyr. For models with $\xi_1 = 1/3$ and $\xi_2 = 2/3$, $t_p$ was adjusted to 0.253 Gyr. For those with $\xi_1 = 2/3$ and $\xi_2 = 1/3$, $t_p$ was chosen to be 0.161 Gyr.

\begin{figure*}
    \centering
    \begin{subfigure}[b]{0.45\textwidth}
        \centering
        \includegraphics[width=\textwidth]{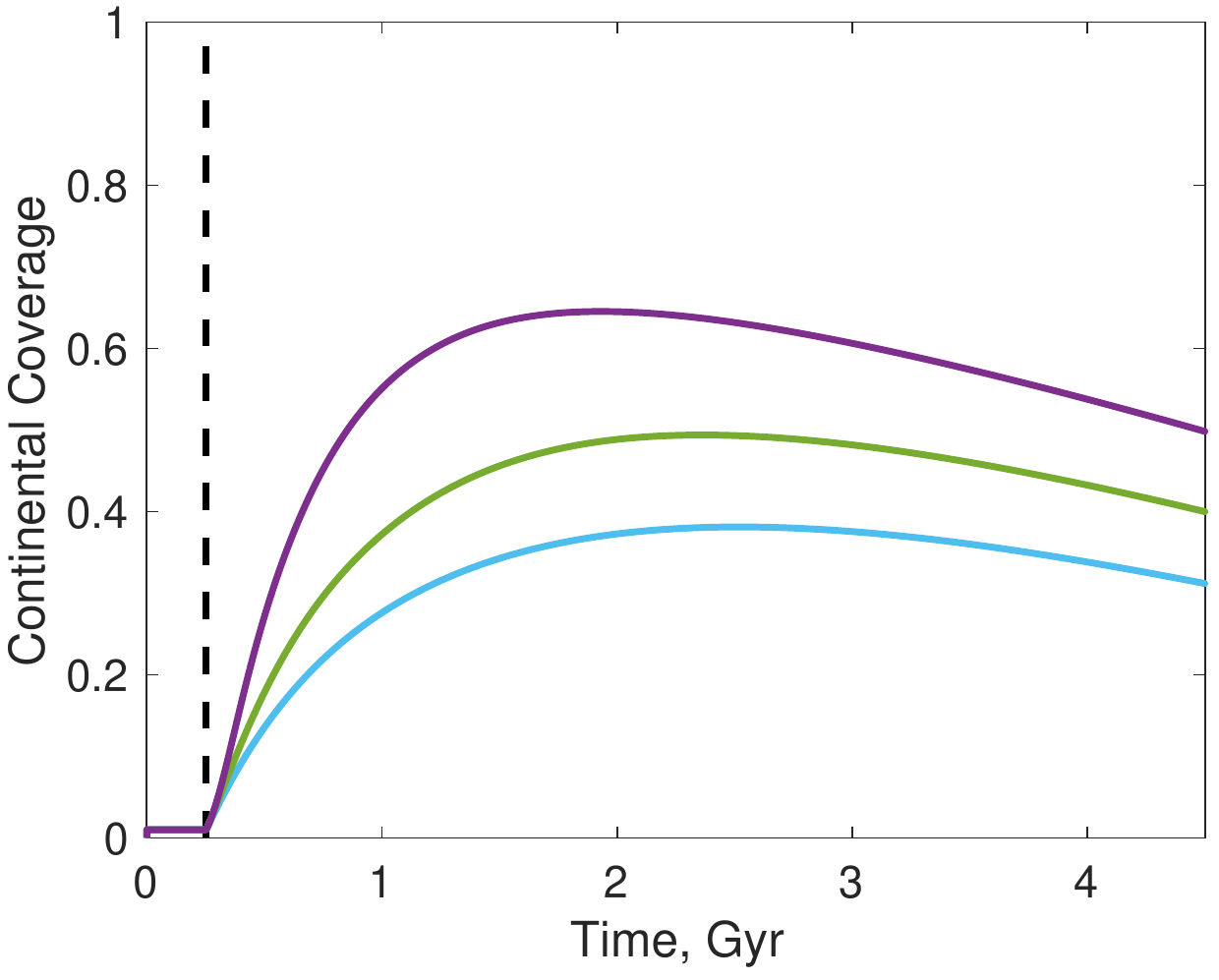}
    \end{subfigure}
    \hfill
    \begin{subfigure}[b]{0.45\textwidth}  
        \centering 
        \includegraphics[width=\textwidth]{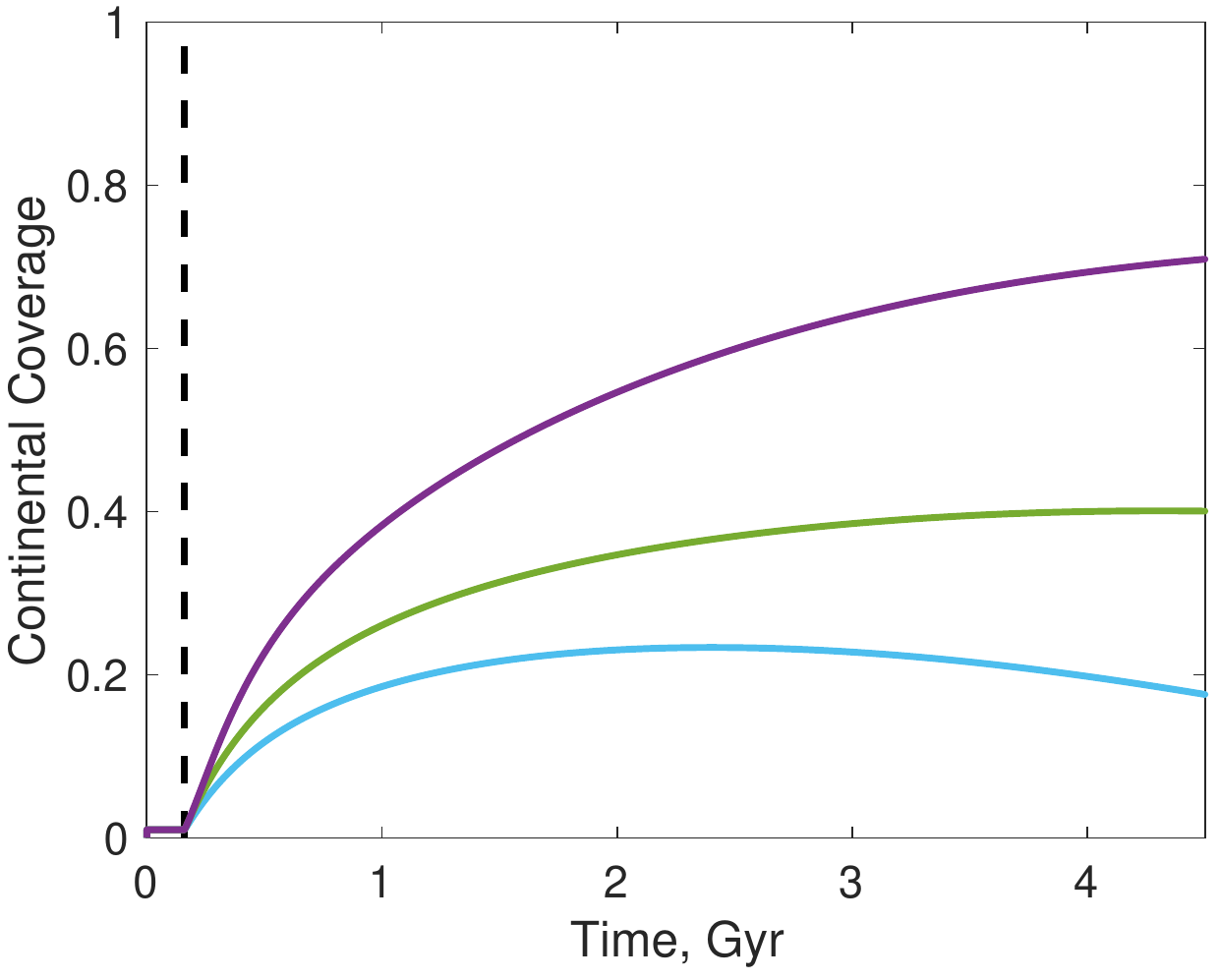}
    \end{subfigure}
    \begin{subfigure}[b]{0.45\textwidth}   
        \centering 
        \includegraphics[width=\textwidth]{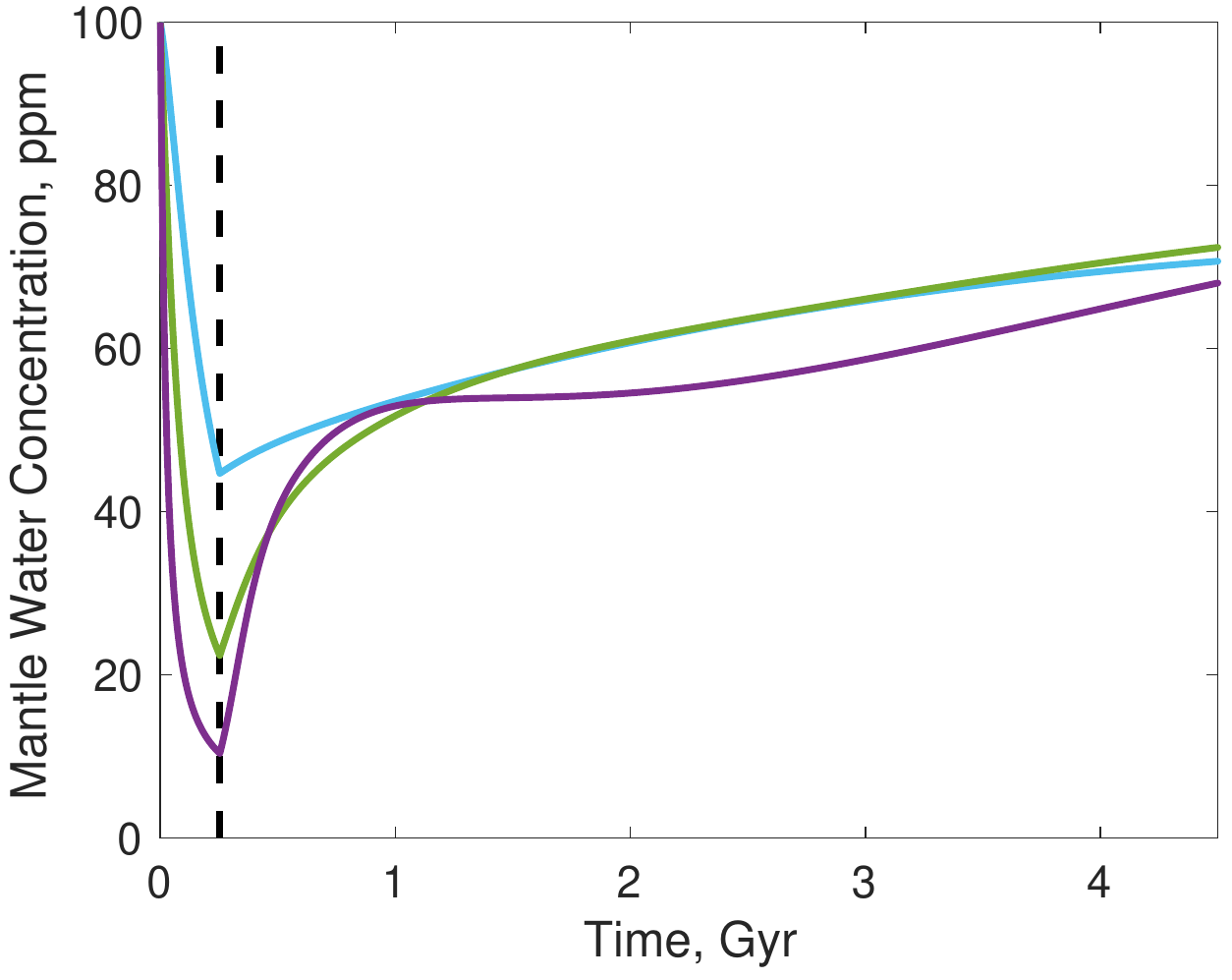}
    \end{subfigure}
    \hfill
    \begin{subfigure}[b]{0.45\textwidth}   
        \centering 
        \includegraphics[width=\textwidth]{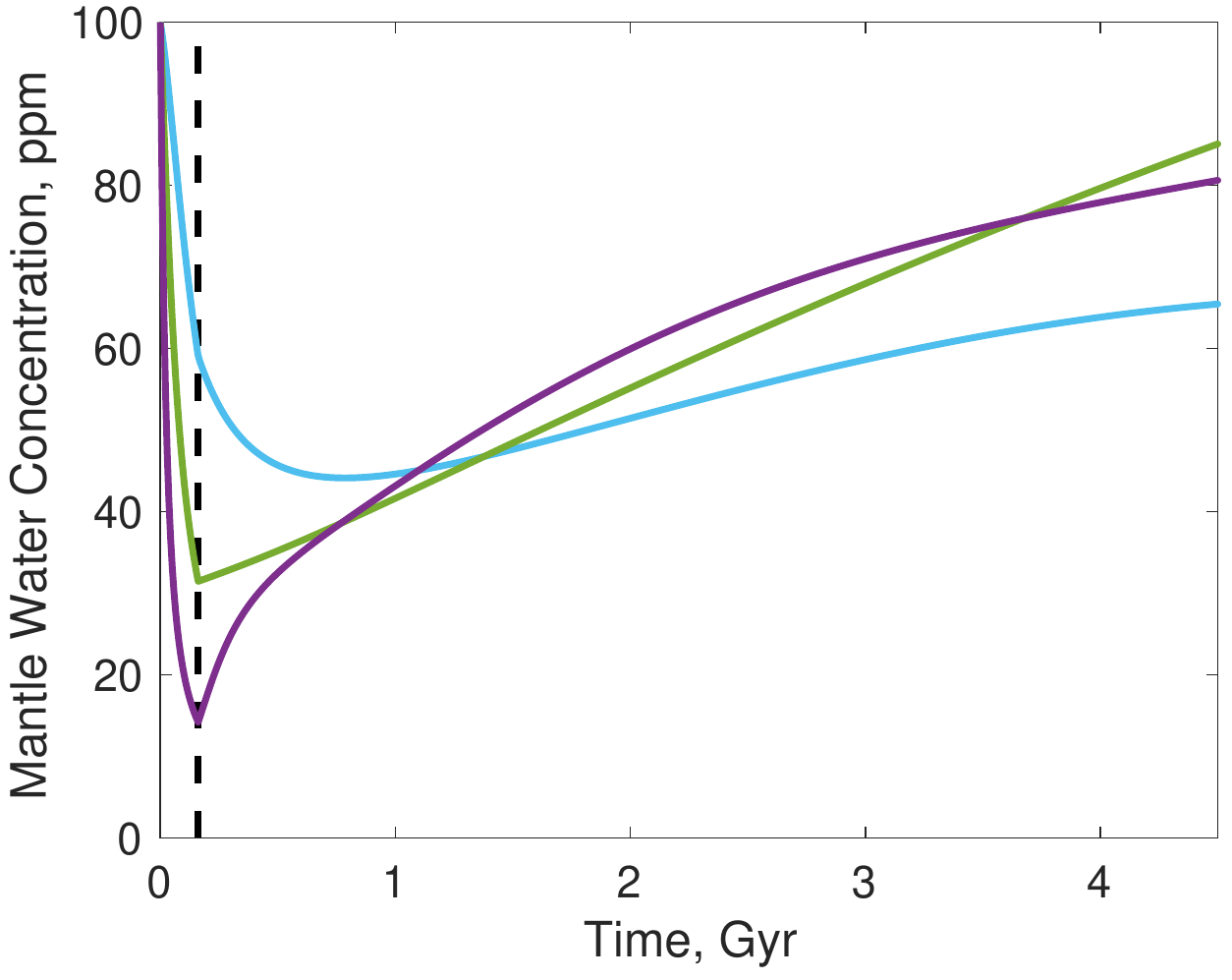}
    \end{subfigure}
    \begin{subfigure}[b]{0.45\textwidth}   
        \centering 
        \includegraphics[width=\textwidth]{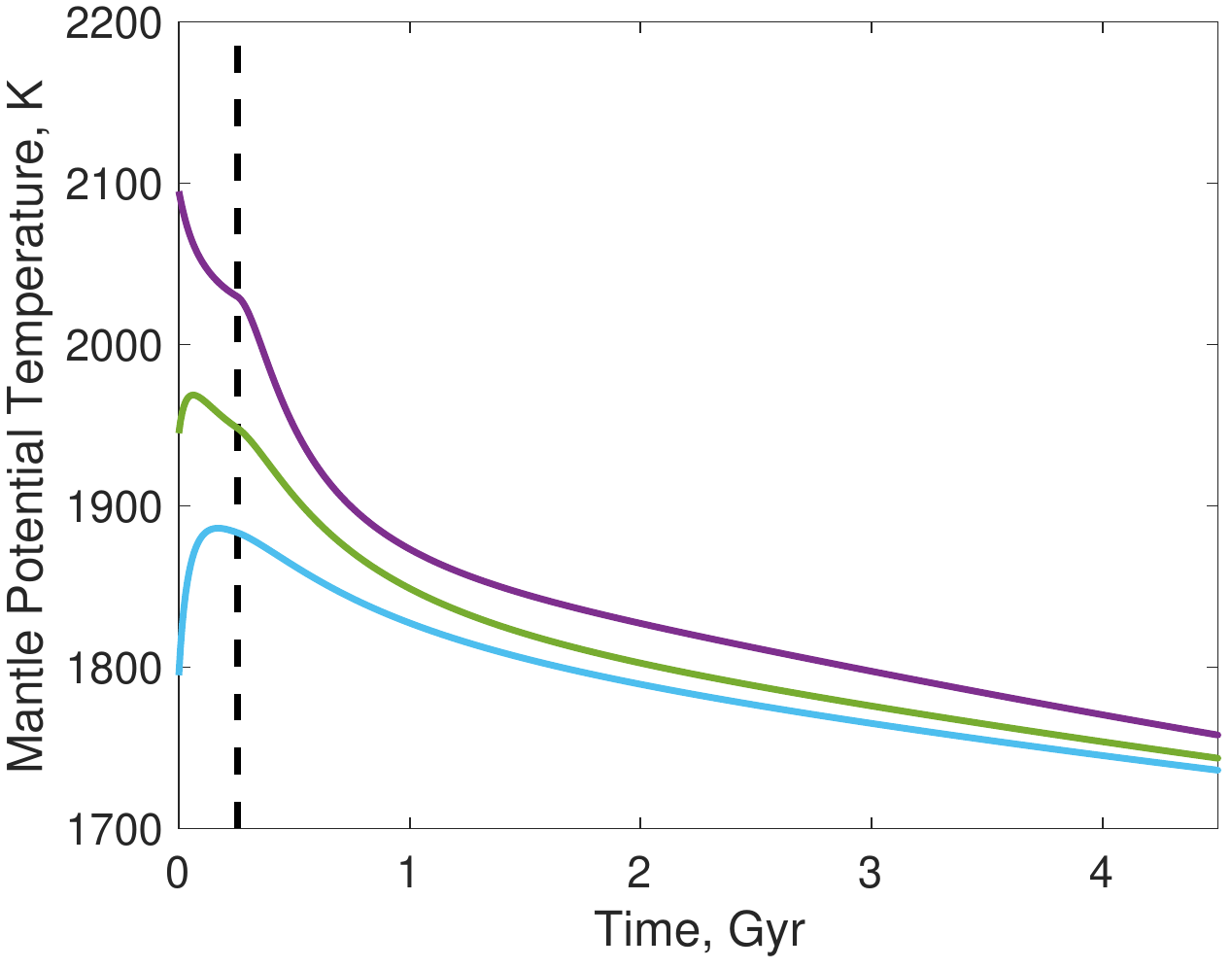}
    \end{subfigure}
    \hfill
    \begin{subfigure}[b]{0.45\textwidth}   
        \centering 
        \includegraphics[width=\textwidth]{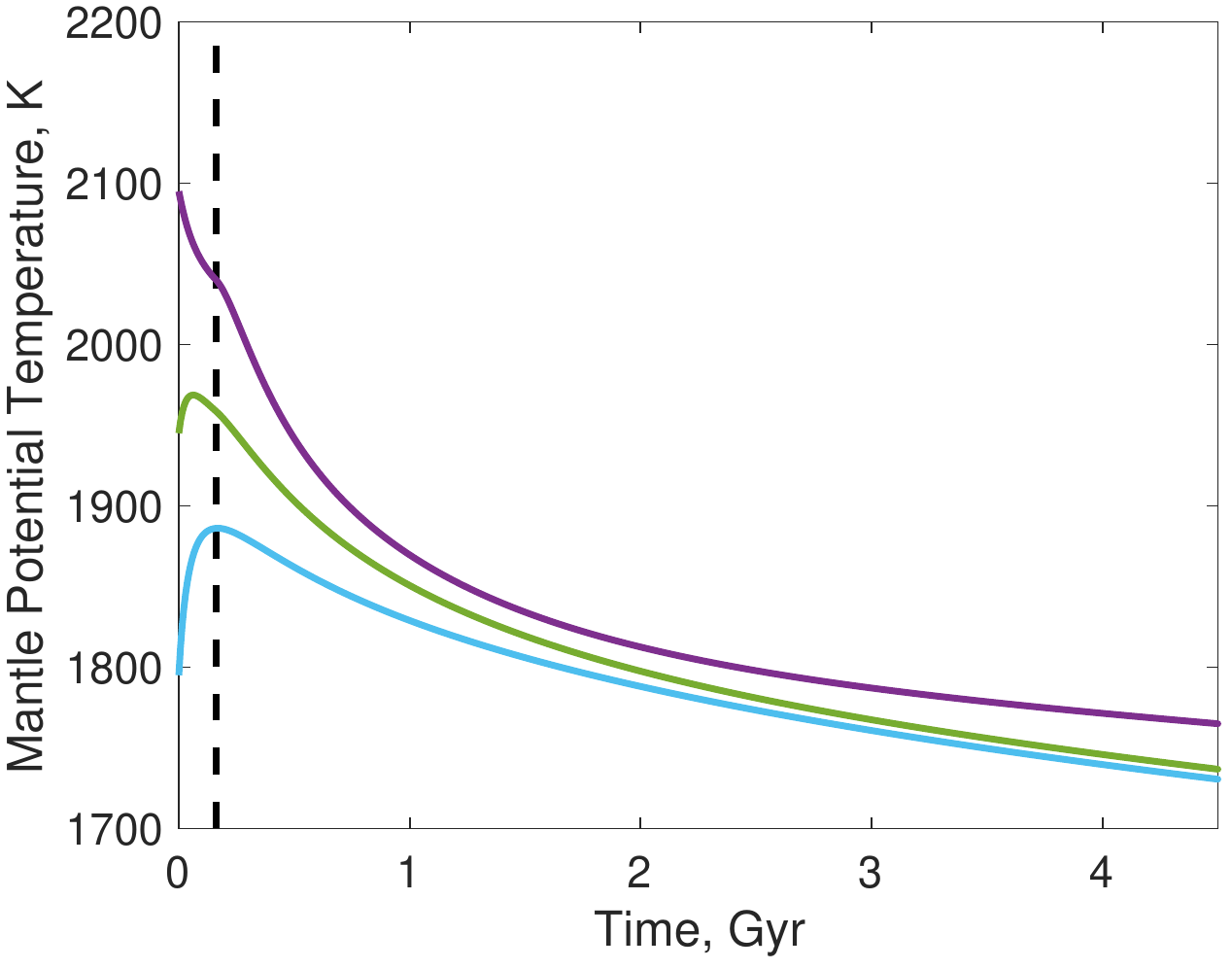}
    \end{subfigure}
    \begin{subfigure}[b]{0.45\textwidth}   
        \centering 
        \includegraphics[width=\textwidth]{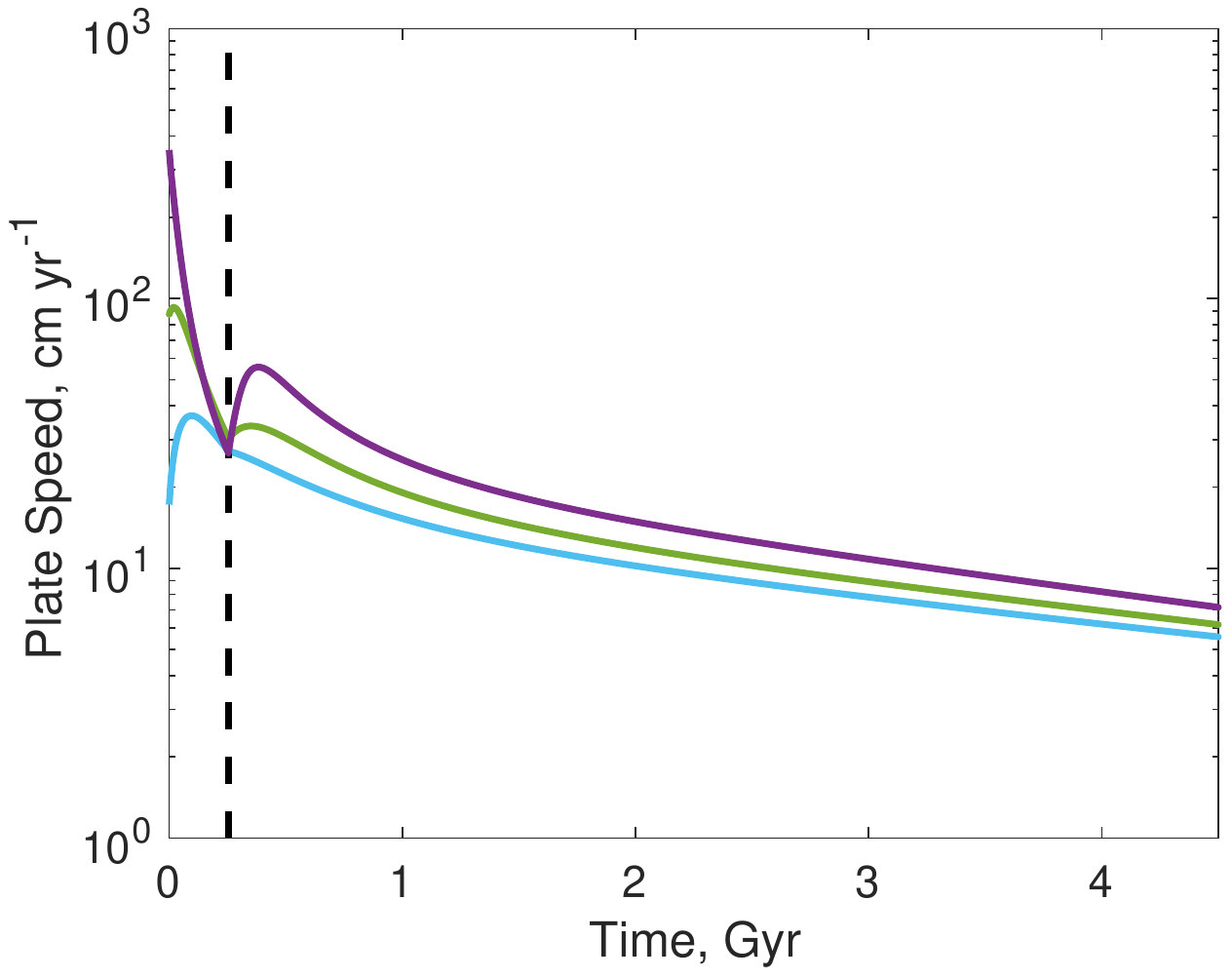}
    \end{subfigure}
    \hfill
    \begin{subfigure}[b]{0.45\textwidth}   
    \centering 
        \includegraphics[width=\textwidth]{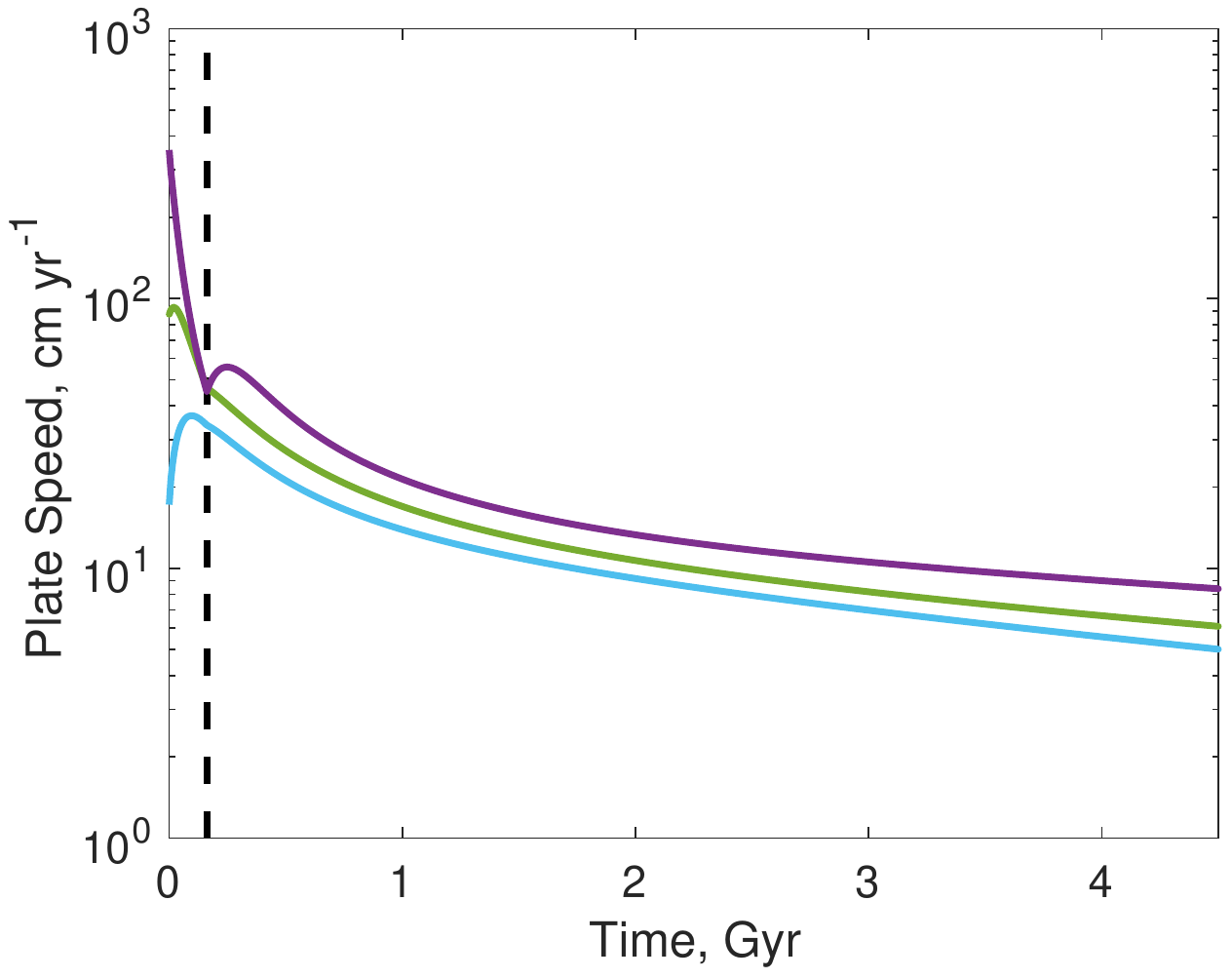}
    \end{subfigure}
    \caption{Continental coverage, mantle water content and temperature and plate speed over time (top to bottom) for the models of Fig. \ref{fig:3dplot_insul}. Left: Negative feedback predominating. Right: Positive feedback predominating.}
    \label{fig:interior_insul}
\end{figure*}

\begin{figure*}
    \centering
    \begin{subfigure}[b]{0.475\textwidth}
        \centering
        \includegraphics[width=\textwidth]{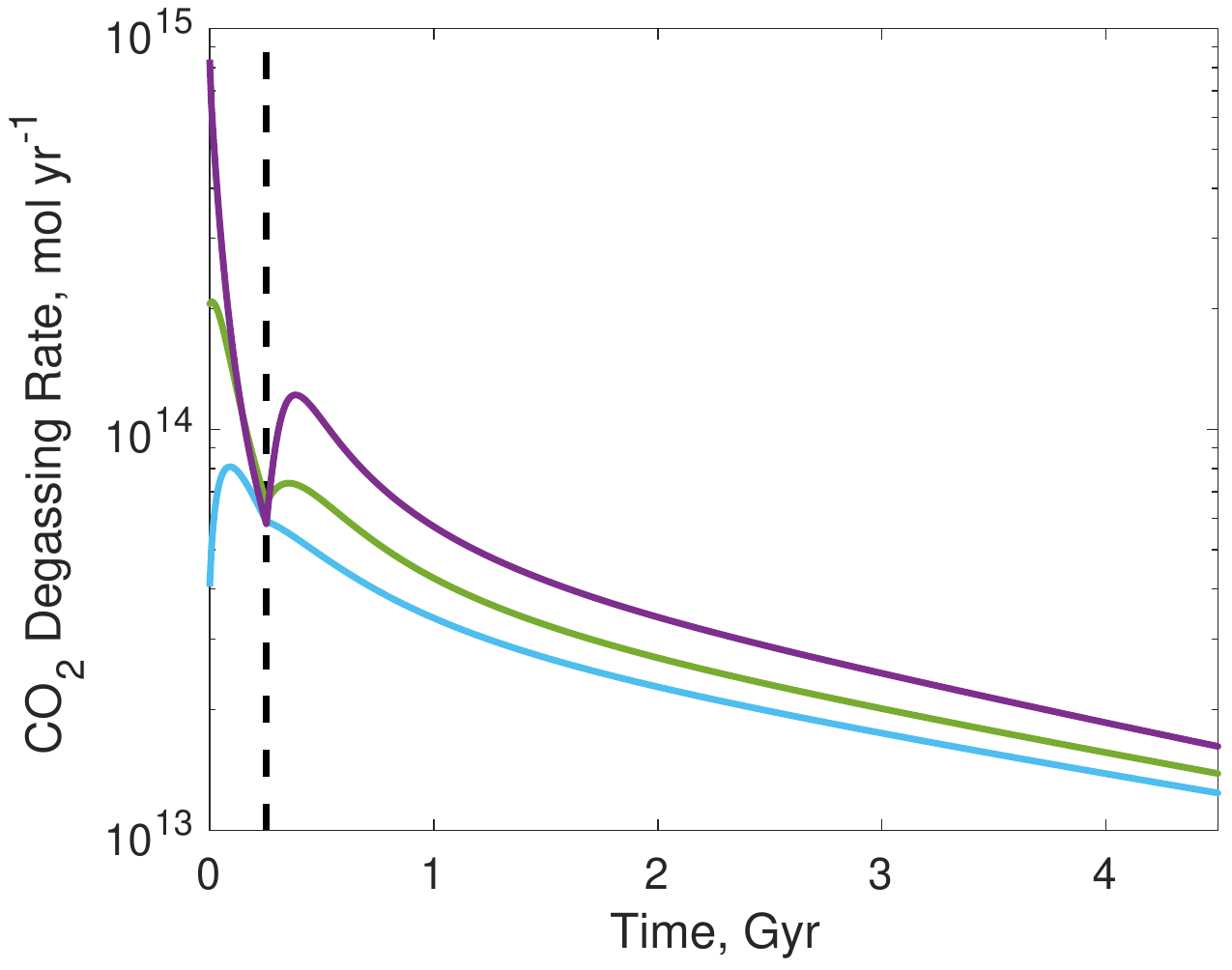}
    \end{subfigure}
    \hfill
    \begin{subfigure}[b]{0.475\textwidth}  
        \centering 
        \includegraphics[width=\textwidth]{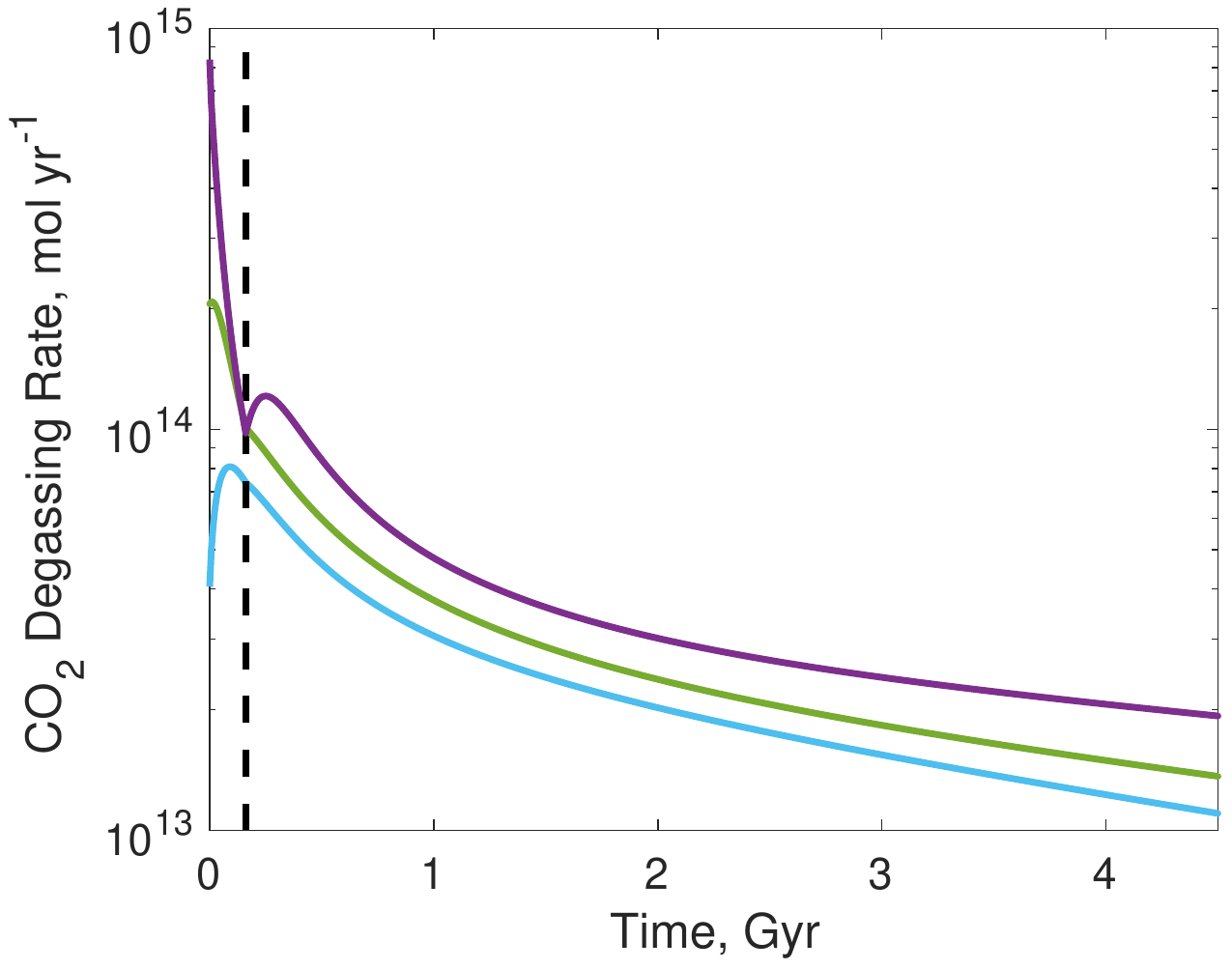}
    \end{subfigure}
    \begin{subfigure}[b]{0.475\textwidth}   
        \centering 
        \includegraphics[width=\textwidth]{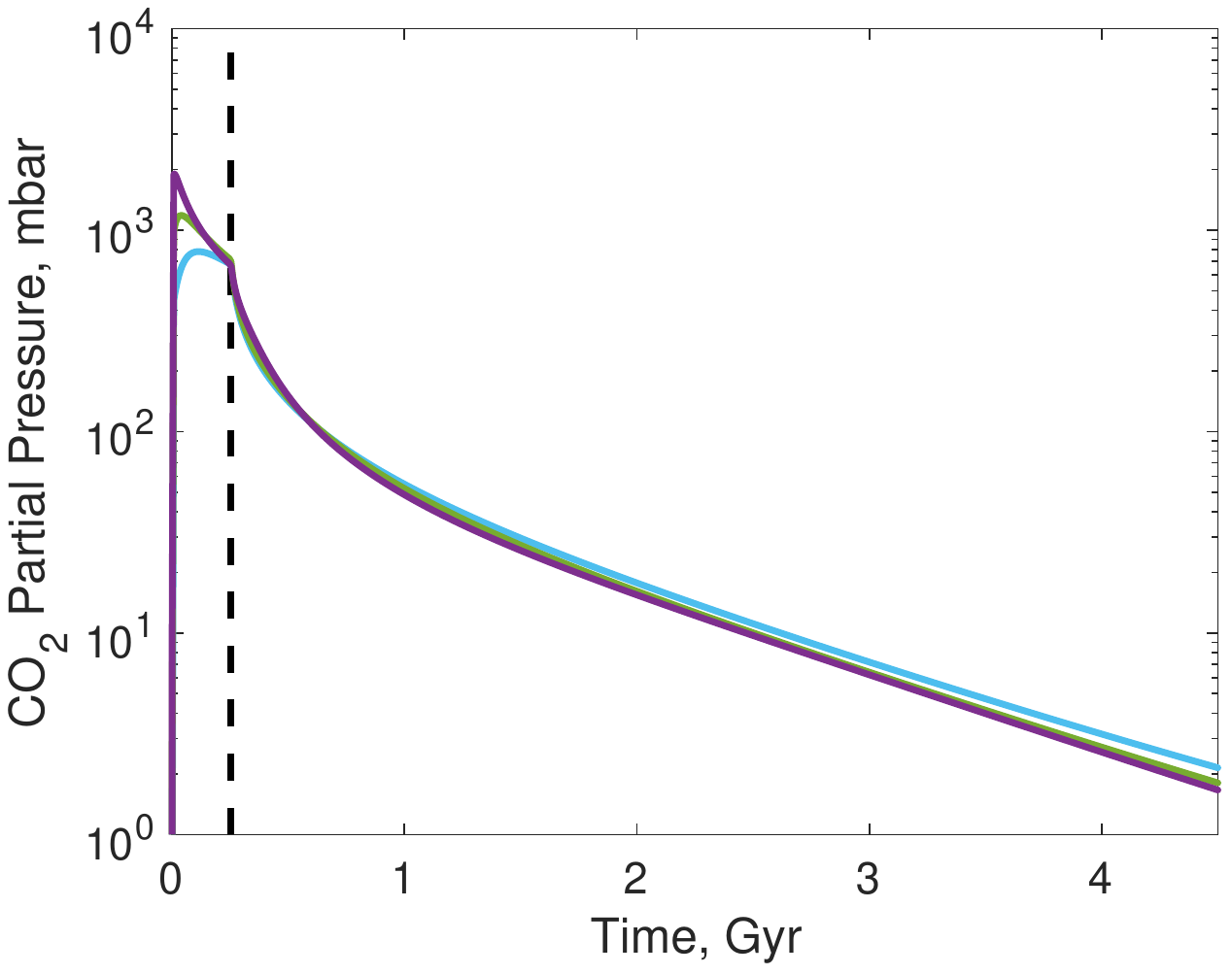}
    \end{subfigure}
    \hfill
    \begin{subfigure}[b]{0.475\textwidth}   
        \centering 
        \includegraphics[width=\textwidth]{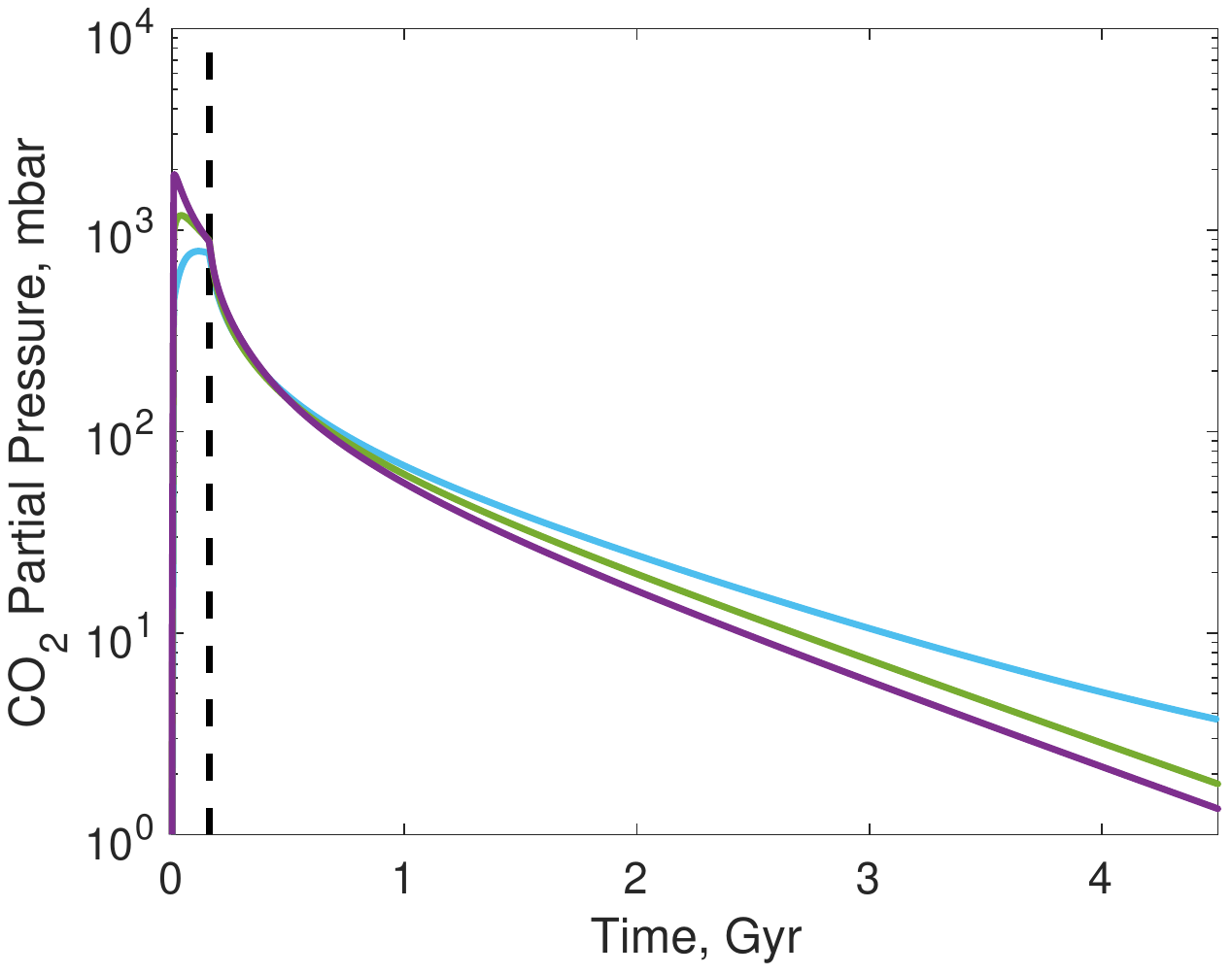}
    \end{subfigure}
    \begin{subfigure}[b]{0.475\textwidth}   
        \centering 
        \includegraphics[width=\textwidth]{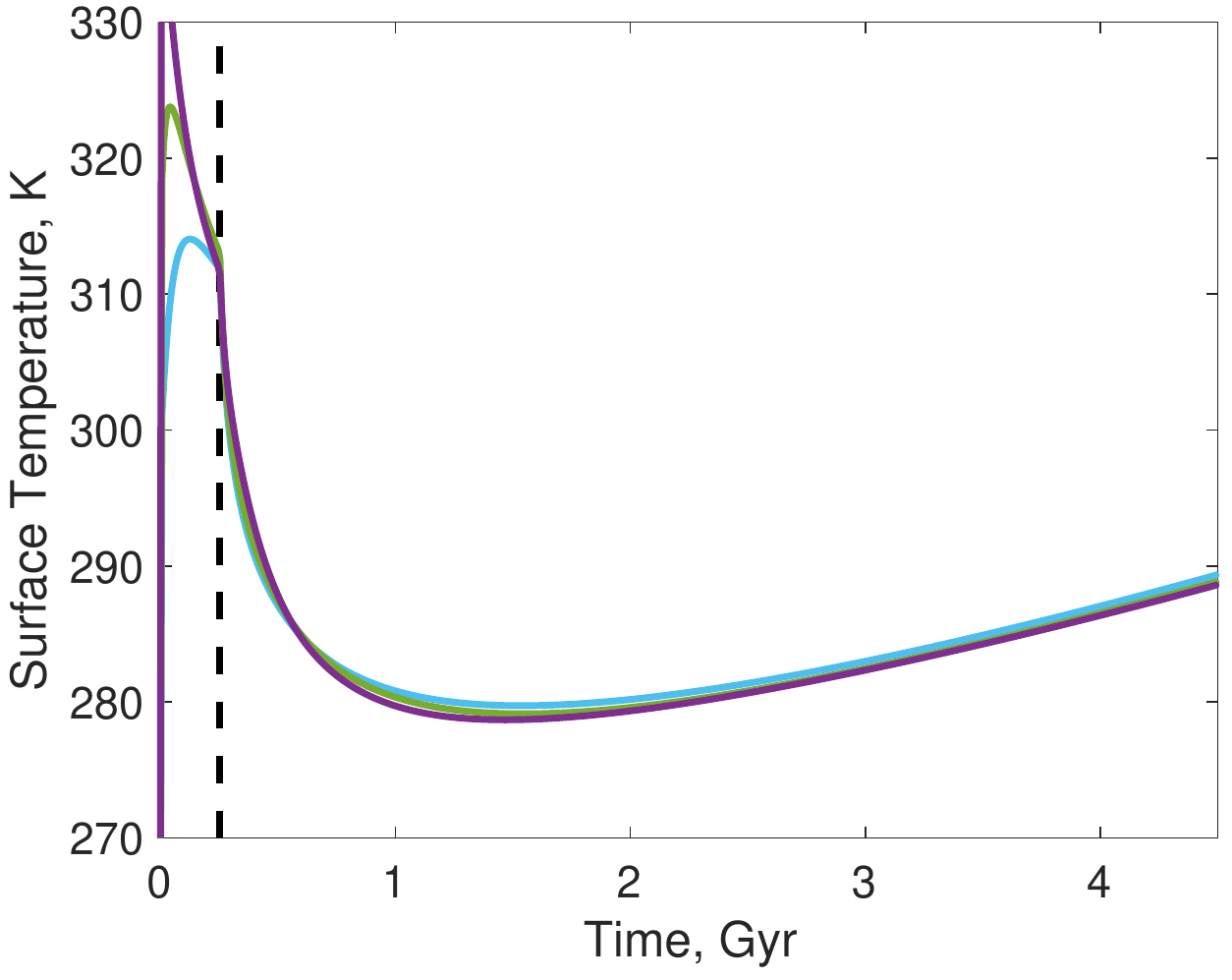}
    \end{subfigure}
    \hfill
    \begin{subfigure}[b]{0.475\textwidth}   
        \centering 
        \includegraphics[width=\textwidth]{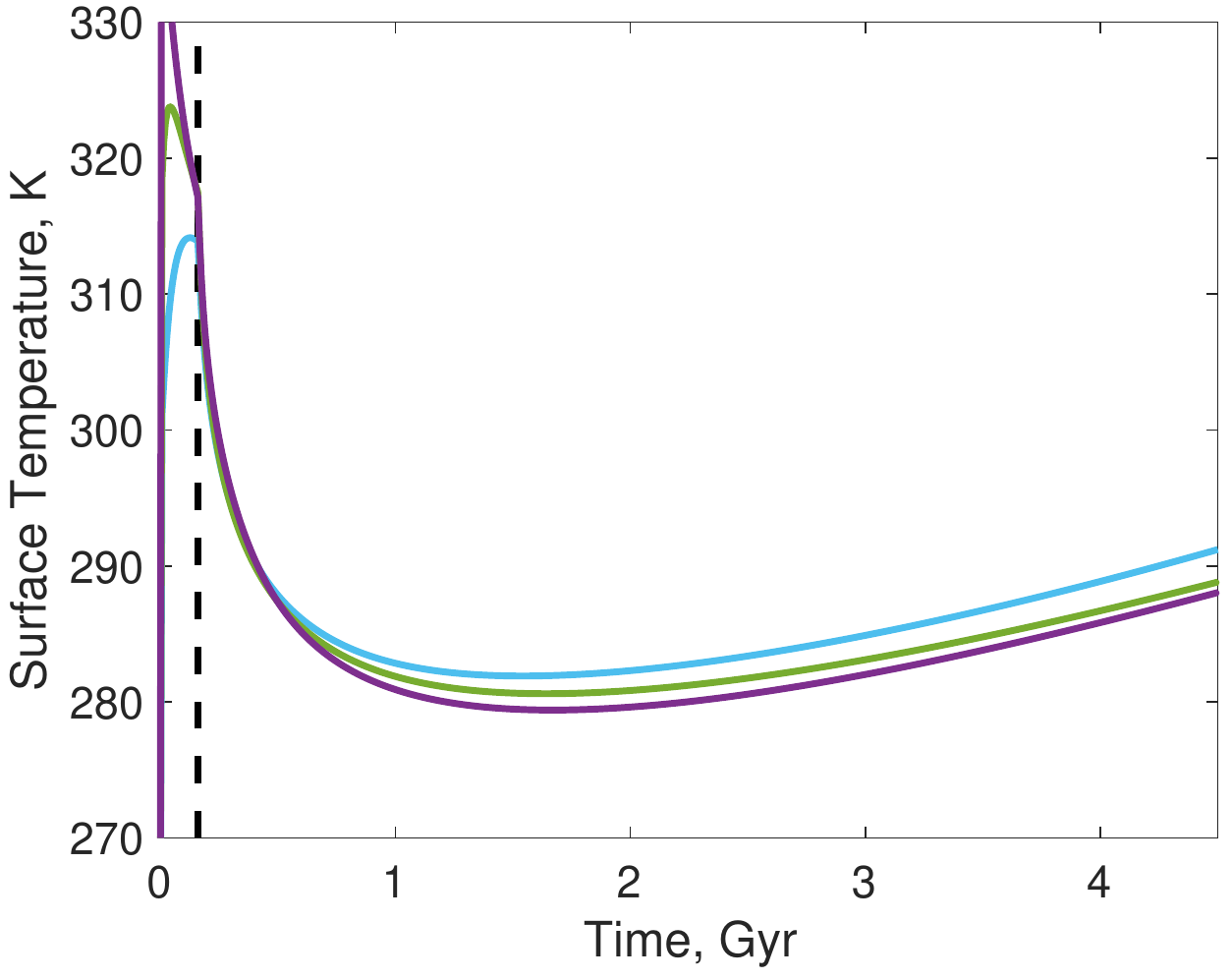}
    \end{subfigure}
    \caption{ 
    Same as in Fig. \ref{fig:climate}, but for models with thermal blanketing and mantle depletion of radiogenic elements upon continental crust growth and $\xi_3=0.85$ and $\xi_4=1/4$. Left panels: Negative feedback predominating. Right panels: Positive feedback predominating.}
    \label{fig:climate_insul}
\end{figure*}

A comparison between Figs. \ref{fig:3dplot} and \ref{fig:3dplot_insul} and between Figs. \ref{fig:interior} and \ref{fig:interior_insul} shows the same general behaviour with the models converging towards the single fixed point when negative feedback dominates. For dominant positive feedback, the models diverge and tend towards three distinct fixed points. There is a larger spread between the continental growth curves for models which include thermal blanketing, though, $0.4 \pm 0.1$ versus $0.4 \pm 0.05$ dimensionless units for negative feedback and -- somewhat more asymmetric -- $0.4^{+0.3}_{-0.22}$ versus $0.4^{+0.27}_{-0.18}$ dimensionless units for dominating positive feedback.

The evolution of the mantle water content is also found to depend on thermal blanketing as can be particularly well seen when comparing the respective panels in Figs. \ref{fig:interior} and \ref{fig:interior_insul}. For positive feedback and blanketing, the model with the largest continental coverage (purple) ends with less mantle water than in the case when blanketing is ignored. It even ends with less mantle water in Fig. \ref{fig:interior_insul} (right column) than the Earth-like model with a present-day continental coverage of 0.4. This can be understood if the mantle temperature is considered, which shows the most prominent differences. Models with no thermal blanketing and heat source differentiation have temperatures converging with time, independent of the choice of the initial temperature. With blanketing and mantle heat source depletion, mantle temperatures tend to converge initially but then tend to diverge with the growth of continents. In particular, the model with dominating positive feedback and with a present-day continental coverage of $>0.6$ (purple) shows the cooling rate to decrease after about 2.5 Gyr and the mantle temperature to start to level off. This explains why the net water gain rate in that model decreases after about that time. Still, the model has a present-day hotter and wetter mantle when compared with the purple model for dominating negative feedback (left column). The difference can be attributed to the larger contribution to water subduction by sediments which explains why the difference in plate speed and continental coverage is particularly pronounced for the purple models. The plate speed, finally, evolves largely in parallel with the mantle temperature as expected. The effect of continental insulation, which tends to reduce cooling, exceeds that of the depletion of mantle radiogenic elements in this model, which tends to enhance cooling.


Interestingly, for the case of dominating negative feedback (Fig. \ref{fig:interior_insul} left), large continents (purple curve) result in comparatively hot and dry mantles, with opposing effects on viscosity and plate speed. For dominating positive feedback (Fig. \ref{fig:interior_insul} right), sediments from the comparatively large continents contribute to water subduction such that a hot and wet mantle results, both increasing the plate speed. This feeds back to continental growth such that the difference in continental coverage at 4.5 Gyr becomes particularly pronounced.

Figure \ref{fig:climate_insul} shows results pertaining to the long-term carbon cycle for the models that include thermally insulating continents. As in previous figures, panels on the left are for models with  predominating negative feedback and those on the right are for models with predominating positive feedback. A larger surface fraction of insulating continents keeps the mantle warmer and therefore results in somewhat larger degassing rates but also in a larger weatherability and rate of CO$_2$ removal. The differences in the CO$_2$ partial pressure and surface temperature between the models remain small and are even a bit smaller than for models neglecting thermal blanketing altogether.

\section{Discussion}
\label{sec7}

In this paper, we modelled three main geologic processes as a nonlinear coupled dynamical system: growth of continental crust, exchange of water between the reservoirs on and above the surface (oceans, atmosphere) and in the mantle, and cooling by mantle convection. These processes are linked through mantle convection and plate tectonics with 
\begin{itemize}
    \item subduction zone related melting and volcanism and continental erosion governing the growth of the continents (Eq. \ref{eq:continent}),
    \item mantle water degassing through volcanism and regassing through subduction governing the water budget (Eq. \ref{eq:water}), and
    \item heat transfer through mantle convection governing the thermal evolution (Eq. \ref{eq:temperature_new})
\end{itemize}
The rates of these processes depend non-linearly on mantle temperature and water concentration through the mantle viscosity (compare Eq. \ref{eq:visc}) and plate speed $v_p(T,w)$ which couples the governing equations \ref{eq:continent} - \ref{eq:temperature_new}. By including a model of the carbonate-silicate cycle, we explored consequences for the atmosphere's CO$_2$ partial pressure and average surface temperature as indicator of the planet's habitability. In addition to the coupling through $v_p(T,w)$, there are immediate feedback mechanisms with the crust production and erosion rates depending on the continental surface area and the length of continental margins and with the mantle water outgassing rate depending on the mantle water concentration. The former is implied by the dependence of the melt production rate in the subduction zone on the water supply through sediments \citep{Honing:2016} which in turn is  directly proportional to the continental surface area. The latter is implied by the dependence of the mantle solidus temperature and the viscosity on the water concentration. The mantle water concentration and the continental surface area feed back on the mantle thermal energy balance; the former through the dependence of $v_p(T,w)$ on the mantle water concentration, the latter through the thermal blanketing effect of the continents.

The conceptual nature of our study motivated us to simplify our model beyond the level of abstraction of our previous work \citep[e.g.,][]{Honing:2016, Honing:2019a} by reducing the number of free parameters to four, $\xi_{1},.., \xi_{4}$, with $0 \leq \xi_{1},.., \xi_{4} \leq 1$. We still use the present-day Earth as a reference and present-day observables as scales. In the following, we will discuss our finding that the spread of continental coverage on Earth-like planets is determined by the respective strengths of positive and negative feedback in continental growth ($\xi_{1}$ vs. $\xi_{2}$, Section \ref{sec:disc_xi}) and by the relationship between thermal blanketing and depletion of radioactive isotopes upon growth of the continental crust ($\xi_{3}$ vs. $\xi_{4}$, Section \ref{sec:disc_insul}). Uncertainty in these parameter values represents the main uncertainty in the model. Our discussion aims to provide a better qualitative understanding of the feedback processes;  we admit to lacking data for a detailed  understanding of quantitative differences. Finally, we discuss the habitability of the model planets and their bioproductivity that result from the various possible evolution pathways in Section \ref{disc_prop}. 

\subsection[Feedback parameters $\xi_1$ and $\xi_2$]{The feedback parameters $\xi_1$ and $\xi_2$ and their implications for continental growth}
\label{sec:disc_xi}

The parameters $\xi_1$ and $\xi_2$ can be regarded  as slide rulers in Eqs. \ref{eq:continent} and \ref{eq:water} to weigh between feedback terms. We have chosen combinations of $\xi_1$ = 1/3 and $\xi_2$ = 2/3, $\xi_1 = \xi_2$ = 0.5, and $\xi_1$ = 2/3 and $\xi_2$ = 1/3 to demonstrate in Figs. \ref{fig:quiver} -- \ref{fig:climate_insul} how the properties of the system vary with these choices. Note that $\xi_1$ and $\xi_2$ do not have to add up to 1, although they do in our parameter choices. The parameter $\xi_1$  weighs between the importance of the continental surface area and plate speed (that is, convection speed), respectively, for both the production rate of continental crust and the mantle water degassing rate of the present-day Earth.

The value of $\xi_1$ becomes particularly important late in the evolution, when the mantle has cooled down sufficiently such that $v_p^*$, the plate speed scaled to its present-day value, will be of the order of unity. Geologically, this would be in the post-Archean. At earlier times, $v_p^*$ will have a greater effect than $A$, unless $\xi_1$ is chosen such that $(1-\xi_1)\ll 1$. As a geological interpretation, $\xi_1$ weighs the importance of sediments  for transporting water in the subducting slab to the source region of continental crust for the present-day Earth. The rate of melting and crust production in the model is directly proportional to the rate of supply of water through oceanic crust and through the sedimentary layer on top of the oceanic crust (see \citealt{Honing:2014} and \citealt{Honing:2016} for details), the share of the sediments in the balance increasing with $\xi_1$.

\cite{Honing:2016} have discussed how continental erosion, subduction of sediments and continental rock formation are related. The ability of hydrous sediments to contribute to water transport with the subducting plate is well established \citep{hacker:2008,deschamps:2012,deschamps:2013} but difficult to quantify. In addition, as described in \cite{Honing:2014}, low permeability sediments can seal the hydrated crust, thereby reducing the rate of shallow dewatering upon subduction. \cite{stern:2011}, \cite{sobolev:2019} and \cite{chen2022} argue that accumulation of sediments in subduction zones is crucial to lubricate the subducting plate, which would effectively introduce another positive feedback to the system that is not considered here, however. \cite{sobolev:2019} even suggest that the persistence of plate tectonics requires a sufficient thickness of the sedimentary layer and that an extended  period of reduced surface erosion and sediment subduction might have resulted in reduced tectonic activity on Earth between 0.75 and 1.75 billion years before present. Note that $L_{oo,oc}(A)$ is a constant for $A \leq 0.4$ (compare Fig. \ref{fig:loc}) and decreases with approximately  $2 A$ for $A > 0.55$ when continents increasingly touch and merge (compare Eq. \ref{eq:derivLoc}).

It may be argued that positive feedback may also originate from the variation of $v_p^*$ with viscosity. After all, an increase in $v_p^*$ will reduce the viscosity and increase the convection speed and, in turn, the water regassing rate. This will be counteracted, however, by increased water outgassing and cooling associated with the convective mantle flow. In the present model, the effect of cooling on increasing the viscosity dominates over the effect of an increasing water concentration. Altogether, the contribution of convection and plate circulation is to provide negative feedback. Other positive feedback mechanisms may exist that are independent of erosion and sediment subduction. We will discuss the contribution of thermal blanketing to positive feedback in the next section. 

The parameter $\xi_2$ weighs between surface erosion and subduction erosion, the latter depending on the length of the ocean-continent subduction zones. A major share ($\approx 1/3$) of the global continental erosion rate has been associated with subduction erosion \citep{stern:2011}. In addition, significant shares of the loss of continental surface area  have been  attributed to continental collision and lower crust delamination, leaving less than a third for surface erosion \cite{stern:2011}. It thus seems reasonable to assume that  $\xi_2 < 0.5$. We note that the rate of subduction erosion may depend on the rate of friction between the subducting slab and the continental keel in addition to its dependence on the length of the subduction zone. Sediments may reduce friction by lubricating the slab \cite{stern:2011}, thereby reducing the rate of subduction erosion for increasing continental coverage. This would effectively introduce another positive feedback to the system that is not considered here, however.

The length of the ocean-continent subduction zones $L_{oc}(A)$ will naturally increase with the size of the continents as long as the continental coverage is smaller than $\approx 0.4$. At larger values of coverage, the subduction zone length decreases because continents impinge upon each other and merge and continental collision zones form. The ocean-continent subduction zones $L_{oc}(A)$ varies non-linearly with the continental surface area in our model (compare \ref{subduction_zones}). In linear approximations, it increases roughly with $3 A$ for $A < 0.3$ and decreases roughly with $2 A$ for $A > 0.55$. Between these two values of $A$ the dependence of $L_{oc}$ on $A$ is weak. Thus, there will be strong negative feedback to $A$ associated with continental erosion  for $A < 0.3$, even for small values of $\xi_2$. The feedback may turn entirely positive, however, for large values of continental coverage, depending on the chosen values of $\xi_1$ and $\xi_2$ if subduction erosion dominates and the overall continental loss rate decreases with increasing $A$.

Our results show how strong negative feedback would lead to an evolution largely independent of the starting conditions and the early history of the planet which would  imply a single stable present-day value of the continental surface area. For strong positive feedback, however, the outcome of the evolution may be quite different depending on starting conditions and the early history. For exoplanets this suggests that Earth-like planets covered largely by land (i.e. \textit{land planets}) are possible as well as planets covered mostly by oceans (i.e. \textit{ocean planets}) (see Fig. \ref{fig:illustration}). \cite{Honing:2016} found that the land planet should be the most likely outcome, followed by cases that end as ocean planets. Only a small percentage end as planets with a balanced Earth-like share of land and ocean surfaces. 

Earth's further evolution can be discussed using the present day phase planes shown in Figs. \ref{fig:3dplot} and \ref{fig:3dplot_insul}. It will depend on the location of the state variable $A$ relative to the fixed point. For values of $A$ close to but larger than the continental coverage at the unstable fixed point, $A$ will increase with time at a small rate beyond its present-day value. While $A$ is growing, the mantle water concentration will increase. With water budget conserved, this will be accompanied by decreasing surface water reservoir and increasing area of emerged land. This would be beneficial for the biosphere allowing for more silicate weathering, provision of nutrients, and partly compensating for an increasing solar luminosity. Eventually, the unstable fixed point will overtake the trajectory and net continental growth will cease. Additional calculations have shown that the upper stable fixed point will move towards smaller values of $A$ together with the unstable fixed point moving towards larger values of $A$. The two fixed points will eventually merge and then disappear with further cooling while the continental area will slowly decrease with time being far away from but evolving towards the remaining lower stable fixed point.

\begin{figure*}
    \centering
    \begin{subfigure}[b]{0.45\textwidth}
        \centering
        \includegraphics[width=\textwidth]{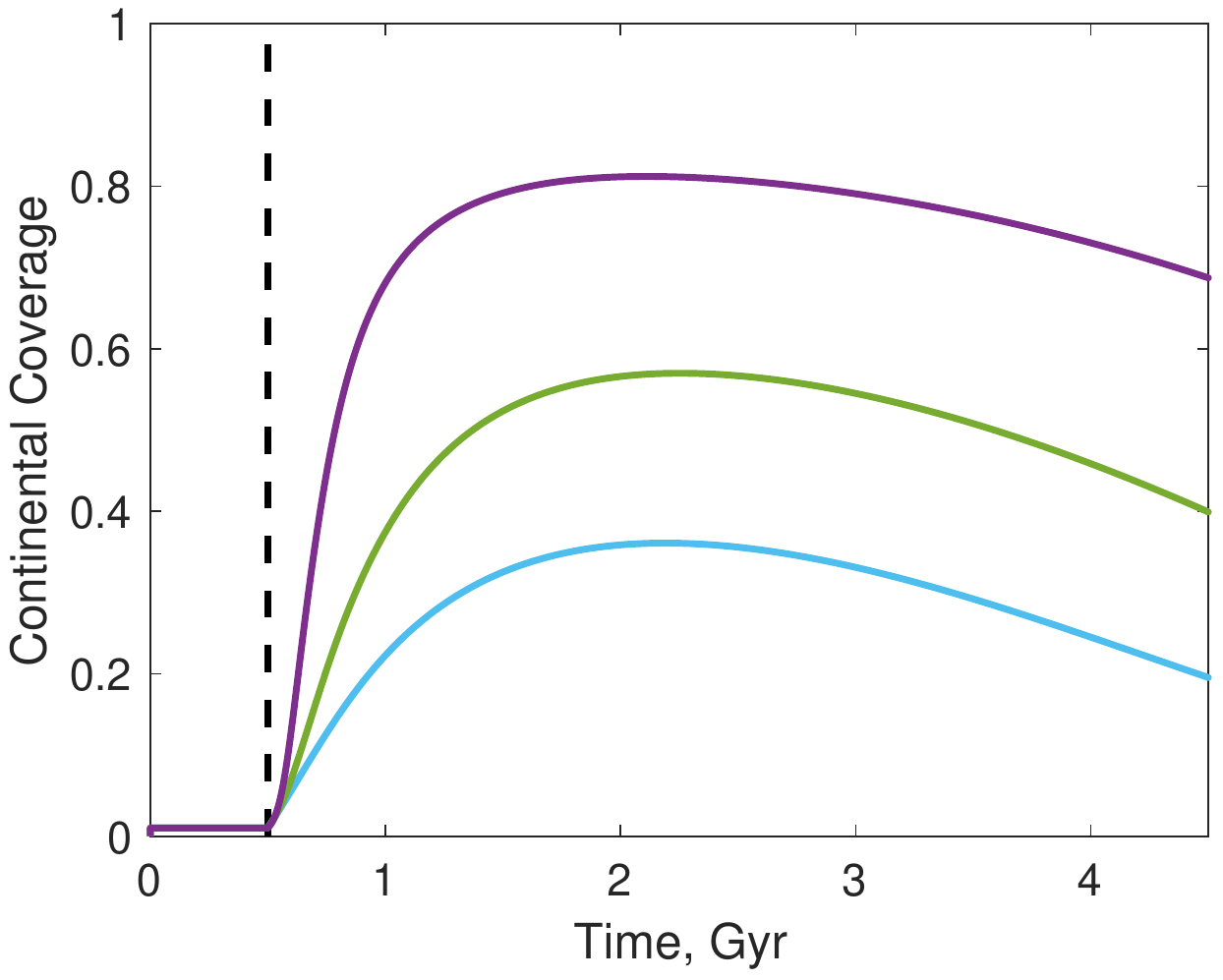}
    \end{subfigure}
    \hfill
    \begin{subfigure}[b]{0.45\textwidth}  
        \centering 
        \includegraphics[width=\textwidth]{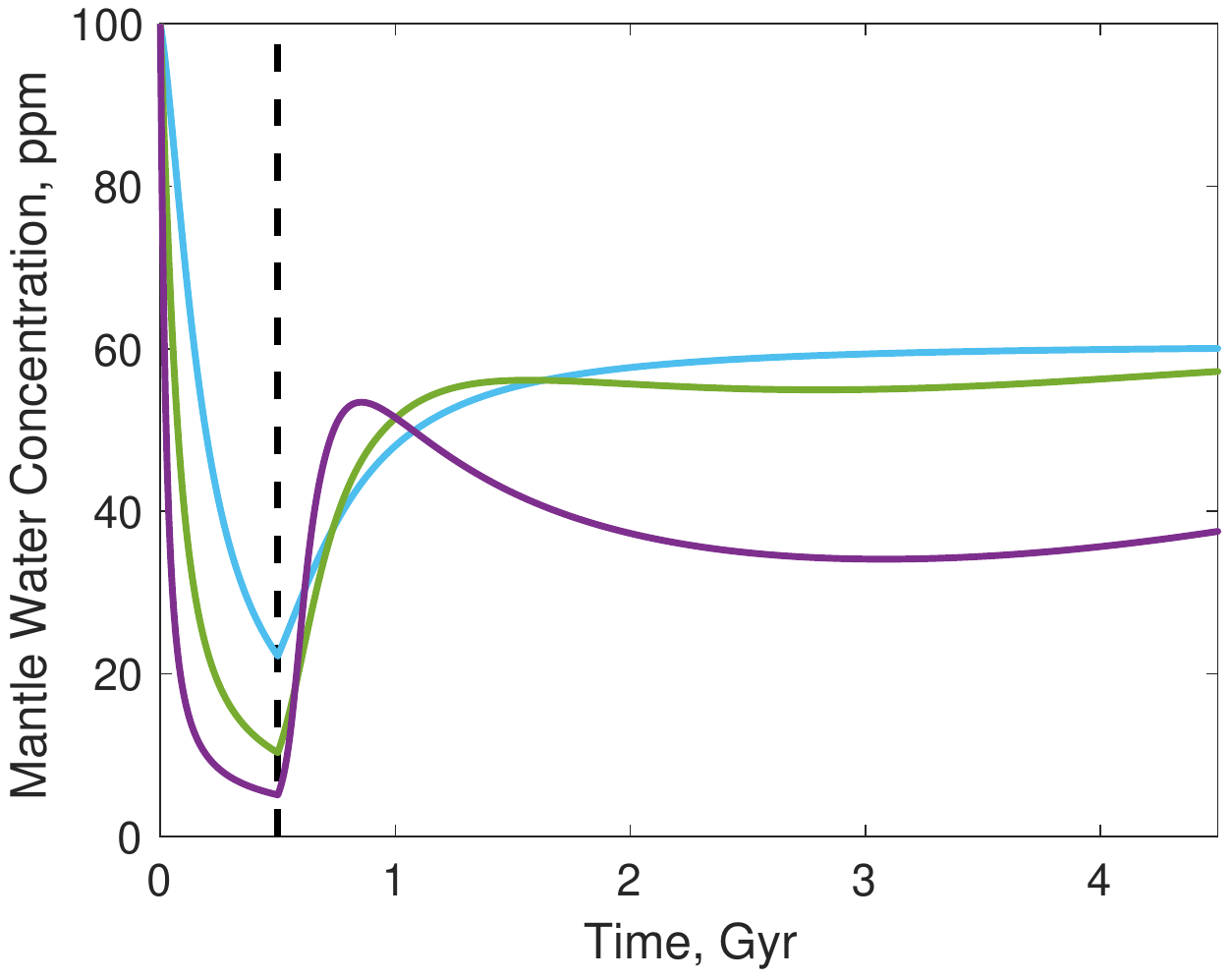}
    \end{subfigure}
    \begin{subfigure}[b]{0.45\textwidth}   
        \centering 
        \includegraphics[width=\textwidth]{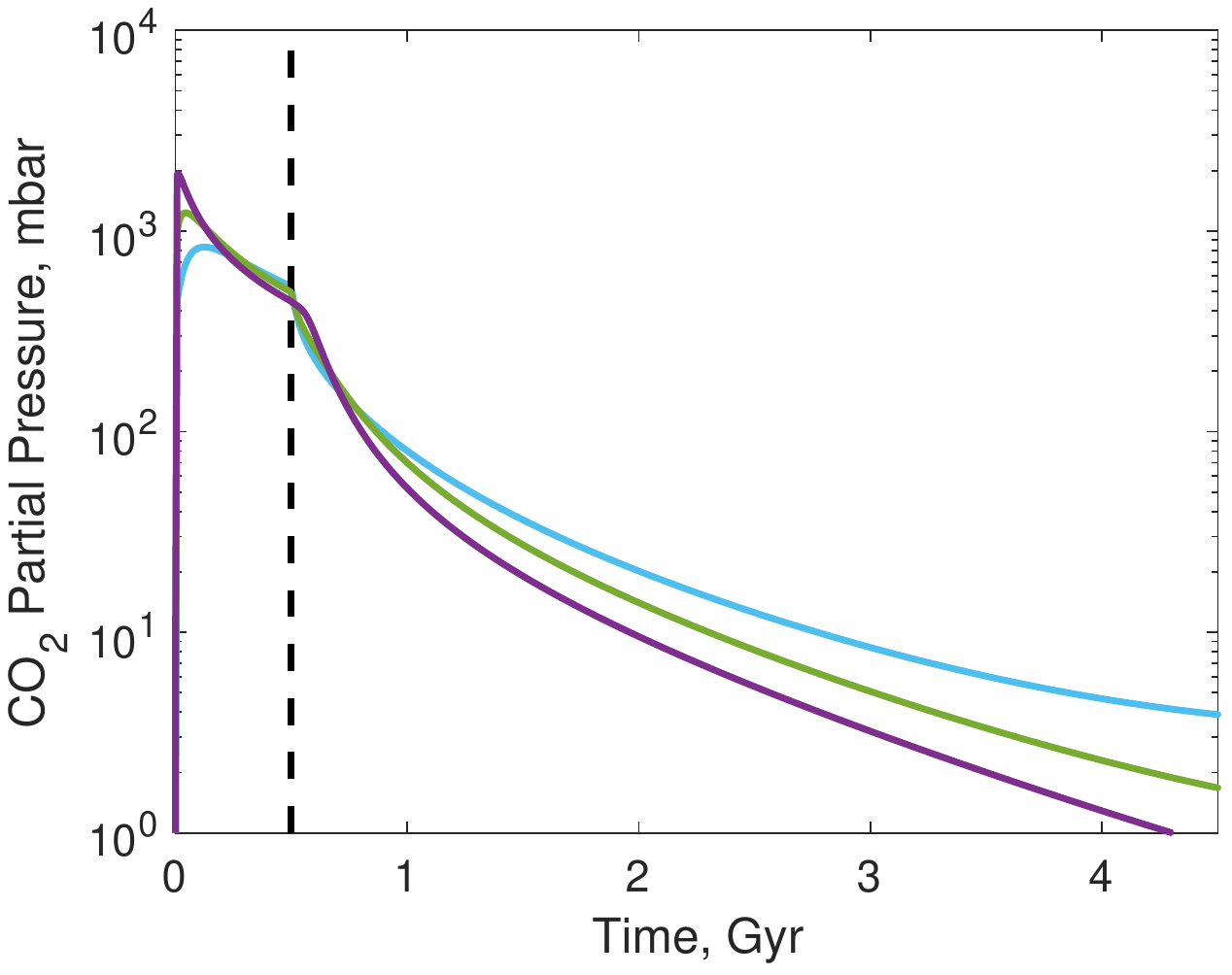}
    \end{subfigure}
    \hfill
    \begin{subfigure}[b]{0.45\textwidth}   
        \centering 
        \includegraphics[width=\textwidth]{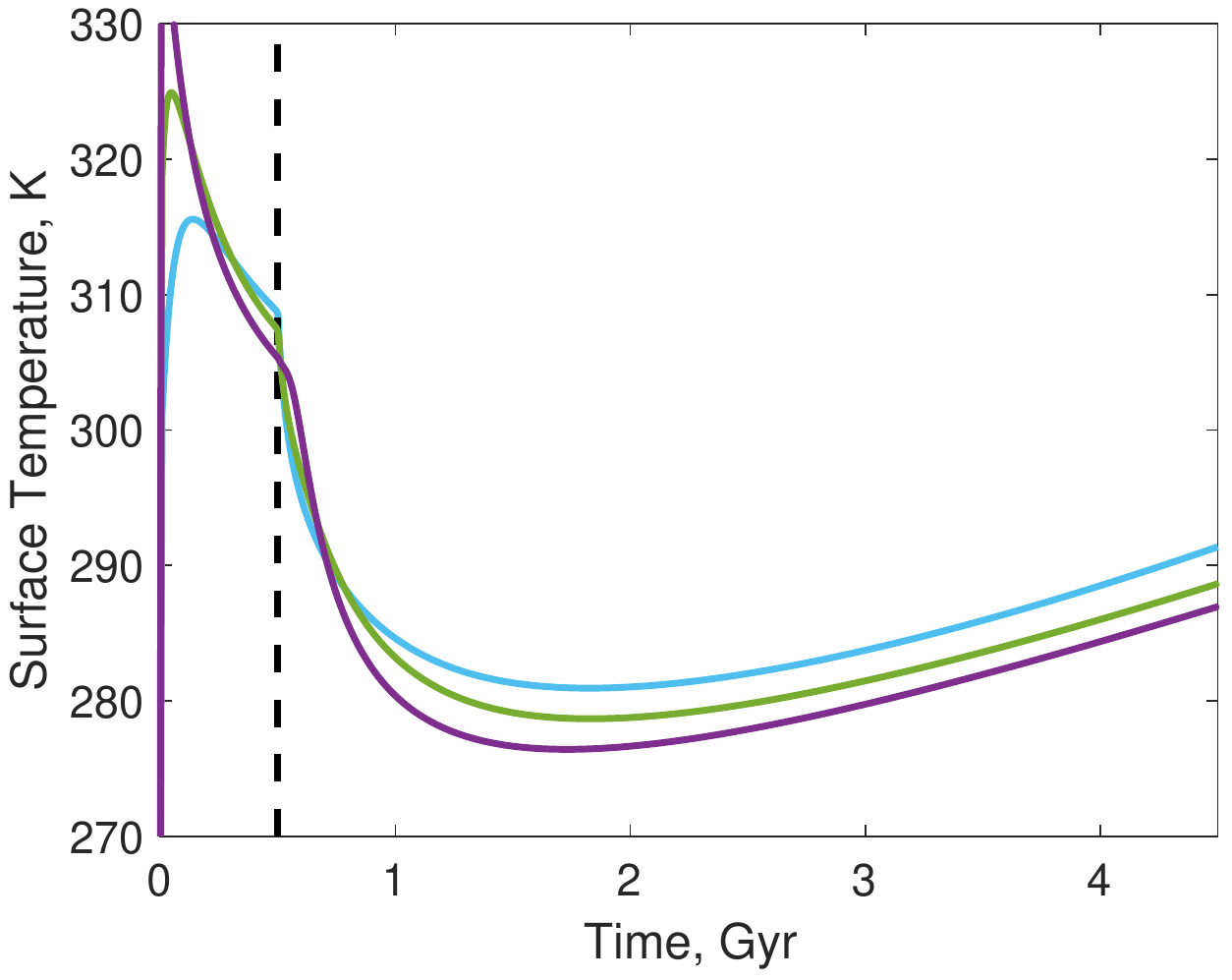}
    \end{subfigure}
    \caption
    {Model scenario that neglects any direct dependence of continental production and erosion on continental area ($\xi_1=0$, $\xi_2=0$). In order for the green curve to reach continental coverage of 0.4 at 4.5 Gyr, we set $f_r=0.603$. Thermal blanketing is neglected and the color code is same as in the previous figures.}
    \label{fig:nofeedb}
\end{figure*}

An extreme case in which continental production and erosion  would be entirely independent of the surface area  can be studied by setting $\xi_1=0$ and $\xi_2=0$. The continent production rate will then depend only on the mantle convection rate and the erosion only on the total length of ocean-continent subduction zones. In addition, we neglect thermal blanketing (as in Fig. \ref{fig:3dplot}) and show the results in Fig. \ref{fig:nofeedb}. Without positive or negative feedback, differences in early continental growth resulting from differences in initial conditions are mostly retained throughout the evolution without being damped or amplified with time as they would be with feedback. Continental coverage at 4.5 Gyr is found to attain similar values to those with dominating positive feedback (compare Fig. \ref{fig:interior}, right panel), though. Mantle water concentrations for the ocean and Earth-like planet are similar to the values found for negative feedback but the land planet (purple curve) of the extreme model has a comparatively dry mantle. This implies a comparatively larger surface water mass and more flooded continental crust and can be explained as a consequence of ignoring the contribution of sediments to water subduction. The  concentrations of atmospheric CO$_2$ and the spread of surface temperatures are similar to the scenario with dominating positive feedback explored above.

It is interesting to note that for planets around M-dwarfs, \cite{Tian:2015} predict a similar bimodal distribution of emerged land area, with most planets either having their surface entirely covered with water or with significantly less surface water than Earth. Their analysis, however, was based on a combination of planet formation and water escape models, not explicitly modelling continental growth.

The small positive net growth rate of the continents since the Archean \cite[e.g.,][]{armstrong:1991,dhuime:2017,rosas:2018,korenaga2021} provides a strong argument for positive feedback in Earth's continental growth history: Positive feedback can sustain a small net growth rate over an extended period of time even in the presence of a slowly decreasing mantle temperature as in the right columns of Figs. \ref{fig:interior} and \ref{fig:interior_insul}. In that case, the unstable fixed point will be evolving slowly towards larger continental coverage and $A$, being located slightly above this fixed point, will follow suit. We note that equilibrium is not expected to be attained as mantle heat production decreases with time and planets necessarily cool as has already been discussed by \cite{Schubert:1979}. With negative feedback prevailing, the continental production rate will decrease with decreasing mantle temperature. As a result, the continental surface area will decrease with time as in the left columns of Figs. \ref{fig:interior} and \ref{fig:interior_insul}. An overall decrease of the continental surface over the past 2 Gyr as predicted by the negative feedback model is not observed, however! Negative feedback evolution trajectories reaching a continental coverage of 0.4 after 4.5 Gyr require a present-day ratio between the rate of continental production and that of continental erosion of $f_r=0.775$, implying an equilibrium continental coverage of only about 0.2 and an evolution of Earth towards what we have termed an ocean planet!

The continents may have grown in the Archean by mechanisms that differ from those of modern plate tectonics \cite[e.g.,][]{condie:2013}. Our analysis would still be applicable, because the bifurcation appears only at post-Archean mantle temperatures and plate speeds. The onset time $t_p$ could be 2.5 Gyr and a value for $A$ at the time would be $>$0 along with assumed values for $T$ and $w$. In such a scenario, the present-day continental coverage would be sensitive to the coverage at $t_p$.

\subsection[Effects of $\xi_3$ and $\xi_4$]
{Thermal blanketing and mantle depletion through crust growth; effects of $\xi_3$ and $\xi_4$}
\label{sec:disc_insul}

Thermal blanketing by continental crust can have two effects that may both act at the same time: Blanketing can result in an overall reduction of the surface heat flow and the cooling rate. And it can result in a redirection of mantle heat flow away from the continents to the oceanic areas. Using stability analysis, \cite{Honing:2019a} have shown that thermal blanketing can lead to runaway continental growth and even mantle heating (see also \citealt{Lenardic:2005} for related findings). The analysis, however, was based on the assumption of a steady state mantle and a constant Urey Ratio. It also did not include the redistribution of radiogenic elements between the mantle and the crust. In their analysis, blanketing caused the heat flow from the mantle to be diverted from the sub-continental to the sub-oceanic areas, resulting in an increasing sub-oceanic mantle temperature, heat flow, plate speed and, finally, continental growth rate. Their result can be considered as an extreme case but suggests that positive feedback between thermal blanketing and continental growth is, in principle, possible.

In the present model, we consider thermal blanketing in the presence of mantle cooling and the decay of radiogenic elements. In addition, we consider the transfer of heat producing elements from the mantle to the crust as a consequence of melting and crust growth. The cooling and the decrease in time of the mantle heating rate will reduce the self-reinforcement of continental growth through thermal blanketing. We do not consider any possible redirection of mantle flow away from sub-continental areas to sub-oceanic areas. Instead, we use $\xi_3$ as a slide ruler to distribute the present-day mantle heat flow $q_{s,E}$ between the oceanic areas and the continents. From its definition in Eq. \ref{eq:def xi3}, $\xi_3$ is the present-day share of the oceanic heat flow in the total heat flow from the mantle. For simplicity, this ratio is kept constant throughout the evolution calculation. Also, we neglect any enrichment of the oceanic crust in radiogenic elements.

For the present-day Earth, the oceanic heat flow per unit area is on average 105.4 mW/m$^2$, while the average continental heat flow is 70.9 mW/m$^2$  and the total average surface heat flow per unit area is 91.6 mW/m$^2$ \citep{Davies:2010}. \cite{Jaupart:2015} give a somewhat smaller value for the continental heat flow of 65 mW/m$^2$ while earlier work resulted in about 60 mW/m$^2$ \citep[e.g.,][]{Williams:1974}. The total heat flow values, accounting for uncertainties, are 31 -- 32 TW, 14 -- 15 TW, and 45 -- 47 TW, respectively. The continental heat flow includes a significant contribution from radiogenic elements in the crust. \cite{Jaupart:2015,jaupart:2016} give values for the basal continental heat flow -- mostly in stable continental shield  areas -- from the mantle of 15 $\pm$ 3 mW/m$^2$, equivalent to 2.4 - 3.7 TW. Subtracting the latter values from the continental heat flow gives 11 -- 12.3 TW. However, this value is much too large to be explained by continental heat production for which \cite{Jaupart:2015,jaupart:2016} give 6 -- 7 TW, the remainder of 3 -- 4 TW to be attributed to continental margins where the heat flow can be significantly higher than in stable continental areas \citep[e.g.,][]{springer:1998}. Altogether, this suggests a global present-day heat flow from the mantle of $\approx$ 38 TW, so that values for $\xi_3$ between 0.8 and 0.85 should be reasonably characterising the heat flow ratios on the present-day Earth.

To calculate a consistent value of $\xi_4$, the share of the heat sources in the present-day continental crust (compare Eq. \ref{eq:xi_4}), we use the observed continental heat flow and subtract the above estimate of the mantle heat flow through the continents including continental margins of 6 -- 7 TW  to obtain 7 -- 8 TW. Unfortunately, the mantle heat production rate is much more uncertain. \cite{dye:2012} cites estimates between 11 and 24 TW for the bulk silicate Earth (mantle + crust) 
as resulting from cosmochemical studies. Using a mantle heat flow estimate of 36 TW and estimates from convection modelling of 30$\%$ to 50$\%$ being attributable to radiogenic heating (\citealt{dye:2012}, the so called Urey ratio), they arrive at 19 to 27 TW, again for the bulk silicate Earth. Note that we do not adopt the case of 80$\%$ radiogenic heating  as we find that the Urey ratio predicted from constant viscosity mantle models is unrealistic. \cite{Jaupart:2015} have compiled mantle and crust heat production rates from the literature and from their own interpretations of global heat flow data and arrive at a representative value of $11 \pm 6$ TW for the mantle heating rate. This value can be considered as a lower limit as it assumes that the (depleted) MORB source extends through the whole mantle. Thus, values for $\xi_4$ between 0.25 and 0.4 should be reasonable to assume.

\begin{figure*}
    \centering
    \begin{subfigure}[b]{0.475\textwidth}
        \centering
        \includegraphics[width=\textwidth]{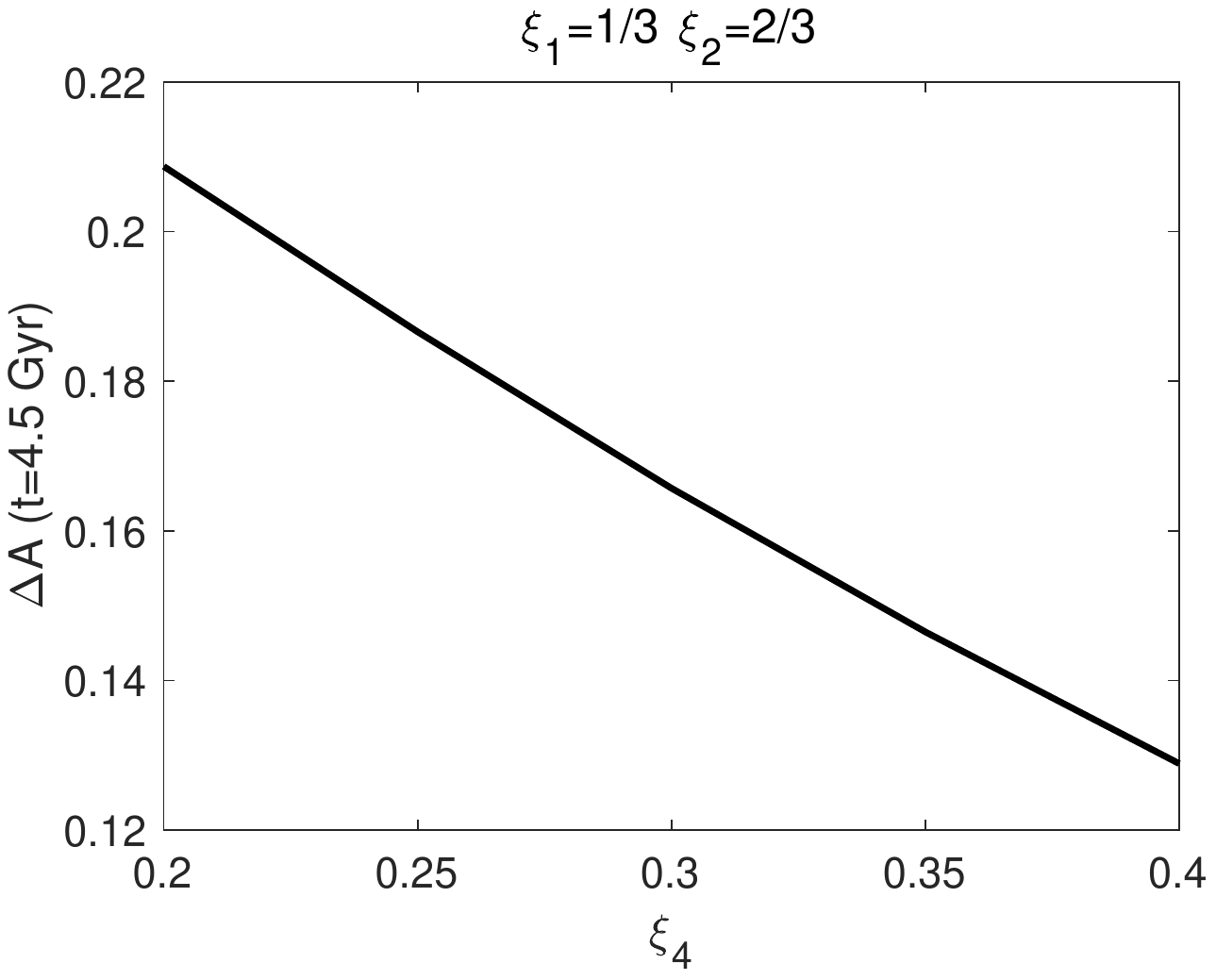}
    \end{subfigure}
    \hfill
    \begin{subfigure}[b]{0.475\textwidth}  
        \centering 
        \includegraphics[width=\textwidth]{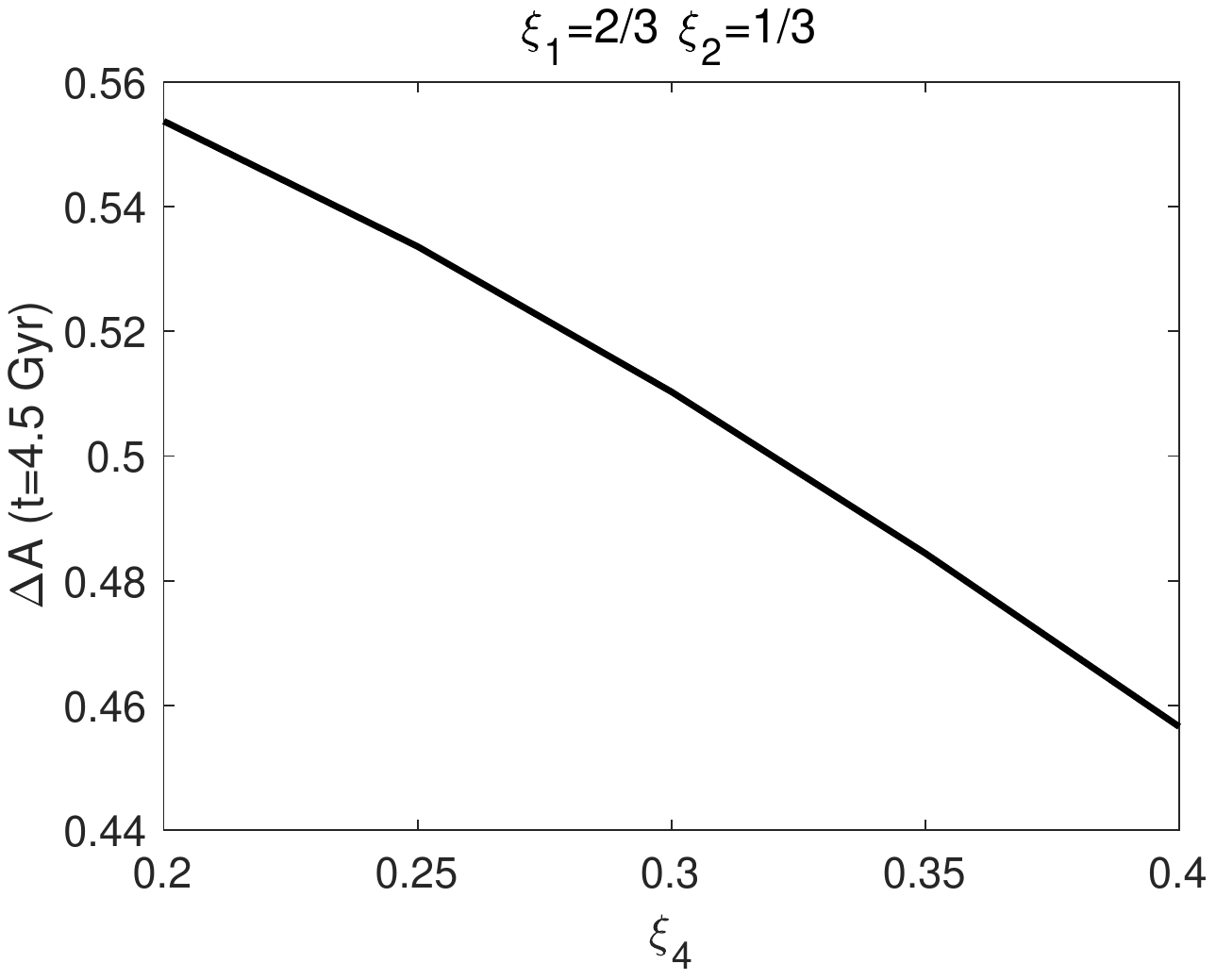}
    \end{subfigure}
    \caption{Difference in continental coverage after 4.5 Gyr for models (light blue and purple in Figs. \ref{fig:3dplot} and \ref{fig:3dplot_insul}) that differ in initial mantle temperature by 300 K  as a function of $\xi_4$ for dominating negative (left) and positive (right) feedback, respectively. The value of $\xi_3$ is kept fixed at 0.85.}
    \label{fig:spread_xi4}
\end{figure*}

In Fig. \ref{fig:3dplot_insul}, we used a combination of values for $\xi_3$ and $\xi_4$ within our range of estimates that maximises the effect of thermal blanketing. For values of $\xi_4 > 0.25$, depletion of heat sources works increasingly against the effect of thermal blanketing such that for $\xi_3 = 0.8$ and $\xi_4 = 0.4$, the results are close to those neglecting blanketing as in Figs. \ref{fig:3dplot} and \ref{fig:interior}. In Fig. \ref{fig:spread_xi4}, we show the difference of the continental surface fraction at 4.5 Gyr resulting from models with a difference in the initial mantle temperature of 300 K (i.e. between the light blue and purple curves in Figs. \ref{fig:3dplot} and \ref{fig:3dplot_insul}) for dominating negative (left) and positive (right) feedback. As expected, the spread in continental surface area at 4.5 Gyr decreases with increasing $\xi_4$ but the spread can be twice as large for cases with dominating negative feedback and $\xi_4=0.25$. 

\begin{figure}
   \centering
  \includegraphics[clip,width=\columnwidth]{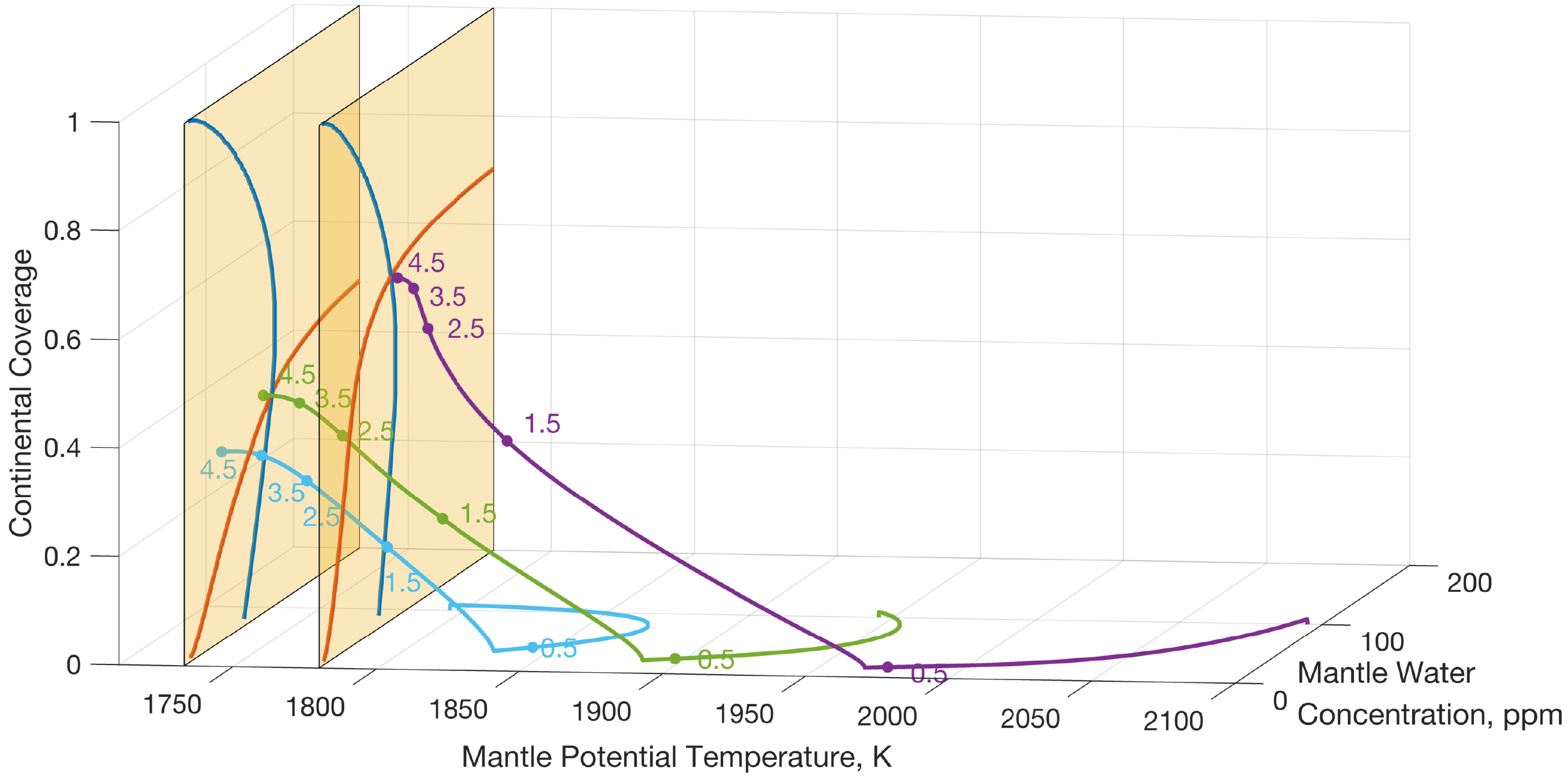}%
  
  \includegraphics[clip,width=\columnwidth]{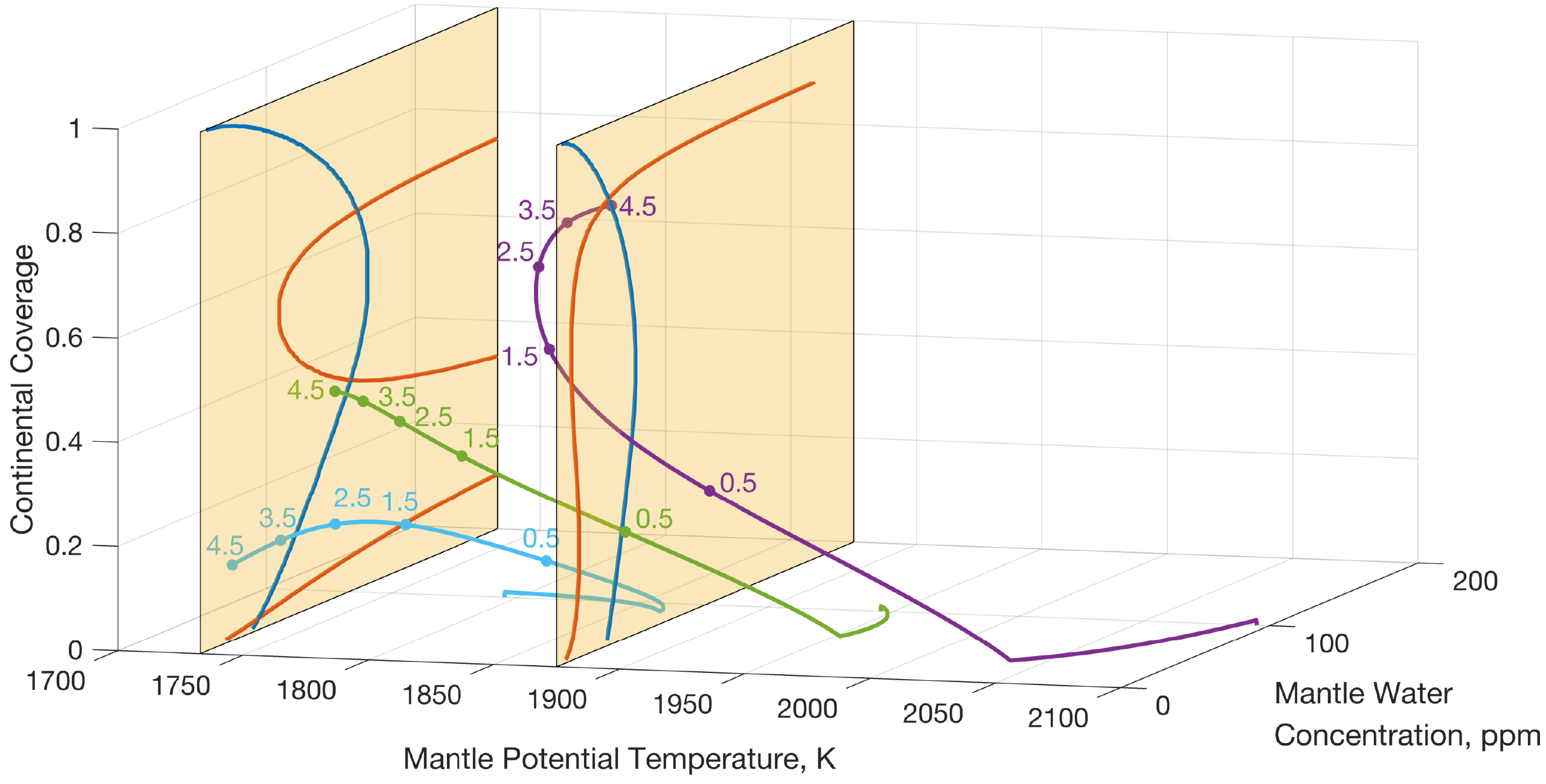}%
   \caption{An extreme case of thermal blanketing with $\xi_3$=1 and  $\xi_4$=1/4. Top panel: $\xi_1$=1/3, $\xi_2$=2/3, $f_r$=1, onset time $t_p$ 0.748 Gyr; bottom: $\xi_1$=2/3, $\xi_2$=1/3, $f_r$=1, onset time $t_p$ 0.0774 Gyr.}
\label{fig:insul1}
\end{figure}

Long-term heating of the mantle can only be obtained if the thermal blanketing is made extreme. In Fig. \ref{fig:insul1}, we show a case with $\xi_3=1$ implying perfectly insulating continents. Here, with dominating positive feedback for a high initial mantle temperature, we obtain heating even though the depletion of radiogenic elements in continental crust is taken into account with $\xi_4$=0.25.

All in all, we conclude that the reduction of cooling by thermal blanketing underneath growing continental crust is partly, potentially largely compensated by the effective cooling of the mantle associated with the transfer of radiogenic elements from the mantle to the crust.

\subsection[CO$_2$ outgassing and weathering]{CO$_2$ outgassing, weathering, surface temperature and bioproductivity}
\label{disc_prop}

Coupling the continental erosion rate and the plate speed to a model of the long-term carbon cycle \ref{sec:carboncycle}, we find that the large spread in continental surface coverage between our models results in differences of surface temperatures of not more than about 5 $K$. The reason for this rather small difference are opposing effects of increasing continental coverage on the atmospheric CO$_2$ concentration: On the one hand, the weatherability and thereby the CO$_2$ sink increases with increasing continental coverage. On the other hand, the mantle remains hotter with larger, insulating continents, which causes an enhanced outgassing rate of CO$_2$. Note that our ocean planet is a shallow water world with an ocean depth of about 4 km for which the carbonate silicate cycle is still operative \cite[e.g.,][]{honing:2019b,Hayworth:2020}, even though the negative feedback associated with silicate weathering would be weaker. It should not be confused with a deep ocean water world with 10 - 1000 Earth ocean masses for which the carbonate-silicate cycle might not work at all due to the formation of high-pressure ice layers and potentially limited magmatic degassing \citep[e.g.,][]{noack:2016,Kite:2018,krissansen2021}.

A 5 K surface temperature difference is considerably smaller than the excursions of the Earth's mean surface temperature over the Phanerozoic \citep[e.g.,][]{Scotese:2016} and might be assumed to be of limited significance for the general discussion of exoplanet habitability. However, \cite{Lingam:2018} and \cite{Schulze-Makuch:2020} have argued that a 5 K higher surface temperature would significantly increase the habitability of an Earth-like world. In combination with the differences in land and water surfaces it could have an impact on climatic zones. Earth climate modelling suggests that the temperature difference on land surfaces, in particular near the equator, may be twice as large \citep{nazarenko:2022}. Furthermore, the polar ice sheets will shrink with increasing temperature. In contrast, the lower surface temperature of the land planet would allow for substantial polar ice caps. These differences should be reflected in the sea-level, which would also differ on account of the non-linear thermal expansion of seawater \citep[e.g.,][]{Widlansky:2020}. In  addition, there may be dry cold deserts in the interiors of large continents. For the present-day Earth, 40\% of the surface is continental crust but a quarter of it (10\% of the Earth's surface) is covered with water in shelf areas, leaving 30\% land surface. Our ocean planet has a continental surface of about 20\% but given the warmer climate predicted a significant part of it may be covered by water, leaving perhaps only 10\% of land surface. On the contrary, the land planet with about 70\% continental surface area could have a land surface twice that of Earth. All in all, we expect the ocean planet to be comparatively warm and wet and the land planet comparatively dry and cold. Future work with comprehensive Earth System and Global Circulation Models is needed in order to better understand the impact of land area on the climate of Earth-like planets together with a varying CO$_2$ concentration in the atmosphere.

We can estimate the combined effects of the land-ocean distribution and the surface temperature on the bioproductivity and the biomass by using the models of \cite{lingam2019} for the effect of the ratio of land to ocean surfaces and of \cite{Lingam:2018} for the surface temperature. While the former model weighs the land and ocean areas based on their differing specific bioproductivities, the latter model considers the temperature effect on the metabolic rate being governed by an activation energy. We again scale to the present Earth and find the land planet to have a net bioproductivity (NPP) of about 43$\%$ of Earth's, corrected down from 50$\%$ when accounting for the about 1.7 K lower surface temperature. The model with thermal blanketing would have a slightly larger productivity of 47$\%$ of Earth's due to the smaller surface temperature correction. The larger land area will lead to a predicted biomass of 67$\%$ of Earths, corrected down from 76$\%$, obtained when the temperature effect is neglected.  The ocean planet has a net bioproductivity of 60$\%$ of Earth's, up from 49$\%$ when accounting for the higher surface temperature. It has a biomass of 53$\%$ of Earth's, up from 43$\%$, again accounting for the higher surface temperature. The indirect effects of thermal blanketing are too small to be relevant for this model. Further using \cite{lingam2019}, the land and the ocean planet would be predicted to both have only a marginal capability of building up O$_2$ for advanced life forms in their atmospheres.

Undoubtedly, the biosphere on worlds with different land fractions would adapt to their environment, and a balanced ratio of ocean-to-land accompanied by large continental shelves is certainly beneficial for an Earth-like biosphere \citep[e.g.,][]{Kallmeyer:2012,Honing:2016,lingam2019,stevenson2021}. Vice versa, life on Earth co-evolved with the environment, shaping the atmosphere and oceans \citep{grenfell2010,lenton2014,algeo2016}. The emergence of pelagic calcifiers in the deep oceans has been argued to have played a role in the stability of Earth's climate \citep{ridgwell2003} and the colonisation of land by plants enhancing continental weathering has presumably been crucial in keeping our planet's biosphere flourishing   in the long term \citep{berner1998,schwartzman2017}. In combination, life on land and in the oceans has been shown to help stabilising Earth's climate on timescales $>$100 kyr with land plants dominating on timescales $>$1 Myr and marine calcifiers being predominantly important on shorter timescales \citep{honing2020}. This particular relative importance of life on land and in the oceans is a direct result of Earth's balanced ratio of oceans to land, however: On planets with a larger continental coverage, land plants would become increasingly important for climate stability, whereas the role of marine calcifiers would become minor \citep{honing2020}.

\section{Conclusions}
\label{sec8}

In this paper, we assessed the coupled evolution of continental growth, mantle water cycling, and thermal evolution of Earth-like planets. In particular, we explored how  differences in initial mantle temperature and the strength of positive and negative feedback would impact their continental growth curves and water budgets. We coupled the model to a model of the carbonate-silicate cycle via the plate speed and the continental surface area to assess the habitability and the bioproductivity of these planets. Our results can be summarised as follows:

\begin{itemize}
\item[$\bullet$] Positive feedback in the coupled continent-water cycle may lead to the emergence of multiple fixed points and to substantially differing land and ocean coverage, with an Earth-like distribution of land and ocean surfaces being the least likely. The average depth of the oceans would be similar, around 4 km, because differences in the ocean surface will be compensated by differences in the distribution of water between the mantle and the surface reservoirs. The final continental coverage and the environment of the planet will depend on the initial conditions, in particular on the initial mantle temperature and the onset time of plate tectonics. If early mechanisms of continental growth differ from plate tectonics then other differences in initial conditions may still matter. An evolution largely independent of the early conditions is possible only if negative feedback mechanisms prevail. 
\item[$\bullet$] 
Positive feedback is mostly provided by the role of sediments in continental growth. The rate of sediment subduction will increase with the continental surface area and the additional water transported will enhance the continental growth rate. In addition, sediments lubricate the subducting plate and facilitate efficient subduction. Thermal blanketing of the mantle by continents may retard mantle cooling, keep an enhanced continental production rate and may cause an even larger difference in final continental coverage. But the effects of thermal blanketing is partly, potentially largely, balanced by the removal of heat sources from the mantle going along with crust growth and with an enrichment of radiogenic elements in the continental crust.
\item[$\bullet$] The carbonate-silicate cycle is coupled to the evolution of the interior through the carbon outgassing rate depending on the plate speed and to continental growth via the weathering rate increasing with the continental surface area. We found differences in the mean global surface temperature between the land planet and the ocean planet of up to $\sim$ 5 K. The climate of all three planets as well as their biospheres may have evolved quite differently, however. The warmer ocean planet's potentially higher specific bioactivity may be limited by the smaller availability of critical nutrients from weathering its smaller land surface. The cooler land planet may have a smaller specific bioactivity further reduced by the limited availability of water but its bioactivity would be spread over a larger area. By using the scalings of \cite{Lingam:2018, lingam2019} we find the bioproductivity and the potential biomass on these planets reduced by a third to half of Earth's. It is questionable whether a biosphere on these planets would be substantial enough to produce more oxygen then is consumed by geologic activity.
\item[$\bullet$] Main model uncertainties are related to quantifying the role of sediments in continental production and to estimating the depletion of radiogenic elements by continental crust production. Future work on these issues is required in order to improve understanding the complexities of our own planet and the prediction of the continental fraction on Earth-like exoplanets.
\end{itemize}

\section*{Conflict of interest}
The authors declare that they have no conflict of interest.

\acknowledgments
We thank two anonymous reviewers for helpful comments and suggestions. We also thank Hendrik Hansen-Goos for valuable insights into stochastic geometry and Thibaut Roger for designing figure\,\ref{fig:illustration}. Part of this research was done during a visitor scientists stay of D.H. at ISSI. Support by ISSI is gratefully acknowledged.

\newpage
\begin{appendix}
\label{sec:appendix}

\section{Coupled continental and mantle water cycles}
\label{continents_water}
Key variables in the model are the surface fraction of continental crust $A$ and its rate of change $\dot{A}$, which depends on the rates of continental production $\dot{A}_g$ (index $g$ for \textit{gain}) and continental erosion $\dot{A}_l$ (index $l$ for \textit{loss})
\begin{linenomath}
\begin{equation}
   \dot{A}=\dot{A_g}-\dot{A_l}.
   \label{eq:V}
\end{equation}
\end{linenomath}
Analogously, for the mass rate of change of the mantle water concentration, we have
\begin{linenomath}
\begin{equation}
   \dot{w}=\dot{w_g}-\dot{w_l}.
   \label{eq:w}
\end{equation}
\end{linenomath}
Continental production depends both on the plate speed and on the rate of sediment subduction, which in turn depends on the continental area (see Section \ref{sec:disc_xi}). A most simple parameterization of continental production that captures the key physics of the system is given by
\begin{linenomath}
\begin{equation}
   \dot{A}_{g}^*=\left(\xi_1 A^*+(1-\xi_1)v_{plate}^*\right)L_{oo,oc}^*,
   \label{eq:r}
\end{equation}
\end{linenomath}
where the asterisk denotes parameters scaled to the present-day Earth, for example 
\begin{linenomath}
\begin{equation}
   \dot{A}_{g}^*=\frac{\dot{A}_{g}}{\dot{A}_{g,E}},
   \label{eq:asterisk}
\end{equation}
\end{linenomath}
and where the index $E$ denotes present-day Earth values. In Eq. \ref{eq:r}, a fraction $\xi_1$ of the continental production rate $\dot{A}_{g}$ is taken proportional to the continental area whereas the rest is taken proportional to the plate speed (derived in Section \ref{sec:interior}). We account for the fact that subduction can only occur at those convergent plate boundaries that are not of continent-continent type by including the function $L_{oo,oc} (A)=L_{oo}(A)+L_{oc}$(A), where $L_{oo}(A)$ and $L_{oc}(A)$ are the total lengths of ocean-ocean type subduction zones and ocean-continent type subduction zones, respectively (see \ref{subduction_zones}).

An essential part of continental erosion takes place on the surface. The rate of surface erosion increases with the surface area but depends, in addition, on the surface relief, the rate of rainfall, and the weatherability of the surface. Of these, we keep for simplicity the dependence on the surface area. There are other mechanisms of continental erosion of which we include subduction erosion for which the subducting plate scrapes off part of the overlying continent. The rate of subduction erosion mainly depends on the total length of ocean-continent type convergent plate boundaries. It would also depend on a friction coefficient but since we scale to the present Earth and assume that parameters are constant the friction coefficient will cancel.  The subduction speed may matter but for plate speeds higher than the present one, the dependence is likely small \citep{stern:2011}. Thus, 
\begin{linenomath}
\begin{equation}
   \dot{A}_l^*=\xi_2 A^*+(1-\xi_2)L_{oc}^*,
   \label{eq:Vl}
\end{equation}
\end{linenomath}
where $\xi_2$ is a constant. 

We assume that the subducted water flux (regassing) to the mantle follows the same functional dependence on plate speed, sediment production and length of subduction zones as the continental production rate (see also \citealt{Karlsen:2019}). This is a reasonable first-order approach, since the production rate of continental crust rock is taken directly proportional to the water subduction rate. The regassing rate should, of course, differ from the water subduction rate as water is consumed upon melting and lost through dehydration. However, we assume for simplicity that the ratio between the two remains constant. The relative mass rates -- scaled to their respective present-day Earth values -- are, therefore, equal:
\begin{linenomath}
\begin{equation}
   \dot{w}_{g}^*=\dot{V}_{g}^*.
   \label{eq:wg}
\end{equation}
\end{linenomath}
Mantle degassing, the release of water to the combined atmosphere-ocean reservoir depends, to first order, on the rate at which mantle material is transported into the melting region beneath mid-ocean ridges and other volcanic units and on the mantle water concentration. We neglect degassing through volcanism in other geologic provinces. 
\begin{linenomath}
\begin{equation}
   \dot{w}_L^*=w^*v_p^*,
   \label{eq:wl}
\end{equation}
\end{linenomath}
where $v_p^*$ is the plate speed scaled by its present-day value.

Equations \ref{eq:continent} and \ref{eq:water} together with Eqs. \ref{eq:V} -- \ref{eq:asterisk} and \ref{eq:Vl} need values of $\dot{A}_{l,E}$ and $\dot{A}_{g,E}$ and $\dot{w}_{l,E}$ and $\dot{w}_{g,E}$ for rescaling. Whereas the exact present-day rates may be of secondary importance, qualitative differences can result from either assuming an equilibrium or disequilibrium between present-day Earth sources and sinks. Even though a present-day equilibrium appears reasonable to assume, we cannot exclude that the sluggish Earth system only slowly converges towards an equilibrium state. We therefore introduce
\begin{linenomath}
\begin{equation}
   f_r=\frac{\dot{A}_{l,E}}{\dot{A}_{g,E}}=\frac{\dot{w}_{l,E}}{\dot{w}_{g,E}},
   \label{eq:xi3}
\end{equation}
\end{linenomath}
and note that a present-day Earth equilibrium implies $f_r=1$.

With Equations \ref{eq:V} -- \ref{eq:xi3}, the coupled continent--water system can qualitatively be described. A combination of these equations result in Eqs. \ref{eq:continent} and \ref{eq:water}. Particularly important are the constants $\xi_1$ and $\xi_2$, which determine the strengths of positive and negative feedback, respectively.

\section{Length of subduction zones}
\label{subduction_zones}

\begin{figure}
    \centering
    \includegraphics[width=0.5\textwidth]{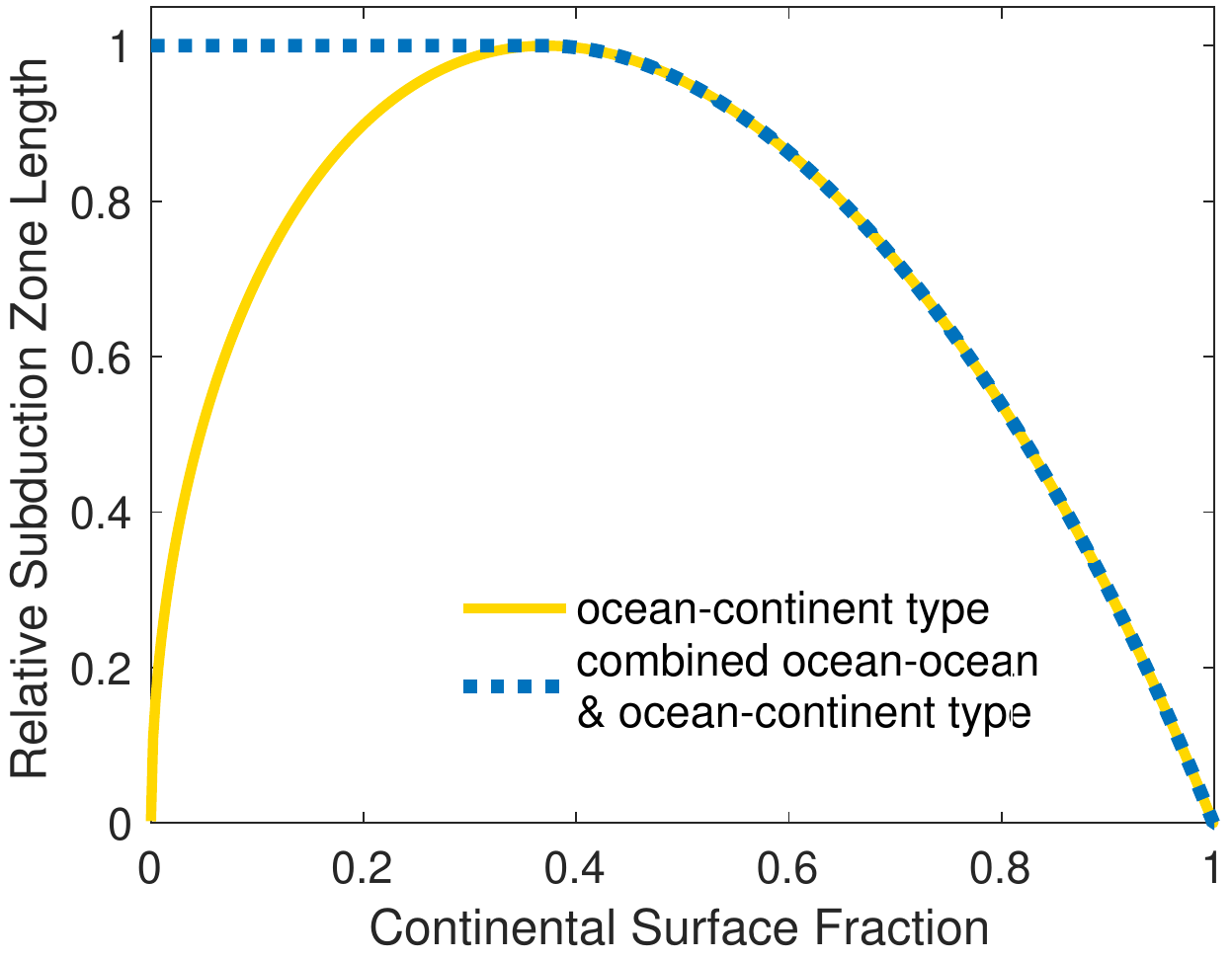}
    \caption{Total length of ocean-continent (yellow) and combined ocean-ocean plus ocean-continent (blue dotted) type subduction zones scaled to the maximum length of subduction zones.}
    \label{fig:loc}
\end{figure}

Calculating the total length of subduction zones, we follow \cite{Honing:2014} and assume that the total length of ocean-continent type subduction zones is proportional to the total length of continental margins ($L_{oc}\sim L_{mar}$). Modeling continents as  randomly distributed spherical caps that may overlap, we find
\begin{linenomath}
\begin{equation}
  \label{eq:Loc}
  \frac{L_{mar}}{4 \pi R_E}= n(1-A)\sqrt{\left(\frac{1}{n}\text{ln}\left[\frac{1}{1-A}\right]\right)-\left(\frac{1}{n}\text{ln}\left[\frac{1}{1-A}\right]\right)^2},
\end{equation}
\end{linenomath}
where $L_{mar}$ is the length of continental margins, $R_E$ is the radius of Earth. For $n$ we use the number of continents on Earth of  $n=7$. \cite{Honing:2016} have shown that $L_{mar}/L_{mar,max}$ shown in Fig. \ref{fig:loc} will not vary significantly with $n$ for $5 \leq n \leq 9$.

Eq. \ref{eq:Loc} has been derived as follows: Assume $n$ spherical caps, each with area $A_0$ and circumference $L_0$, Poisson randomly distributed on a sphere with surface area $A_{surf}$. For a sufficiently large number $n$, the area covered by spherical caps $A_c$ is \citep{Mecke:2000,Schneider:2008}
\begin{linenomath}
\begin{equation}
   \frac{A_c}{A_{surf}}=1-e^{-\nu},
   \label{eq:SW1}
\end{equation}
\end{linenomath}
where $\nu=n\cdot\frac{A_0}{A_{surf}}$, and the total length of margins $L_{mar}$ is
\begin{linenomath}
\begin{equation}
   \frac{L_{mar}}{n}=L_0 e^{-\nu}.
   \label{eq:SW2}
\end{equation}
\end{linenomath}
Using the definition $A=\frac{A_c}{A_{surf}}$, Eq. \ref{eq:SW1} yields
\begin{linenomath}
\begin{equation}
   \frac{A_0}{A_{surf}}=\frac{1}{n} \ln\left(\frac{1}{1-A}\right),
   \label{eq:SW3}
\end{equation}
\end{linenomath}
and combining Eq. \ref{eq:SW1} and Eq. \ref{eq:SW2} yields
\begin{linenomath}
\begin{equation}
   \frac{L_{mar}}{n}=L_0 (1-A).
   \label{eq:SW4}
\end{equation}
\end{linenomath}
Furthermore, consider the radius of the base of the spherical cap $r_c$ and its height $h_c$. The radius of the sphere $R_E$ is then
\begin{linenomath}
\begin{equation}
   R_E^2=r_c^2+(R_E-h_c)^2.
   \label{eq:SW5}
\end{equation}
\end{linenomath}
Since $L_0=2\pi r_c$ and $A_0=2\pi R_E h_c$, we get
\begin{linenomath}
\begin{equation}
   L_0=2\pi\sqrt{R_E^2-\left(R_E-\frac{A_0}{2\pi R_E}\right)^2}.
   \label{eq:SW6}
\end{equation}
\end{linenomath}
Combining Eq. \ref{eq:SW4} and Eq. \ref{eq:SW6} yields
\begin{linenomath}
\begin{equation}
\begin{split}
   \frac{L_{mar}}{n}&=2\pi R_E (1-A) \sqrt{1-\left(1-\frac{A_0}{2\pi R_E^2}\right)^2}\\&=4\pi R_E (1-A) \sqrt{\frac{A_0}{A_{surf}}-\frac{A_0^2}{A_{surf}^2}},
   \label{eq:SW7}
\end{split}
\end{equation}
\end{linenomath}
and combining Eq. \ref{eq:SW3} and Eq. \ref{eq:SW7} results in Eq. \ref{eq:Loc}.

With increasing continental area, continents increasingly overlap, which in turn reduces the total length of subduction zones. Following our assumption of $L_{oc}\sim L_{mar}$, the total length of ocean-continent-type subduction zones has its maximum at a continental surface fraction of $\approx0.37$. For a smaller (larger) fraction we assume that the remaining length of subduction zones is of ocean-ocean (continent-continent) type (Fig. \ref{fig:loc}).

The derivative of Eq. \ref{eq:Loc} with respect to A is
\begin{linenomath}
\begin{equation}
\label{eq:derivLoc}
\frac{d}{dA}\frac{L_{mar}}{4\pi R_E} =  \dfrac{n + 2 \left(n+1\right)\ln\left(1-A\right)+ \ln^2\left(1-A\right)}{\sqrt{-\ln^2\left(1-A\right)- n\ln\left(1-A\right)}}
\end{equation}
\end{linenomath}

\section{Interior thermal evolution}
\label{sec:interior}

The calculation of the thermal evolution of the mantle using Eq. \ref{eq:temperature_new} is based on the conventional thermal history models of terrestrial planets \citep[e.g.,][]{Schubert:2001} and includes terms describing the heat flow though the surface, heat production  within the mantle, and the heat flow from the core.

For a constant (present-day Earth) continental crust coverage $A_E$, the mantle heat production rate can be calculated following \cite{korenaga:2008} 
\begin{linenomath}
\begin{equation}
\label{eq:heatprod}
Q_{A_E}(t)=Q_0\sum_i q'_i \exp\left\lbrace-\frac{\log(2)}{h'_i}(t-4.5)\right\rbrace,
\end{equation}
\end{linenomath}
where $t$ is time in Gyr, $Q_0$ is the present-day Earth mantle heat production rate and $q'_i$ and $h'_i$ are the relative present-day heat production rates and half-lifes of the individual radiogenic elements $^{238}$U, $^{235}$U, $^{232}$Th, and $^{40}$K. Eq. \ref{eq:heatprod} accounts for radiogenic decay but not for the transfer of radiogenic elements to the continental crust upon growth of the latter.  
We account for the depletion of the mantle due to continental crust growth by writing
\begin{linenomath}
\begin{equation}
\label{eq:fd}
\frac{Q_m (A,t)}{Q_{A_E}}=1+\xi_4(1-A^*),
\end{equation}
\end{linenomath}
where $Q_m (A,t)$ is the mantle radiogenic heat production rate as a function of time and continental coverage and where  
\begin{linenomath}
\begin{equation}
\label{eq:xi_4}
\xi_4 = \frac{Q_{C_E}}{Q_{A_E}+Q_{C_E}}
\end{equation}
\end{linenomath}
with Q$_{C_E}$ the present-day heat production rate in the continental crust. 
In a scenario where we neglect the transfer of radiogenic elements, we set $\xi_4=0$.

The mantle heat flow is given by
\begin{linenomath}
\begin{equation}
\label{eq:qm}
q_m = A_{surf} k \frac{T_u-T_s}{\delta_u},
\end{equation}
\end{linenomath}
where $T_u$ and $T_s$ are the temperatures of the upper mantle and the surface, respectively, $k$ is the thermal conductivity, $A_{surf}$ is the surface area of Earth, and $\delta_u$ is the thickness of the upper boundary layer and given by
\begin{linenomath}
\begin{equation}
\label{eq:bld}
\delta_u = \left(\frac{\kappa \eta Ra_c}{g \alpha_x \rho_m \Delta T}\right)^\beta,
\end{equation}
\end{linenomath}
where $\kappa$, $g$, $\alpha_x$, $\rho$, are the thermal diffusivity, gravitational acceleration, thermal expansivity, and density of the mantle, $Ra_c$ is the critical Rayleigh Number, $\Delta T$ is the temperature difference across the boundary layer, and $\eta$ is the mantle viscosity  given by
\begin{linenomath}
\begin{equation}
\label{eq:visc}
\eta_m=f_w^{-1} \, \eta_0 \, \text{exp}\left(\frac{E_a+p_m V_a}{R_g} \left(\frac{1}{T_m}-\frac{1}{T_{ref}}\right)\right),
\end{equation}
\end{linenomath}
where $\eta_0$ is a reference viscosity, $E_a$ and $V_a$ are the activation energy and volume, respectively, $R_g$ is the universal gas constant, $T_m$ and $p_m$ the temperature and pressure half-way through the mantle, $T_{ref}$ a reference temperature, and $f_w$ accounts for the dependence of the mantle viscosity on the concentration of water $w$. $f_w$ is given by
\begin{linenomath}
\begin{equation}
\label{eq:fw}
f_w=\text{exp}\left(c_0+c_1 \text{log}(w)+c_2 \text{log}^2(w)+c_3 \text{log}^3(w)\right),
\end{equation}
\end{linenomath}
where $c_0$ through $c_4$ are constants. Following boundary layer theory \citep[e.g.,][]{Schubert:2001}, the plate speed is a function of the boundary layer thickness
\begin{linenomath}
\begin{equation}
\label{eq:vp}
v_p = 5.38 \kappa \frac{d_m}{\delta^2},
\end{equation}
\end{linenomath}
where $d_m$ is the thickness of the mantle.
Scaled to the present-day Earth, the mantle heat flow can, therefore, be written as
\begin{linenomath}
\begin{equation}
q_m = q_{m,E} \Delta T_u^{* (1+\beta)} \sqrt{v_p^*},
\end{equation}
\end{linenomath}
where $\Delta T_u=T_u-T_s$ and
\begin{linenomath}
\begin{equation}
\label{eq:q_me}
q_{m,E}=A_{surf} \frac{k \Delta T_{u,E}}{\delta_{u,E}}.
\end{equation}
\end{linenomath}

The parameterisation of the effects of thermal blanketing by the continental crust will complete the derivation of  Eq. \ref{eq:temperature_new} and is introduced in \ref{sec:blanketing}. 

\section{Continental insulation}
\label{sec:blanketing}

Considering thermal blanketing of the mantle by (partly) insulating continents, we begin by  defining $\chi$ as the ratio of the present-day specific heat flow from the mantle through continental crust to that through oceanic crust
\begin{linenomath}
\begin{equation}
\chi \equiv \frac{q_{c}/A}{q_{o}/(1-A)} = \frac{q_{c,E}/A_E}{q_{o,E}/(1-A_E)} .
\end{equation}
\end{linenomath}
where $q_o$ is the heat flow from the mantle through the oceanic crust and $q_{c}$ the heat flow from the mantle through the continental crust. We assume that $\chi$ remains constant throughout the evolution as it depends on the heat transfer properties of the oceanic and continental crust rather than on the thermal state of the mantle. We acknowledge, however, that $\chi$ may depend on the areal extent of the crust and the ratio between continental margins and old stable continental shield areas. 

The heat flow from the mantle to the surface $q_s$ can then be written as
\begin{linenomath}
\begin{equation}
\label{eq:q_s}
q_s=q_o+q_c=q_{m,E} \left((1-A)+A\chi\right) \Delta T_u^{*(1+\beta)} \sqrt{v_p^*},
\end{equation}
\end{linenomath}
The mantle heat flow  $q_{m,E}$ has been defined by Eq. \ref{eq:q_me}.

Since by definition
\begin{linenomath}
\begin{equation}
\label{eq:def xi3}
\xi_3\equiv\frac{q_{o,E}}{q_{o,E}+q_{c,E}},
\end{equation}
\end{linenomath}
or $q_{o,E}=\xi_3 q_{s,E}$ and $q_{c,E}=(1-\xi_3)q_{s,E}$, we can write 
\begin{linenomath}
\begin{equation}
\chi=(1/A_E-1) (1/\xi_3-1)=1/(A_E \xi_3) - 1/A_E - 1/\xi_3 +1.
\end{equation}
\end{linenomath}
For the surface heat flow (Eq. \ref{eq:q_s}), we get
\begin{linenomath}
\begin{equation}
q_s=q_{m,E} \left(1+A^*/\xi_3 - A^* - A/\xi_3 \right) \Delta T_u^{*(1+\beta)} \sqrt{v_p^*}.
\end{equation}
\end{linenomath}
The present-day Earth surface heat flow is analogously:
\begin{linenomath}
\begin{equation}
q_{s,E}=q_{m,E}\left((1-A_E)+A_E\chi\right)=q_{m,E}\left(\frac{1-A_E}{\xi_3}\right).
\end{equation}
\end{linenomath}
The relative heat surface heat flow is then
\begin{linenomath}
\begin{equation}
\begin{split}
\frac{q_s}{q_{s,E}}&=\Delta T_u^{*(1+\beta)} \sqrt{v_p^*} \frac{ \left(1+A^*/\xi_3 - A^* - A/\xi_3 \right)  }{\left(\frac{1-A_E}{\xi_3}\right)}\\&=\Delta T_u^{*(1+\beta)} \sqrt{v_p^*} \frac{\left(\xi_3+A^* - A^*\xi_3 - A \right)  }{\left(1-A_E\right)}\\&=\Delta T_u^{*(1+\beta)} \sqrt{v_p^*} \left[(1-A)^*\xi_3+A^*(1-\xi_3)\right] .
\end{split}
\end{equation}
\end{linenomath}

\section{Long-term carbon cycle}
\label{sec:carboncycle}
The plate speed determines the flux of mantle material into the melting region beneath mid-ocean ridges and thereby the mass rate of carbon outgassing from the mantle into the atmosphere
\begin{linenomath}
\begin{equation}
\label{eq:F_out}
F_{out}=\frac{R_{man}}{V_m}f_{deg} 2 v_p L_{mor} d_{melt},
\end{equation}
\end{linenomath}
where $R_{man}$ is the carbon reservoir in the mantle, $V_m$ the mantle volume, $f_{deg}$ the fraction of upwelling mantle material that degasses, $L_{mor}$ the total length of mid-ocean ridges, and $d_{melt}$ the depth below mid-ocean ridges where melting begins.

The weathering flux is calculated by  \citep{Foley:2015}
\begin{linenomath}
\begin{equation} \label{eq:F_w}
    F_w=F_{ws}\left(1-\exp \left(-f_{land}^*\frac{F_{w,E}}{F_{ws}}\left(P_{CO_2}^*\right)^b \left(P_{sat}^*\right)^a \exp\left(\frac{E_{a,s}}{R_g}\left(\frac{1}{T_{s,E}}-\frac{1}{T_s}\right)\right)\right)\right),
\end{equation}
\end{linenomath}
where $P_{CO_2}$ is the partial pressure of CO$_2$ in the atmosphere, $P_{sat}$ is the saturation pressure, $E_{a,s}$ is the activation energy of silicate weathering, $T_s$ is the surface temperature, $f_{land}$ is the land fraction, and $F_{ws}$ is the supply limit to weathering. Since we explicitly calculate surface erosion in this model by setting it proportional to the surface area of continental crust, we use $f_{land}^*=A^*$. We note that this is a significant simplification in order to obtain first-order results without introducing additional equations. In reality, the emerged land fraction would also depend on additional parameters such as topography and surface water budget. The supply limit to weathering is given by
\begin{linenomath}
\begin{equation} \label{eq:F_ws}
    F_{ws}=\frac{A_{surf} f_{land}E_{max}\psi\rho_r}{\overline{m}_c},
\end{equation}
\end{linenomath}
where $E_{max}$ is the maximum physical erosion rate (in thickness per time), $\psi$ is the fraction of reactable cations, $\overline{m}_c$ is its average molar mass, and $\rho_r$ is the density of regolith \citep{Foley:2015}.

Seafloor weathering is calculated as \citep{Sleep:2001,Foley:2015}
\begin{linenomath}
\begin{equation}
\label{eq:F_sfw}
    F_{sfw}^*=v_p^* \left(P_{CO_2}^*\right)^\alpha.
\end{equation}
\end{linenomath}
Following classic long-term carbon cycle models \citep{Sleep:2001,Foley:2015}, the weathering fluxes transfer carbon from the atmosphere to the subduction zones considering half of that carbon is released during carbonate precipitation on the seafloor. Of the carbon that is subducted, we assume that another half is returned to the mantle while the rest is degassed at volcanic arcs back into the atmosphere.

The saturation pressure is calculated as
\begin{linenomath}
\begin{equation}
\label{eq:P_sat}
    P_{sat}=P_{sat0}\exp\left(-\frac{m_w L_w}{R_g}\left(\frac{1}{T_s}-\frac{1}{T_{s0}}\right)\right),
\end{equation}
\end{linenomath}
where $P_{sat0}$ and $T_{s0}$ are reference values for the saturation pressure and surface temperature and $m_w$ and $L_w$ are the molar mass and latent heat of water, respectively. The surface temperature is calculated as
\begin{linenomath}
\begin{equation}
\label{eq:T_s}
   T_s=T_{s,E}+2(T_e-T_{e,E})+4.6\left(\left(P_{CO_2}^*\right)^{0.346}-1\right),
\end{equation}
\end{linenomath}
where $T_e$ is the effective temperature and given by
\begin{linenomath}
\begin{equation}
\label{eq:T_e}
   T_e=\left(\frac{S(1-\lambda)}{4\sigma}\right)^{\frac{1}{4}},
\end{equation}
\end{linenomath}
where $\lambda$ is the planetary albedo, $\sigma$ is the Stefan-Boltzmann constant, and $S$ is the solar irradiation, which is assumed to increase by a factor of 1/3 throughout 4.5 Gyr \citep{Gough:1981}. Parameter values are given in Tables \ref{tab1}, \ref{tab2}, and \ref{tab3}.

\section{Parameter Values}
\label{sec:tables}

\begin{table}
\caption[]{Parameter values used in the model. Carbon cycle parameters from \cite{Foley:2015}.}
\label{tab1}
\centering
\begin{tabular}{l l l l}
\hline
Symbol & Description & Value\\
\hline
$c_p$ & Specific heat capacity & 1200 J kg$^{-1}$ K$^{-1}$ \\
$V_m$ & Mantle volume & $9.1\cdot10^{20}$ m$^3$ \\
$k$ & Thermal conductivity & 4.2 W m$^{-1}$ K$^{-1}$ \\
$\kappa$ & Thermal diffusivity & 10$^{-6}$ m$^2$ s$^{-1}$ \\
$Ra_c$ & Critical Rayleigh Number & 1100 \\
$g$ & Gravitational acceleration & 9.81 \\
$\alpha_x$ & Thermal expansivity & $2\cdot10^{-5}$ \\
$d_m$ & Mantle thickness & 2900 km \\
$L_{mor}$ & Total length mid-ocean ridges & 6$\cdot10^7$ m \\
$d_{melt}$ & Maximum depth of melting beneath mid-ocean ridges & 120 km \\
$A_{surf}$ & Earth surface area  & 5.1$\cdot10^{14}$ m$^2$ \\
$E_a$ & Activation energy (viscosity) & 335 kJ mol$^{-1}$ \\
$V_a$ & Activation volume (viscosity) & 4 cm$^3$ mol$^{-1}$\\
$R_g$ & Universal gas constant & 8.314 J mol$^{-1}$ K$^{-1}$ \\
$R_g$ & Stefan-Boltzmann constant & 5.67$\cdot10^{-8}$ W m$^{-2}$ K$^{-4}$ \\
$E_{a,s}$ & Activation energy (weathering)  & 42 kJ mol$^{-1}$\\
$E_{max}$ & Maximum physical erosion rate & 0.01 m yr$^{-1}$ \\
$\rho_r$ & Density regolith & 2500 kg m$^{-3}$ \\
$\psi$ & Fraction reactable cations & 0.08 \\
$\overline{m}_c$ & Average molar mass reactable cations & 32 g mol$^{-1}$ \\
$m_w$ & Molar mass of water & 44 g mol$^{-1}$ \\
$L_w$ & Latent heat of water & 2469 J g$^{-1}$ \\
$f_{deg}$ & Fraction of upwelling mantle that degasses & 0.32 \\
\hline
\end{tabular}
\end{table}

\begin{table}
\caption[]{Present-day Earth parameters and reference parameters used in the model. F15: \cite{Foley:2015}}
\label{tab2}
\centering
\begin{tabular}{l l l l}
\hline
Symbol & Description & Value & Reference \\
\hline
$P_{CO_2,E}$ & Partial pressure of CO$_2$ & 33 Pa \\
$T_{s,E}$ & Surface temperature & 285 K \\
$S_E$ & Solar irradiation & 1360 W m$^{-2}$ \\
$T_{ref}$ & Reference temperature (Eq. \ref{eq:visc}) & 2050 K & \\
$\eta_0$ & Reference viscosity (Eq. \ref{eq:visc}) & 10$^{21}$ Pa s & \\
$P_{sat0}$ & Reference saturation pressure (Eq. \ref{eq:P_sat}) & 610 Pa & F15\\
$T_{s0}$ & Reference surface temperature (Eq. \ref{eq:P_sat}) & 273 K & F15\\
$q_{m,E}$ & Mantle heat flow (Eq. \ref{eq:temperature_new}) & 35 TW & $A_{surf}\frac{k \Delta T_{u,E}}{\delta_{u,E}}$ \\
\hline
\end{tabular}
\end{table}

\begin{table}
\caption[]{Scaling constants used in the model. F15: \cite{Foley:2015}, L08: \cite{li2008}}
\label{tab3}
\centering
\begin{tabular}{l l l l}
\hline
Symbol & Equation & Value & Reference\\
\hline
$\alpha$ & Eq. \ref{eq:F_sfw} & 0.25 & F15 \\
$\beta$ & Eq. \ref{eq:bld} & 1/3 &\\
$a$ & Eq. \ref{eq:F_w} & 0.3 & F15\\
$b$ & Eq. \ref{eq:F_w} & 0.55 & F15\\
$c_0$ & Eq. \ref{eq:fw} & -7.9859 & L08\\
$c_1$ & Eq. \ref{eq:fw} & 4.3559 & L08\\
$c_2$ & Eq. \ref{eq:fw} & -0.5742 & L08\\
$c_3$ & Eq. \ref{eq:fw} & 0.0227 & L08\\
\hline
\end{tabular}
\end{table}
\end{appendix}
\newpage

\bibliography{references}

\end{document}